\documentclass[review,authoryear,3p,times,10pt]{elsarticle}

\usepackage{multirow,setspace,times,amssymb,amsmath,graphicx,color,rotating,subfigure,url}
\usepackage{lineno,color}
\usepackage{natbib}
\usepackage{booktabs}%
\usepackage{longtable}%
\usepackage{rotating}
\usepackage{lscape}
\usepackage{ulem}
\usepackage{threeparttable}
\graphicspath{{Figures/}}
\usepackage[table]{xcolor}
\usepackage{tabularx}
\usepackage{graphicx} 
\usepackage{epstopdf}
\usepackage{mathrsfs}
\usepackage{makecell}
\usepackage[bookmarks=true,colorlinks,linkcolor=blue,anchorcolor=blue,citecolor=blue,unicode]{hyperref}
\usepackage{bookmark}

\usepackage{todonotes}
\usepackage{ragged2e}
\usepackage[font=small]{caption}

\hypersetup{CJKbookmarks=true}%
\makeatletter
\def\ps@pprintTitle{%
\let\@oddhead\@empty 
\let\@evenhead\@empty
}

\begin{document}

\begin{frontmatter}

\title{Multiscale risk spillovers and external driving factors: Evidence from the global futures and spot markets of staple foods}

\author[SB,UVSQ,EMLV]{Yun-Shi Dai}
\author[WHUT1,WHUT2]{Peng-Fei Dai}
\author[UVSQ]{St{\'e}phane Goutte}
\author[EMLV]{Duc Khuong Nguyen}
\author[SB,RCE,DM]{Wei-Xing Zhou\corref{CorAuth}}
\ead{wxzhou@ecust.edu.cn}
\cortext[CorAuth]{Corresponding author.} 

\address[SB]{School of Business, East China University of Science and Technology, Shanghai 200237, China}
\address[UVSQ]{University Paris-Saclay, UMI SOURCE, UVSQ, IRD, France}
\address[EMLV]{De Vinci Research Center, L{\'e}onard de Vinci P{\^o}le Universitaire, Paris La D{\'e}fense, France}
\address[WHUT1]{School of Management, Wuhan University of Technology, Wuhan 430070, China}
\address[WHUT2]{Research Institute of Digital Governance and Management Decision Innovation, Wuhan University of Technology, Wuhan 430070, China}
\address[RCE]{Research Center for Econophysics, East China University of Science and Technology, Shanghai 200237, China}
\address[DM]{School of Mathematics, East China University of Science and Technology, Shanghai 200237, China}

\begin{abstract}
Stable and efficient food markets are crucial for global food security, yet international staple food markets are increasingly exposed to complex risks, including intensified risk contagion and escalating external uncertainties. This paper systematically investigates risk spillovers in global staple food markets and explores the key determinants of these spillover effects, combining innovative decomposition-reconstruction techniques, risk connectedness analysis, and random forest models. The findings reveal that short-term components exhibit the highest volatility, with futures components generally more volatile than spot components. Further analysis identifies two main risk transmission patterns, namely cross-grain and cross-timescale transmission, and clarifies the distinct roles of each component in various net risk spillover networks. Additionally, price drivers, external uncertainties, and core supply-demand indicators significantly influence these spillover effects, with heterogeneous importance of varying factors in explaining different risk spillovers. This study provides valuable insights into the risk dynamics of staple food markets, offers evidence-based guidance for policymakers and market participants to enhance risk warning and mitigation efforts, and supports the stabilization of international food markets and the safeguarding of global food security.
\end{abstract}


\end{frontmatter}

\section{Introduction}

Food is the cornerstone of national stability and the foundation of human survival. However, global and local food security is becoming increasingly precarious, with numerous nations and regions grappling with mounting risks and challenges. According to {\textit{The State of Food Security and Nutrition in the World 2024}}\footnote{\url{https://openknowledge.fao.org/items/d8f47624-8b43-412a-bbc2-18d2d830ad5b}}, published by the Food and Agriculture Organization of the United Nations, an estimated 9.1\% of the global population remained undernourished in 2023, and approximately 2.33 billion people experienced moderate or severe food insecurity. The report underscores the intensifying severity of key challenges, including armed conflicts, extreme weather events, and economic slowdowns. Furthermore, persistent structural issues, such as increasing food costs, heightened food price volatility, and pronounced regional imbalances in supply and demand, exacerbate the current food crisis. These adverse factors are not only occurring with greater frequency and intensity but are also interacting and compounding, resulting in a growing number of people affected by hunger and food insecurity across the globe. 

Addressing global food insecurity has thus become an urgent priority, with the stability and efficiency of food markets playing a pivotal role. Futures and spot markets for agricultural commodities are critical components of the global food system. They facilitate global food trade and production through price discovery and risk management functions while simultaneously influencing food security through resource allocation and transmission of external shocks \citep{Li-Chavas-2023-AmJAgrEcon}. Futures prices, reflecting market expectations of future supply and demand conditions, are vital indicators for global food trade and policy-making, as they promptly capture external changes impacting supply and demand \citep{Noussair-Tucker-Xu-2016-JEconBehavOrgan,Arzandeh-Frank-2019-AmJAgrEcon}. Spot prices, on the other hand, directly represent current market conditions, guiding decisions on trade, storage, and consumption. However, the increasing interconnections among agricultural commodities and the deepening financialization of staple crops have exposed food futures and spots to significant market volatility and systemic risk \citep{Basak-Pavlova-2016-JFinanc,Bianchi-Fan-Todorova-2020-IntRevFinancAnal}. In particular, risks associated with major staple foods, including wheat, maize, soybean, and rice, may have far-reaching implications for both food security in vulnerable regions and the stability of the world economy.

A critical concern lies in developing a comprehensive understanding of how risks transmit within international staple crop markets, spanning food futures and spots, across different commodities, and over varying timescales. These spillover effects are considered to magnify market instability, disrupt supply chains, and increase uncertainties for producers, traders, and policymakers \citep{Huynh-Burggraf-Nasir-2020-ResourPolicy, DeJong-Sonnemans-Tuinstra-2022-JEconBehavOrgan}. Additionally, external shocks and internal drivers, such as climate policy uncertainty, energy prices, and transportation costs, have the potential to exacerbate these risk spillover effects \citep{Wright-2011-ApplEconPerspectPolicy,Zmami-BenSalha-2023-AgricEcon}. Therefore, examining the impact of these possible factors on different dimensions of risk spillovers within the global food market provides valuable insights for developing effective risk-hedging strategies, formulating targeted food policies and regulations, and identifying and forecasting market risks to safeguard global and local food security.

This paper systematically investigates risk contagion within the international food market and identifies the underlying determinants of spillover effects. By employing a novel decomposition-reconstruction methodology, futures and spot returns of four staple grains—wheat, maize, soybean, and rice—are decomposed into short-, medium-, and long-term components. The findings reveal that short-term components exhibit the highest volatility, with short- and medium-term components generally fluctuating around long-term components, and futures return components are more volatile than spot return components. Risk connectedness analysis uncovers two main patterns of risk transmission: stronger spillovers occur between same-timescale components of different grains and between same-grain components at different timescales. Both static and dynamic analyses consistently demonstrate significant and robust cross-grain and cross-timescale spillovers, highlighting the distinct roles of various components in different spillover networks. Notably, the level of risk transmission in global food markets has shown a marked upward trend in recent years. The examination of influencing factors confirms that the selected price drivers, external uncertainties, and core supply-demand indicators significantly affect these risk spillovers, serving as reliable determinants of risk contagion. Furthermore, the key drivers of different types of risk transmission vary substantially, emphasizing the importance of tailored strategies for effective risk mitigation.

From a dual perspective of cross-grain and cross-timescale, this study utilizes advanced and integrative methods to clarify the mechanisms of risk transmission and identify critical drivers of risk spillovers in the global staple food market. Unlike previous research, which predominantly relies on original prices or returns \citep{Dai-Dai-Zhou-2023-JIntFinancMarkInstMoney,Zhu-Dai-Zhou-2024-JFuturesMark}, this paper innovatively introduces decomposition and reconstruction techniques to reveal the intrinsic characteristics of grain futures and spot returns, thereby proposing a novel analytical framework for examining risk contagion across multiple dimensions. Furthermore, this study delves into the heterogeneous roles of diverse components in shaping risk spillovers of different dimensions, offering crucial insights into the risk dynamics of the global staple crop submarkets. The findings contribute to determining underlying sources of risk transmission, developing effective early-warning systems, and formulating targeted risk management strategies, as well as providing valuable guidance for policymakers, investors, and market participants.

The remainder of this paper is organized as follows: Section~\ref{S1:LitRev} reviews related literature; Section~\ref{S1:Methodology} outlines the research methodology; Section~\ref{S1:Data} provides the data sources and statistical description; Section~\ref{S1:EmpAnal} presents and discusses the empirical results; and Section~\ref{S1:Conclude} concludes with policy implications.

\section{Literature review}
\label{S1:LitRev}

The phenomenon of risk transmission across financial markets or assets, often referred to as risk spillover, has garnered significant academic attention. With the deepening globalization and integration of financial markets, the interconnections between different markets have intensified, enabling localized risk events to propagate rapidly through various channels, potentially triggering systemic financial crises \citep{Bekaert-Harvey-Ng-2005-JBus}. Notable instances of risk contagion during the 2008 global financial crisis and the 2020 COVID-19 pandemic underscore the importance of understanding risk spillover effects. Against this backdrop, research on risk spillovers has become a critical focus in finance, exploring transmission pathways within complex financial systems \citep{Reboredo-RiveraCastro-Ugolini-2016-JBankFinanc,Finta-Aboura-2020-JFinancMark,Tian-Alshater-Yoon-2022-EnergyEcon} and identifying their driving factors \citep{Li-Wei-2018-EnergyEcon,Ji-Liu-Zhao-Fan-2020-IntRevFinancAnal}.

Risk spillover measures have evolved from traditional linear analysis to sophisticated dynamic modeling, reflecting the growing complexity of financial markets. Early studies relied on classic statistical tools, such as Granger causality tests and linear regression models, to examine causality and correlation in time series \citep{Engle-Ito-Lin-1990-Econometrica}. While straightforward, these methods struggled to capture asymmetric and nonlinear relationships, as well as dynamic risk transmission. To address these limitations, advanced methods emerged, including the CoVaR approach proposed by \cite{Adrian-Brunnermeier-2016-AmEconRev} and the dynamic connectedness framework by \cite{Diebold-Yilmaz-2009-EJ,Diebold-Yilmaz-2012-IntJForecast}. CoVaR, concerning tail risks, measures the additional risk stemming from interdependence between markets. It is recognized as an effective risk management tool and has been widely adopted in finance \citep{Abuzayed-Bouri-AlFayoumi-Jalkh-2021-EconAnalPolicy}. However, CoVaR is based on the bilateral estimation of variable interdependence, whereas the generalized connectedness framework allows for capturing risk transmission dynamics across multiple time series simultaneously.

The connectedness method, leveraging generalized forecast error variance decomposition within vector autoregressive models, innovatively proposes return and volatility spillover indices \citep{Diebold-Yilmaz-2009-EJ,Diebold-Yilmaz-2012-IntJForecast}. With the matrix generated by the forecast error variance decomposition, this approach constructs a total connectedness index (measuring overall risk spillovers of the system) and directional connectedness indices (capturing risk transmission between individuals in the system), and combines the rolling window to dynamically analyze risk spillovers. Widely applied to multi-asset and cross-market studies, the connectedness framework provides a unified theoretical basis for investigating risk spillovers \citep{Chen-Zheng-Hao-2022-JBusRes,Diebold-Yilmaz-2023-JEconom,BenAmeur-Ftiti-Louhichi-2024-AnnOperRes}.

Recent developments have addressed limitations in the original connectedness approach, such as standardization issues and interpretability challenges \citep{Balli-Balli-Dang-Gabauer-2023-FinancResLett}. The frequency-domain connectedness analysis, for instance, differentiates between short-term shocks and long-term volatility \citep{Barunik-Krehlik-2018-JFinancEconom}. The high-dimensional connectedness method based on sparse networks improves computational efficiency in scenarios with numerous variables \citep{Giglio-Kelly-Pruitt-2016-JFinancEcon}. Additionally, the $R^{2}$ decomposed connectedness approach developed by \cite{Naeem-Chatziantoniou-Gabauer-Karim-2024-IntRevFinancAnal} enhances model interpretability and avoids biases introduced by standardization techniques, offering a robust depiction of risk transmission pathways. These innovations have facilitated the integration of novel factors into connectedness analysis. For example, \cite{Zhang-Hong-Ding-2023-Energy} incorporate climate policy uncertainty into the framework, revealing heterogeneous relationships between oil prices and clean energy market volatility. The flexibility of connectedness models has enabled dynamic analysis, high-dimensional data processing, and multidisciplinary applications, providing insights into risk propagation mechanisms within complex financial systems.

The increasing financialization of commodities has heightened price volatility and market linkages in agricultural markets \citep{Tang-Xiong-2012-FinancAnalJ,Basak-Pavlova-2016-JFinanc,Bianchi-Fan-Todorova-2020-IntRevFinancAnal}. Given the key role of agricultural markets in ensuring global food security, research on risk spillovers in agricultural markets has gained prominence. Existing studies primarily focus on the interplay between agricultural and other commodity markets. \cite{Ji-Bouri-Roubaud-Shahzad-2018-EnergyEcon} point out that energy market spillovers significantly exacerbate risk exposures in agricultural markets. Further analysis shows that risk spillovers from the oil market to the agricultural market are particularly pronounced during the financial crisis \citep{Kumar-Tiwari-Raheem-Hille-2021-ResourPolicy} and exhibit notable asymmetries in upside and downside risks \citep{Hanif-Hernandez-Shahzad-Yoon-2021-QRevEconFinanc}. Research based on static and dynamic connectedness indicates that agricultural futures are more susceptible to shocks, and that inter-commodity connectedness intensifies during turbulent periods \citep{Xiao-Yu-Fang-Ding-2020-JFuturesMark}. Studies incorporating time- and frequency-domain analyses further reveal the heterogeneous spillover effects between energy, agricultural raw materials, and food markets, with energy markets exerting the most significant influence on food markets \citep{Adeleke-Awodumi-2022-JApplEcon}. Extreme spillover effects among energy, metals, and agricultural products have also been investigated using quantile connectedness methods \citep{Iqbal-Bouri-Grebinevych-Roubaud-2023-AnnOperRes}.

A small body of literature has attempted to explore the factors influencing price volatility in the food market. \cite{Headey-Fan-2008-AgricEcon} attribute the 2007–2008 global food price crisis to factors such as rising biofuel demand, climate-induced yield reductions, energy price hikes, declining global inventories, and export-restriction policies. \cite{Gilbert-Morgan-2010-PhilosTransRSocB-BiolSci} discuss the impact of supply-demand fundamentals, policy changes, and oil price volatility on food price fluctuations. \cite{Wright-2011-ApplEconPerspectPolicy} emphasizes the importance of stock levels as a buffer against price volatility and suggests that policy uncertainty and extreme weather events are also key drivers of increased price volatility. \cite{Zmami-BenSalha-2023-AgricEcon} examine the influences of oil prices, fertilizer prices, economic activities, and geopolitical risks, finding that geopolitical risks are the most critical factor affecting all food products. In addition, recent studies have paid particular attention to the exacerbating effects of extreme events on risk spillovers, such as the COVID-19 pandemic \citep{Cao-Cheng-2021-ResourPolicy} and the Russia–Ukraine conflict \citep{Zhou-Dai-Duong-Dai-2024-JEconBehavOrgan}.

While existing studies have explored various aspects of risk spillovers in agricultural markets, several key gaps remain. First, the transmission of risks across grain markets at different time scales is an understudied area. Time-scale analysis, such as empirical mode decomposition, has proven effective in examining short-, medium-, and long-term market dynamics in financial markets \citep{Luo-Liu-Wang-2021-NAmEconFinanc,Cheng-Li-Cui-Wei-Wang-Hong-2024-IntRevFinancAnal}, but its application to grain markets remains underexplored. Second, the ability of potential drivers to explain risk spillovers of different dimensions, along with their relative importance, has received limited attention. This study aims to address these gaps by systematically examining the risk contagion in international staple food markets from a dual perspective of cross-grain and cross-timescale, as well as identifying the determinants of various risk spillovers by quantifying the respective impacts of price drivers, external uncertainties, and core supply-demand indicators.

\section{Methodology}
\label{S1:Methodology}

\subsection{Empirical mode decomposition} 

Empirical mode decomposition (EMD), initially introduced by \cite{Huang-Shen-Long-Wu-Shih-Zheng-Yen-Tung-Liu-1998-PRSA}, is a signal decomposition method designed to handle nonlinear and non-stationary data. With stabilizing, a given signal can be decomposed by the EMD method into a time-frequency spectrogram consisting of physically significant frequencies for further analysis. Other widely-used signal decomposition methods, such as wavelet decomposition and Fourier decomposition, come with certain constraints, primarily because they rely on predefined basis functions, which must be chosen or constructed to match the signal characteristics. EMD, in contrast, adapts to the intrinsic timescale features of the data, making it theoretically applicable to any type of signal.

The core concept behind EMD is the Hilbert-Huang Transform (HHT). The fundamental principle of HHT involves extracting the original time series step by step, based on different time periods, to generate several data sequences with distinct characteristic scales. Each of these sequences can be regarded as an intrinsic mode function (IMF). To qualify as an IMF, a sequence must satisfy two specific conditions. First, the number of extrema must be equal to, or differ by at most one from, the number of zero crossings throughout the entire sequence. Second, at any given time, the mean value of the upper and lower envelopes that bound the sequence must always be zero. Typically, an IMF represents an oscillatory mode similar to that of a simple harmonic function.

According to \cite{Huang-Shen-Long-Wu-Shih-Zheng-Yen-Tung-Liu-1998-PRSA}, any complex time series $x(t),\ t=1,\dots,T$ can be decomposed into a series of IMFs through a process known as sifting, which involves the following procedures. First, identify all the local extrema of $x(t)$, including both local maxima and minima. Second, use an interpolation method to construct the envelope lines $emax(t)$ and $emin(t)$, respectively, and calculate their mean values by
$m(t)=\left(emax(t)+emin(t)\right)/2$. Third, subtract $m(t)$ from the original signal $x(t)$ to obtain an intermediate signal $d(t)$, and check if $d(t)$ satisfies the two conditions of being an IMF. If it does, then $d(t)$ is classified as an IMF. If not, use $d(t)$ to replace $x(t)$ and repeat the above steps. Fourth, once an IMF is identified, subtract it from $x(t)$ and continue the decomposition process until the predefined stopping criterion is met. The remaining part of the signal, $u(t)$, should be a monotonic residue.

The stopping criterion for extracting IMFs is based on iterating a predefined number of times after the residual sequence meets the requirement that the number of extrema and zero crossings differ by no more than one. The sifting process can be terminated by any of the following predetermined conditions: when the IMF or residue becomes sufficiently small, falling below a threshold of substantive significance, or when the residue becomes monotonic that it is impossible to extract any more IMFs.

By executing the sifting process, the original signal $x(t)$ can be decomposed into a sum of multiple IMFs and a residue, given by
\begin{equation}
    x(t) = \sum_{n=1}^{N} c_{n}(t) + u(t),\ t=1,\dots,T,
    \label{Eq:function_EMD}
\end{equation}
where $N$ is the number of extracted IMFs, and $c_{n}$ and $u$ represent the $n$-th IMF and the residue, respectively.

In EMD, two commonly used interpolation methods are Hermite interpolation and cubic spline interpolation. Hermite interpolation offers good convergence and stability, but it cannot guarantee smoothness across all connection points along the curve. In contrast, cubic spline interpolation provides a higher degree of smoothness due to its continuous second derivative. Consequently, in this study, we employ cubic spline interpolation for data decomposition.

Although EMD offers significant advantages over traditional time-frequency analysis techniques, it also faces certain challenges in practical applications, such as mode mixing, sifting stoppage criteria, and end effect reduction \citep{Cheng-Yu-Yang-2006-MechSystSignalProc,Hu-Peng-Hwang-2012-IEEETransSignalProcess}. To address these issues, numerous enhanced EMD-based models have been developed, with one of the latest being the improved complete ensemble EMD with adaptive noise (ICEEMDAN) proposed by \cite{Colominas-Schlotthauer-Torres-2014-BiomedSignalProcessControl}. Let $M(\cdot)$ denote an operator that computes the local mean of a signal through local envelope averaging. The steps of the ICEEMDAN algorithm are as follows:

First, the first IMF $E_{1}(w^{i})$ obtained by EMD decomposition of white noise is added to the original signal $x$, denoted as $x^{i} = x + \beta_{0}E_{1}(w^{i})$, and its local envelope average $M(x^{i})$ is computed. Through ensemble averaging, the first residue can be obtained as follows:
\begin{equation}
    u_{1} = \frac{1}{I} \sum\limits_{i=1}^{I} M(x^{i}),
    \label{Eq:function_ICEEMDAN_residue1}
\end{equation}
where $w^{i}$ is a white noise sequence with zero mean and unit variance, and $I$ denotes the number of trials with added noise. $\beta$ represents the noise amplitude, typically set as a multiple of the signal's standard deviation, given by $\beta = \varepsilon \times \text{std}(x)$. In most cases, $I$ is set to 100, and $\varepsilon$ is taken to be 0.2.

Second, the first IMF of the original signal is calculated by
\begin{equation}
    \tilde{c}_{1} = x - u_{1}.
    \label{Eq:function_ICEEMDAN_IMF1}
\end{equation}

Third, the second IMF $E_{2}(w^{i})$ obtained by EMD decomposition of white noise is added to $u_{1}$, that is, $u_{1} + \beta_{1}E_{2}(w^{i})$. Then, its local envelope average and ensemble average are computed to obtain the second residue $u_{2}$. The second IMF of the original signal can be derived by subtracting the second residue from the first residue as follows:
\begin{equation}
    \tilde{c}_{2} = u_{1} - u_{2} = u_{1} - \frac{1}{I} \sum\limits_{i=1}^{I}  M\left (u_{1} + \beta_{1}E_{2}(w^{i}) \right).
    \label{Eq:function_ICEEMDAN_IMF2}
\end{equation}

Fourth, the $n$-th residue $u_{n}$ is computed, and the $n$-th IMF of the original signal is obtained by subtracting $u_{n}$ from $u_{n-1}$ as follows:
\begin{equation}
    \tilde{c}_{n} = u_{n-1} - u_{n} = u_{n-1} - \frac{1}{I} \sum\limits_{i=1}^{I} M\left (u_{n-1} + \beta_{n-1}E_{n}(w^{i}) \right).
    \label{Eq:function_ICEEMDAN_IMFn}
\end{equation}

Fifth, the above steps are repeated until the stopping criterion is satisfied. After extracting all IMFs, the original signal $x$ can be expressed as
\begin{equation}
    x = \sum\limits_{n=1}^{N}\tilde{c}_{n} + u_{N},
    \label{Eq:function_CEEMDAN_signal}
\end{equation}
where $N$ is the number of IMFs, and $u_{N}$ represents the final residue.

By incorporating local envelope averaging, ICEEMDAN reduces residual noise and sequentially calculates IMFs, ensuring complete decomposition and minimizing reconstruction error.

\subsection{Mode reconstruction method} 

After decomposing the data into multiple independent intrinsic mode functions and a residual term using the ICEEMDAN method, we improve the K-means clustering of the run-length number (KMC-RLN) reconstruction method developed by \cite{Ding-Chen-Zhou-Wang-2022-ExpertSystAppl}, and propose the Gaussian mixture model of the run-length number (GMM-RLN) reconstruction method. This enhanced approach is utilized to perform mode reconstruction.

The run-length number is an indicator of the volatility of a sequence, with higher values implying greater volatility. The procedures for calculating the run-length number of the $n$-th IMF are as follows. First, compute the mean of the $n$-th IMF sequence, denoted as $m_{n}$. Next, classify the observations in the IMF sequence as either ``$-$'' or ``$+$'' based on whether they are less than or greater than $m_{n}$, creating a symbolic sequence. Finally, define consecutive sequences of the same symbol as a single run, thereby obtaining the run-length number for the $n$-th IMF.

With the run-length number for each decomposed IMF, we apply the Gaussian mixture model for clustering. The steps for GMM clustering are as follows. First, determine the number of clusters and randomly initialize the parameters of each Gaussian distribution, including mean, covariance, and mixing coefficients. Second, calculate the posterior probability that each point belongs to each cluster with the current parameter estimates, where the closer the point is to the center of the Gaussian distribution of a cluster indicates that the more likely it is for the point to belong to this cluster. Third, optimize the parameters by the expectation maximization algorithm, in which the parameters of each Gaussian distribution are recalculated based on posterior probabilities. This process iterates until the likelihood of the observed data is maximized, thereby maximizing the probability of each point within its assigned cluster.

Compared to existing reconstruction methods, our proposed GMM-RLN reconstruction method offers distinct advantages. For one thing, this approach performs reconstruction based on both data fluctuation characteristics and data length, which is objective and conducive to uncovering the evolution law and economic interpretability of the reconstructed components. For another, this method enhances flexibility and efficiency in clustering, as it allows for customizable cluster numbers and adapts to cases with fuzzy boundaries or clusters of various shapes. In this study, we set the number of clusters to 3 and, based on the GMM clustering results, reconstruct modes with higher run-length numbers as short-term (high-frequency) components, followed by medium-term (medium-frequency) and long-term (low-frequency) components.

\subsection{$R^{2}$ decomposed connectedness approach} 

Next, we utilize the $R^{2}$ decomposed connectedness approach proposed by \cite{Naeem-Chatziantoniou-Gabauer-Karim-2024-IntRevFinancAnal} to investigate risk transmission within the international futures and spot markets of staple foods.

A vector autoregression model (VAR) is first constructed and represented in terms of an infinite-order vector moving average (VMA) form:
\begin{subequations}
   \begin{equation}
    \mathbf{y}_{t} = \sum\limits_{i=0}^{p} \mathbf{B}_{i} \mathbf{y}_{t-i} + \mathbf{v}_{t} = \sum\limits_{i=0}^{\infty} \mathbf{A}_{i} \mathbf{v}_{t-i},\ v \sim N(0,\mathbf{\Sigma}),
   \label{Eq:function_VAR}
   \end{equation}
with
   \begin{equation}
   \mathbf{A}_{i} = \mathbf{B}_{1}\mathbf{A}_{i-1} + \mathbf{B}_{2}\mathbf{A}_{i-2} + \cdots + \mathbf{B}_{p}\mathbf{A}_{i-p},
   \label{Eq:function_VMA}
   \end{equation}
   \label{Eq:function_VAR_VMA}
\end{subequations}
where $p$ is the lag order of the VAR model. $\mathbf{y}_{t}$ and $\mathbf{v}_{t}$ represent the $k \times 1$ vectors of endogenous variables and error terms, respectively. $\mathbf{B}$, $\mathbf{A}$, and $\mathbf{\Sigma}$ denote the $k \times k$ VAR coefficient matrix, VMA coefficient matrix, and covariance matrix, respectively. $\mathbf{A}_{0}$ is the identity matrix, and $\mathbf{A}_{i}$=0 when $i<0$.

For forecasting $H$ steps ahead, the generalized forecast error variance decomposition (GFEVD) quantifies the directional risk spillover from variable $j$ to variable $i$, as follows:
\begin{equation}
    \phi_{i \leftarrow j}^{gen}(H) = \frac{\sum_{h=0}^{H-1} \left( \mathbf{e}_{i}^{\prime} \mathbf{A}_{h} \mathbf{\Sigma} \mathbf{e}_{j} \right)^{2} }{ \left( \mathbf{e}_{j}^{\prime} \mathbf{\Sigma} \mathbf{e}_{j} \right)\sum_{h=0}^{H-1} \left( \mathbf{e}_{i}^{\prime} \mathbf{A}_{h} \mathbf{\Sigma} \mathbf{A}_{h}^{\prime} \mathbf{e}_{i} \right) },
    \label{Eq:function_GFEVD}
\end{equation}
where $\mathbf{e}_{i}$ represents a $k \times 1$ selection vector, in which the $i$-th element is 1 and all other elements are 0. When assuming all variables follow a random walk, the GFEVD, denoted as $\phi_{i \leftarrow j}^{gen}(H)$, simplifies to:
\begin{equation}
    \phi_{i \leftarrow j}^{gen} = \left( \frac{ \mathbf{\Sigma}_{ij} }{ \sqrt{\mathbf{\Sigma}_{jj} \mathbf{\Sigma}_{ii}} } \right)^{2} = R_{ij}^{2},
    \label{Eq:function_GFEVD_R2}
\end{equation}
where $R_{ij}^{2}$ denotes the goodness-of-fit measure $R^{2}$ from a bivariate regression model between variable $i$ and variable $j$, hence $R_{ii}^{2}=1$ and $R_{ij}^{2}=R_{ji}^{2}$. It is noteworthy that, under these assumptions, the GFEVD no longer varies with the forecast horizon $H$.

According to \cite{Naeem-Chatziantoniou-Gabauer-Karim-2024-IntRevFinancAnal}, normalizing GFEVD by dividing by each row sum to address cases where the row sum may exceed 1 is suboptimal, because it may overestimate or underestimate the influence of variable $j$ on variable $i$. Therefore, we adopt the $R^{2}$ decomposed connectedness approach, which uses a multivariate regression in place of $k-1$ bivariate regressions to evaluate the impact of other variables on variable $i$, yielding a goodness-of-fit measure $R^{2}$ constrained between 0 and 1.

The estimation of $k$ multivariate linear regression models can be expressed as:
\begin{equation}
    \mathbf{y}_{t} = \mathbf{B} \mathbf{x}_{t} + \mathbf{v}_{t},\ v \sim N(0,\mathbf{\Sigma}),
    \label{Eq:function_MLR}
\end{equation}
where $\mathbf{B}$ is a $k \times k$ coefficient matrix with zeros on the diagonal.

In general, when all right-hand side variables in a multivariate linear regression model are uncorrelated, the goodness-of-fit measure $R^{2}$ is equal to the sum of the corresponding $R^{2}$ values from all bivariate linear regression models. Therefore, to measure the influence of variable $j$ on variable $i$, we employ $R^{2}$ decomposition and construct $k$ orthogonal variables based on the right-hand side variables using principal component analysis. The calculation of $R^{2}$ decomposition for each multivariate linear regression model is as follows:
\begin{subequations}
\begin{equation}
    \mathbf{R}_{xx} = \mathbf{V} \mathbf{\Lambda} \mathbf{V}^{\prime} = \mathbf{R}_{xz}\mathbf{R}_{xz}^{\prime},
    \label{Eq:function_R2_MLR_Rxx}
\end{equation}
\begin{equation}
    \mathbf{R}_{xz} = \mathbf{V} \mathbf{\Lambda}^{1/2} \mathbf{V}^{\prime},
    \label{Eq:function_R2_MLR_Rxz}
\end{equation}
\begin{equation}
    \mathbf{R}^{2G} = \mathbf{R}_{xz}^{2} \left( \mathbf{R}_{xz}^{-1} \mathbf{R}_{yx} \right)^{2},
    \label{Eq:function_R2_MLR_R2G}
\end{equation}
\label{Eq:function_R2_MLR}
\end{subequations}
where $\mathbf{V}$ and $\mathbf{\Lambda}=\text{diag}\left( \lambda_{1}, \dots, \lambda_{k(p+1)-1} \right)$ are the eigenvectors and eigenvalues of the correlation matrix of the right-hand side variables $\mathbf{R}_{xx}$, respectively. $\mathbf{R}_{yx}$ and $\mathbf{R}_{xz}$ represent the correlation matrices between the left-hand side variables and the right-hand side variables, and between the right-hand side variables and the orthogonal variables, respectively. $\mathbf{R}^{2G}$ denotes the $R^{2}$ contribution vector, with its sum equal to the goodness-of-fit measure $R^{2}$ of the corresponding multivariate linear regression model.

We replace GFEVD with $R_{ij}^{2G}$ and calculate the connectedness indices proposed by \cite{Diebold-Yilmaz-2012-IntJForecast,Diebold-Yilmaz-2014-JEconom}, including total directional connectedness from variable $i$ to all other variables ($TO_{i}$), total directional connectedness into variable $i$ from all other variables ($FROM_{i}$), net total directional connectedness for variable $i$ ($NET_{i}$), net pairwise directional connectedness from variable $j$ to variable $i$ ($NPDC_{ij}$), and the total connectedness index ($TCI$). These indices allow us to examine the risk transmission and spillover effects between variables, and their respective expressions are given by
\begin{subequations}
\begin{equation}
    TO_{i} = \sum_{j=1}^{k}R_{ji}^{2G},
    \label{Eq:function_R2_TO}
\end{equation}
\begin{equation}
    FROM_{i} = \sum_{j=1}^{k}R_{ij}^{2G},
    \label{Eq:function_R2_FROM}
\end{equation}
\begin{equation}
    NET_{i} = TO_{i} - FROM_{i},
    \label{Eq:function_R2_NET}
\end{equation}
\begin{equation}
    NPDC_{ij} = R_{ij}^{2G} - R_{ji}^{2G},
    \label{Eq:function_R2_NPDC}
\end{equation}
\begin{equation}
    TCI = \frac{1}{k} \sum_{i=1}^{k} R_{i}^{2},
    \label{Eq:function_R2_TCI}
\end{equation}
\label{Eq:function_R2_Connectedness_Measures}
\end{subequations}
where $NET_{i}>0$ ($NET_{i}<0$) indicates that variable $i$ is a net transmitter (receiver) of risk, which means that the risk spillover from variable $i$ to others is larger (smaller) than the risk spillover into variable $i$ from others. Similarly, $NPDC_{ij}>0$ ($NPDC_{ij}<0$) suggests that the risk spillover from variable $j$ to variable $i$ is greater (less) than the risk spillover in the opposite direction. In other words, variable $j$ explains a larger proportion of the variation in variable $i$ than variable $i$ explains the variation in variable $j$. Besides, the TCI index measures the average risk spillover in the network and is considered a proxy for market risk.

\subsection{Random forest model}

Random forest (RF), first proposed by \cite{Breiman-2001-MachLearn}, is a non-parametric statistical method designed for both classification and regression problems. Due to its outstanding predictability and flexibility, as well as less restriction on data properties, random forest has become one of the primary methods in machine learning and has been widely applied across diverse fields \citep{Athey-Tibshirani-Wager-2019-AnnStat, Lundberg-Erion-Chen-DeGrave-Prutkin-Nair-Katz-Himmelfarb-Bansal-Lee-2020-NatMachIntell, Podgorski-Berg-2020-Science}.

The random forest model is an ensemble learning algorithm that realizes regression by combining multiple decision trees. The basic steps are as follows. First, a subset of samples is randomly selected as the training set. Second, a subset of features is randomly chosen as the feature set. Third, a decision tree is constructed based on the random training set and feature set, continuing until a predetermined number of leaf nodes is reached or further splitting is not possible. Fourth, the above steps are repeated to build multiple decision trees. Fifth, a new sample is input into each decision tree to get multiple predictions. Sixth, the final prediction is obtained by averaging the individual predictions. Accordingly, the output of the random forest model for the regression problem is expressed as
\begin{equation}
    Y = \frac{1}{K}\sum\limits_{k=1}^{K}f_{k}(x),
    \label{Eq:function_RF}
\end{equation}
where $K$ represents the number of decision trees and $f_{k}(x)$ denotes the output of the $k$-th decision tree.

Furthermore, the commonly used metrics, namely mean absolute error (MAE) and mean square error (MSE), are adopted to assess the prediction accuracy, where smaller values indicate better predictive performance. The expressions for MAE and MSE are as follows:
\begin{equation}
    MAE = \frac{1}{Q}\sum\limits_{q=1}^{Q} \mid y_{q}-\hat{y}_{q} \mid,
    \label{Eq:function_RF_MAE}
\end{equation}
\begin{equation}
    MSE = \frac{1}{Q}\sum\limits_{q=1}^{Q} \left( y_{q}-\hat{y}_{q}\right)^{2}.
    \label{Eq:function_RF_MSE}
\end{equation}
where $Q$ denotes the sample size, and $y_{q}$ and $\hat{y}_{q}$ refer to the true value and the predicted value of the $q$-th sample, respectively.

We apply the random forest machine learning algorithm to explore the relationship between the total connectedness index TCI of each risk spillover network and a series of explanatory variables. Referring to \cite{Wei-Gephart-Iizumi-Ramankutty-Davis-2023-NatSustain}, we divide the data into 80\% for the training set and 20\% for the test set, and employ the grid search method proposed by \cite{Bergstra-Bengio-2012-JMachLearnRes} to tune the hyperparameters in the model, including the number of decision trees, the maximum depth of the tree, and the number of features for splitting. Compared to traditional predictive models, the random forest model offers high accuracy in dealing with complex datasets and demonstrates robustness against noise, missing data, and overfitting. Moreover, this approach is capable of capturing nonlinear relationships among variables and evaluating the importance of each explanatory variable.

\section{Data description}
\label{S1:Data}

\subsection{Data sources}

We select four major staple crops—wheat, maize, soybean, and rice—as the focus of this study. These staples provide the majority of energy and nutrients essential for human survival, making it crucial to investigate their market risks in relation to food security. Considering the differing roles of futures and spot markets, we retrieve the daily closing prices of the continuous contracts for wheat, corn, soybean, and rough rice futures traded on the Chicago Board of Trade (CBOT) through the Wind database, and collect daily price indices for wheat, maize, soybean, and rice from the International Grains Council (IGC) website. CBOT agricultural futures prices are regarded as global pricing benchmarks for agricultural commodities, while IGC grain price indices reflect spot export quotations from major global producers. Thus, the two data sources together effectively capture the price movements in both futures and spot markets for these major staples. Since the IGC spot indices use January 3, 2000, as the base date, our sample period spans from January 3, 2000, to December 29, 2023.

In addition to grain price data, twelve variables closely related to food are chosen to investigate the potential drivers of risk spillovers within the international markets of staple foods. The global price of agricultural raw material index (ARMI) represents the benchmark prices of agricultural inputs such as seeds, fertilizers, pesticides, and agricultural machinery. The global price of food index (FPI) captures international trade prices for a basket of food commodities, including grains. The global price of energy index (EPI) reflects global prices for energy sources such as oil and natural gas, with existing literature highlighting the strong interconnection between energy and grain markets \citep{Ji-Bouri-Roubaud-Shahzad-2018-EnergyEcon, Tiwari-Abakah-Adewuyi-Lee-2022-EnergyEcon}. Grains are dry bulk commodities, and maritime transport is a critical channel for global food trade. The Baltic dry index (BDI), a key measure of dry bulk shipping costs, is regarded as a barometer of global economic activity and a leading indicator of commodity trends \citep{Lin-Sim-2013-EurEconRev, Lin-Chang-Hsiao-2019-TranspResPte-LogistTranspRev}. 

The geopolitical risk index (GPR) quantifies time-varying risks of adverse geopolitical events, providing a reliable measure of geopolitical instability and its fluctuations \citep{Caldara-Iacoviello-2022-AmEconRev}. The economic policy uncertainty index (EPU) tracks uncertainties surrounding economic policies in major economies worldwide \citep{Baker-Bloom-Davis-2016-QJEcon}. Similarly, the climate policy uncertainty index (CPU) reflects uncertainty related to climate policies, such as carbon taxes and emission reduction initiatives \citep{Gavriilidis-2021-SSRN}. The global natural disasters index (GND), derived from the emergency events database (EM-DAT), represents the monthly occurrence of climatological, hydrological, and meteorological disasters, which are relevant to agriculture. According to the FAO release, uncertainties such as geopolitical risk, economic situation, climate change, and natural disasters have a non-negligible impact on global food security by influencing grain production and trade. 

Additionally, the global total production (PROD), total imports (IMP), total domestic consumption (CONS), and total ending stocks (ES) of the four staple foods serve as statistical indicators of global grain production, supply, and demand. Of these variables, BDI and GPR are available as daily data; ARMI, FPI, EPI, EPU, CPU, and GND are reported monthly; and PROD, IMP, CONS, and ES are provided annually. Data alignment is achieved through forward-filling. To categorize these variables, ARMI, FPI, EPI, and BDI represent price-driven factors; GPR, EPU, CPU, and GND capture external shock factors; and PROD, IMP, CONS, and ES reflect core supply-demand indicators. With these diverse variables, we aim to provide a comprehensive understanding of the sources of risks in grain markets, as well as to explore the internal and external drivers of their risk spillovers from multiple perspectives.

\subsection{Statistical description}

\begin{figure}[!h]
  \centering
  \includegraphics[width=0.475\linewidth]{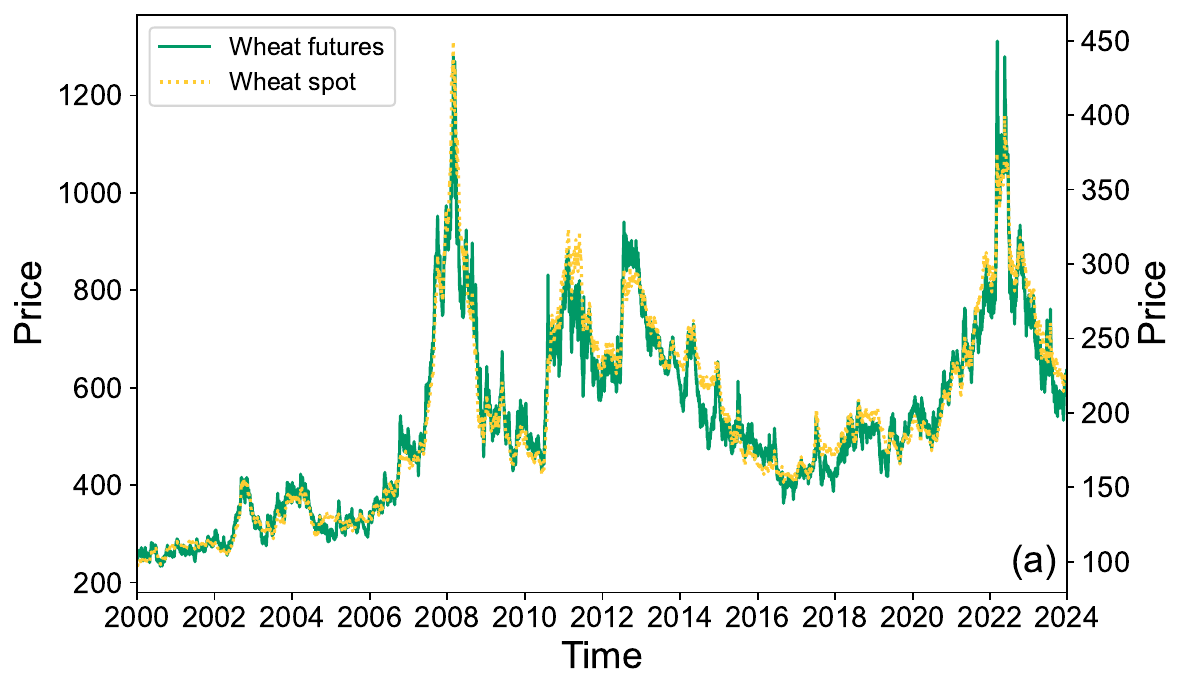}
  \includegraphics[width=0.475\linewidth]{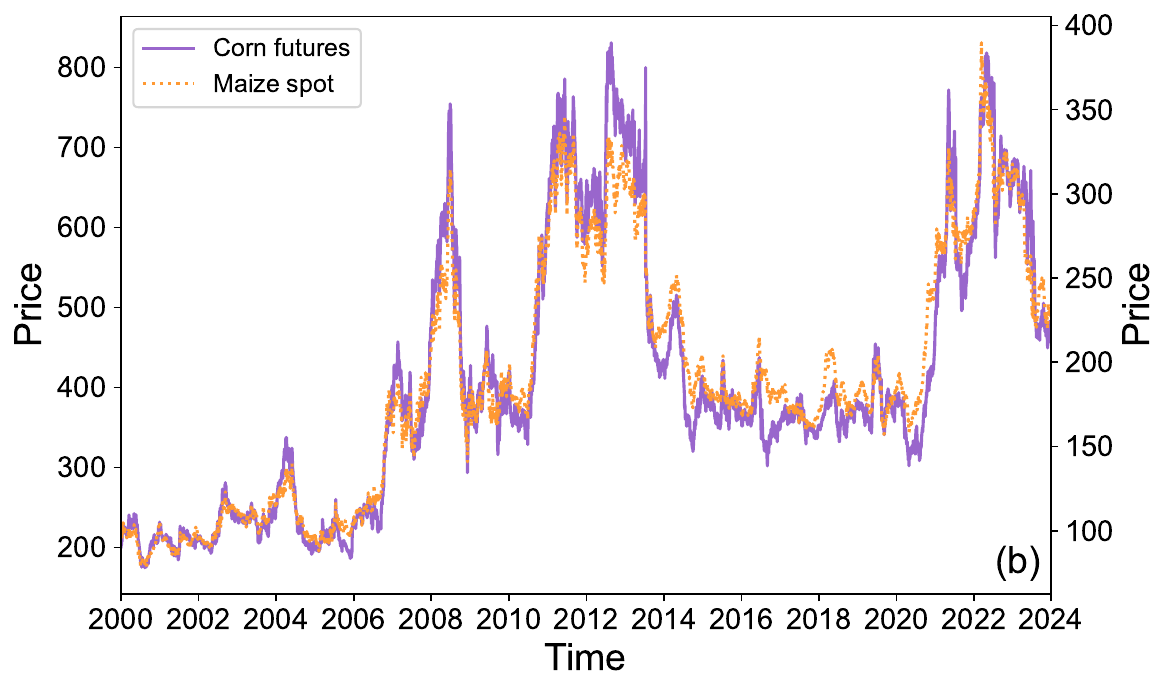}\\
  \includegraphics[width=0.475\linewidth]{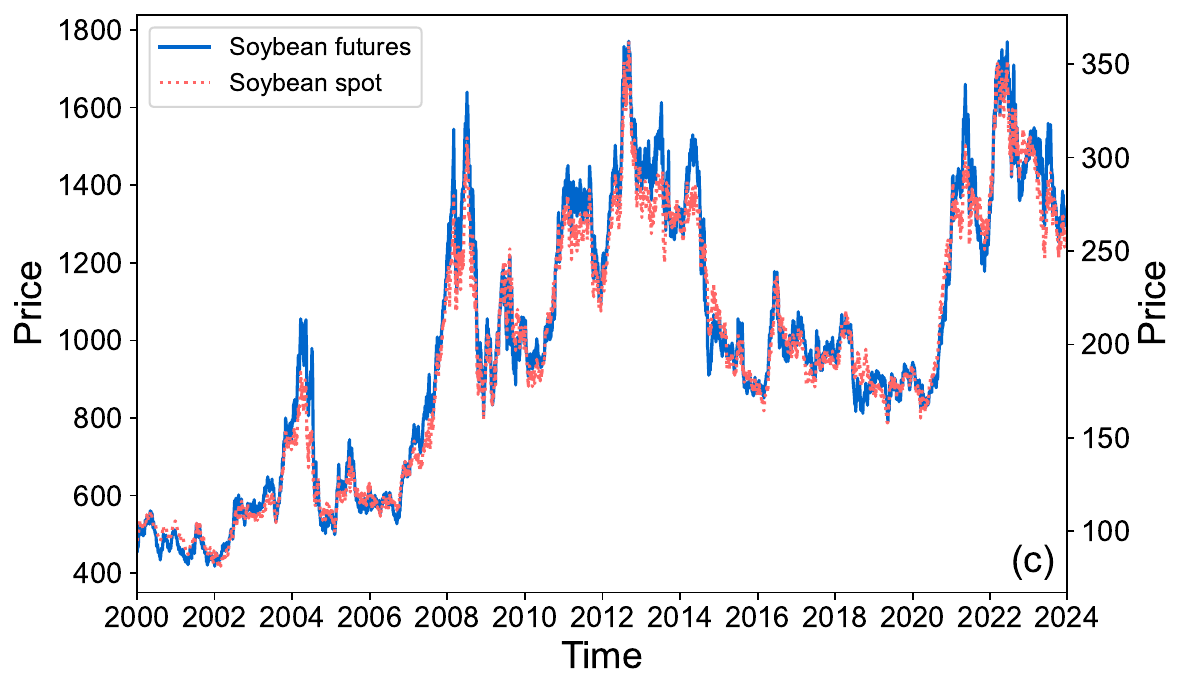}
  \includegraphics[width=0.475\linewidth]{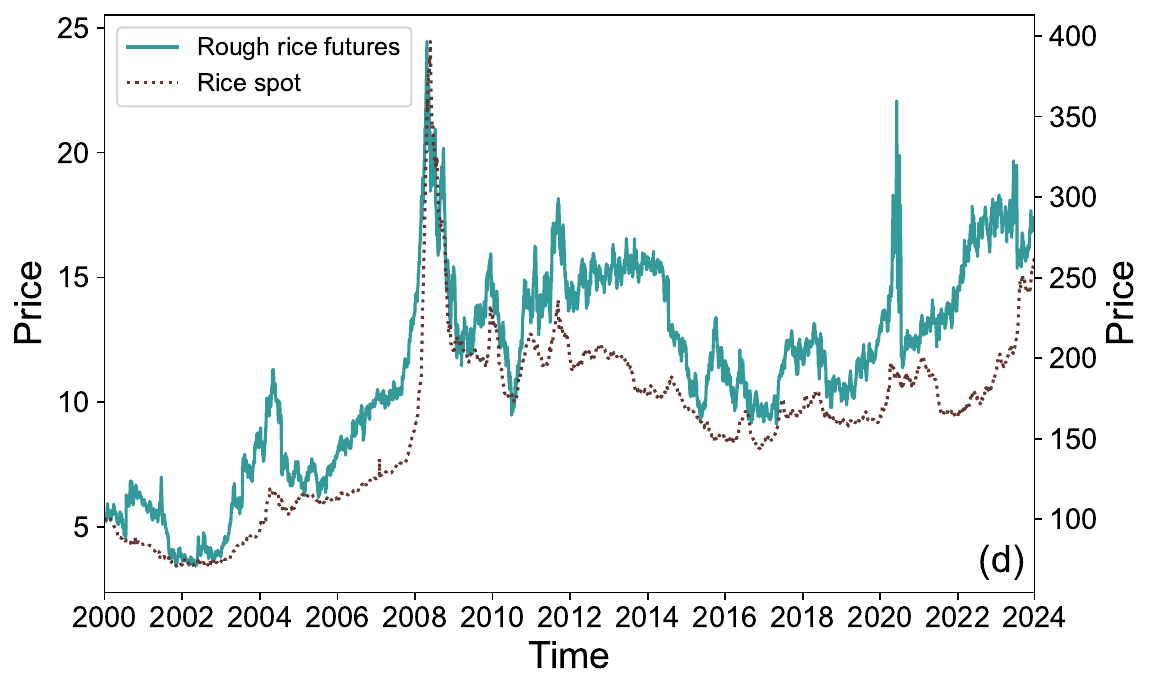}
  \caption{Evolution of futures and spot prices for wheat (a), maize (b), soybean (c), and rice (d).}
\label{Fig:AgroPrice_evolution}
\end{figure}

Figures~\ref{Fig:AgroPrice_evolution}(a)–(d) illustrate the evolution of the futures and spot prices for wheat, maize, soybean, and rice, respectively, from 2000 to 2023. A comparison reveals that each crop exhibits a unique evolution process, yet the overall price trajectories of its futures and spot remain aligned, with statistically significant positive correlations. Additionally, all four staples display notable price volatility during the three global food crises of the 21st century, namely, from 2006 to 2008, from 2010 to 2012, and from 2020 onward.

\begin{figure}[!h]
  \centering
  \includegraphics[width=0.475\linewidth]{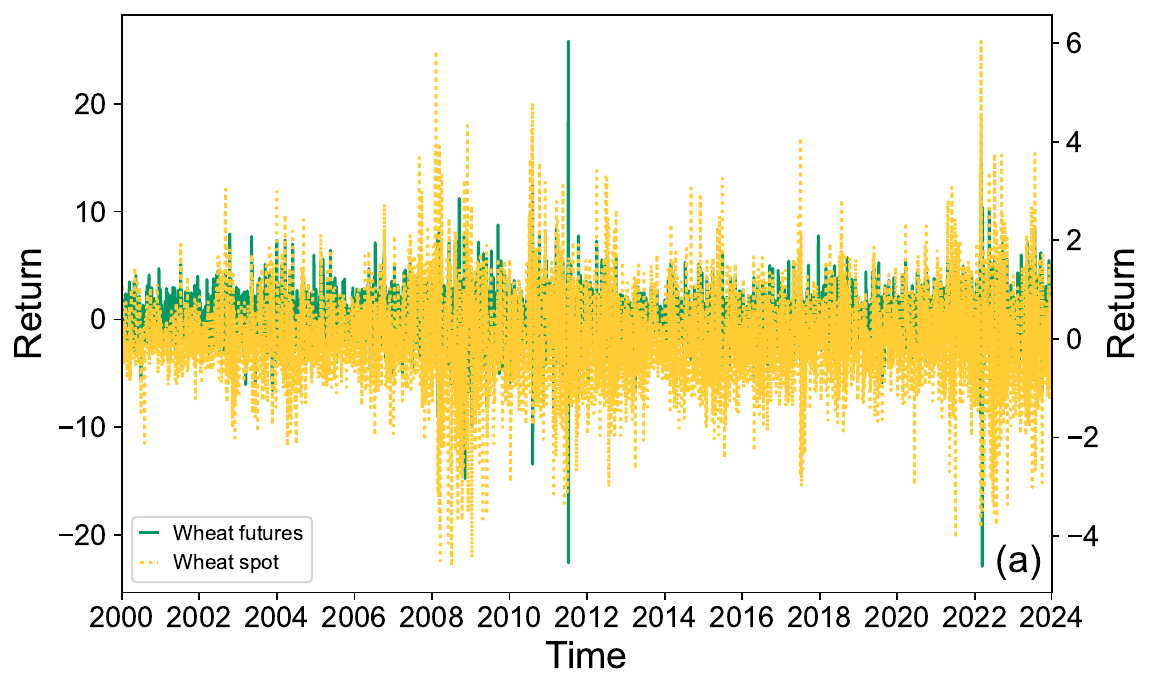}
  \includegraphics[width=0.475\linewidth]{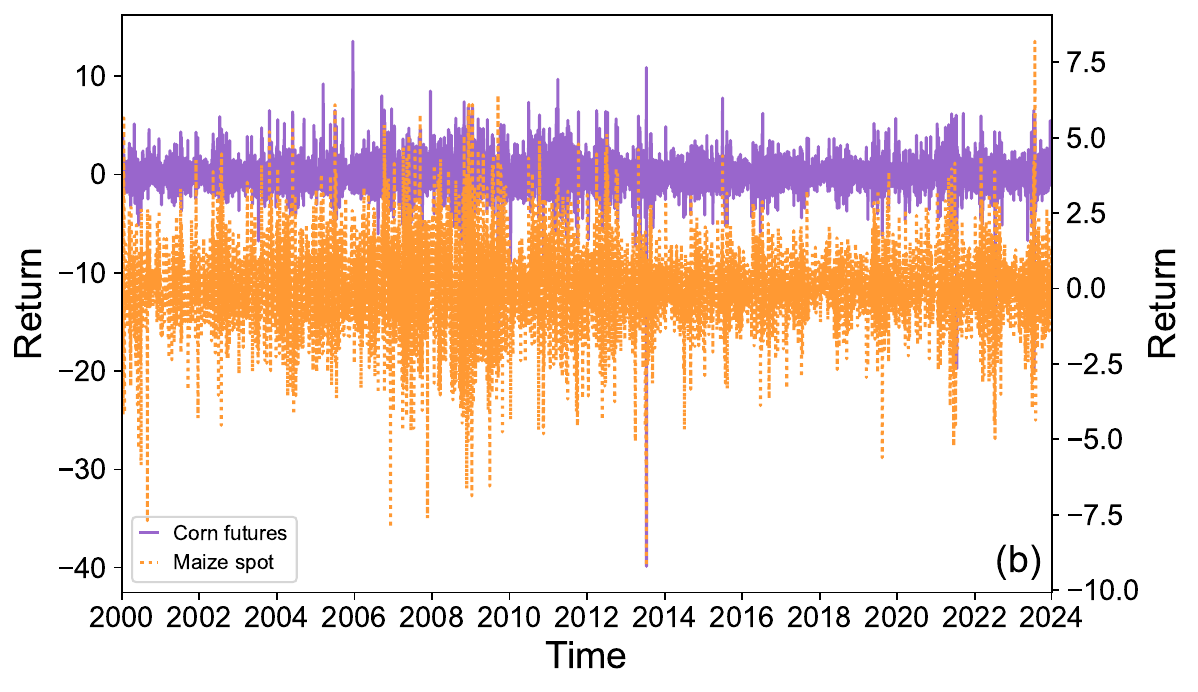}\\
  \includegraphics[width=0.475\linewidth]{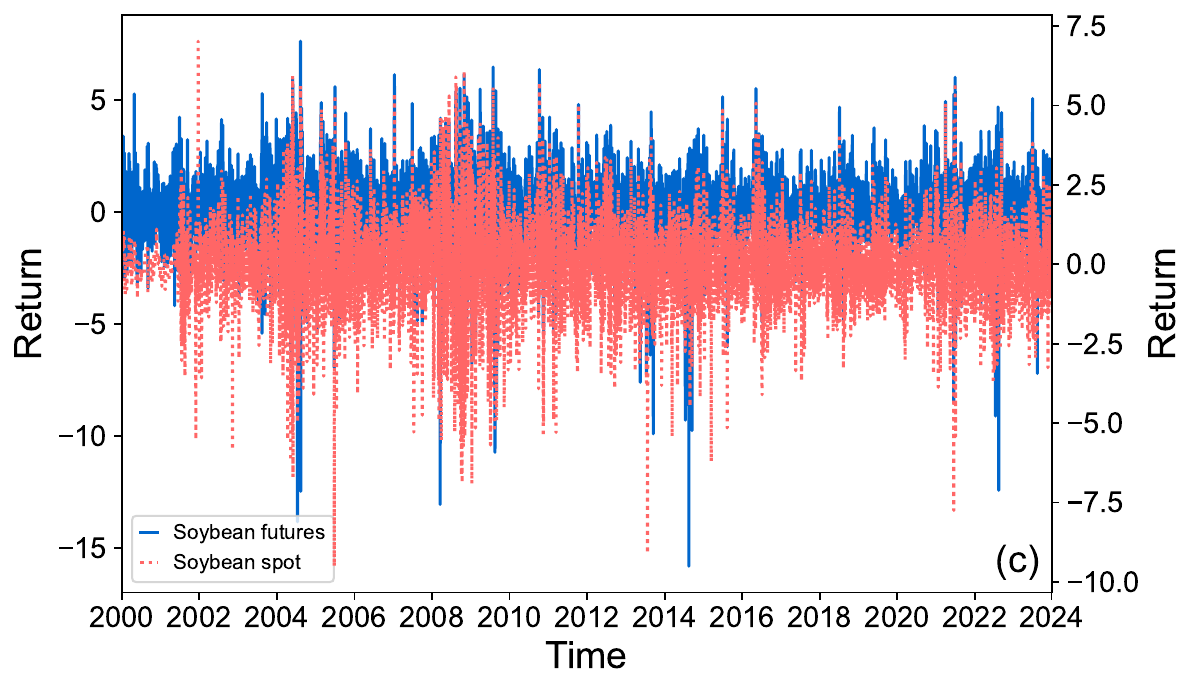}
  \includegraphics[width=0.475\linewidth]{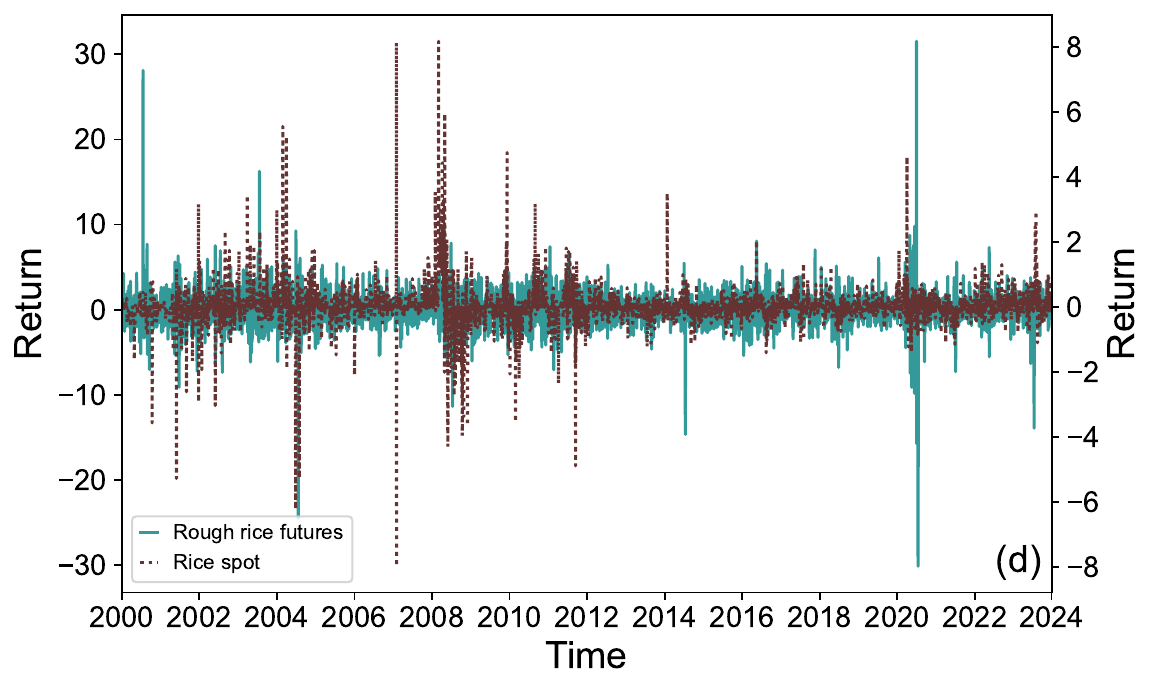}
  \caption{Evolution of futures and spot returns for wheat (a), maize (b), soybean (c), and rice (d).}
\label{Fig:AgroReturn_evolution}
\end{figure}

The daily logarithmic returns for the four staple crops are calculated by $r(t) = \ln P(t) - \ln P(t - 1)$ and scaled by a factor of 100 to facilitate subsequent analysis. Figures~\ref{Fig:AgroReturn_evolution}(a)–(d) depict the evolution of the futures and spot returns for wheat, maize, soybean, and rice, respectively, from 2000 to 2023. It is evident that the fluctuations of futures returns exceed those of spot returns, and all the return series exhibit significant volatility clustering.

Table~\ref{Tab:Agro_Stat_Test} presents the descriptive statistics for each return series. As can be seen, the futures returns have larger maxima, smaller minima, and larger standard deviations than the corresponding spot returns, indicating greater volatility in grain futures markets. Moreover, the skewness of each series is nonzero, which represents the asymmetric distribution. Specifically, the futures and spot returns for wheat and rice are right-skewed, while maize and soybean returns are left-skewed. Each series also displays kurtosis higher than 3, implying a leptokurtic distribution with fat tails. In addition, the statistics of both the Jarque-Bera test and the ADF test are significant at the 1\% level, which means that none of the return series follow a normal distribution and all series are stationary.

\begin{table}[!ht]
  \centering
  \setlength{\abovecaptionskip}{0pt}
  \setlength{\belowcaptionskip}{10pt}
  \caption{Descriptive statistics of grain futures and spot returns}
  \setlength\tabcolsep{3pt}   \resizebox{\textwidth}{!}{ 
    \begin{tabular}{l r@{.}l r@{.}l c r@{.}l r@{.}l c r@{.}l r@{.}l c r@{.}l r@{.}l}
    \toprule
         & \multicolumn{4}{c}{Wheat} && \multicolumn{4}{c}{Maize} && \multicolumn{4}{c}{Soybean} && \multicolumn{4}{c}{Rice}  \\
    \cline{2-5} \cline{7-10} \cline{12-15} \cline{17-20}
         & \multicolumn{2}{c}{Futures} & \multicolumn{2}{c}{Spot} && \multicolumn{2}{c}{Futures} & \multicolumn{2}{c}{Spot} && \multicolumn{2}{c}{Futures} & \multicolumn{2}{c}{Spot} && \multicolumn{2}{c}{Futures} & \multicolumn{2}{c}{Spot}  \\
    \midrule
    Max & 25&8053 & 6&0397 && 13&5589 & 8&1969 && 7&6292 & 7&0204 && 31&5128 & 8&1817 \\
    Min & $-$22&8880 & $-$4&6166 && $-$39&8614 & $-$9&2090 && $-$15&8290 & $-$9&5068 && $-$30&1080 & $-$7&9870 \\
    Mean & 0&0154 & 0&0139 && 0&0141 & 0&0141 && 0&0172 & 0&0162 && 0&0198 & 0&0161 \\
    Std. Dev. & 2&0999 & 0&8614 && 1&8642 & 1&3295 && 1&5623 & 1&3361 && 1&8154 & 0&4918 \\
    Skew. & 0&1068 & 0&1927 && $-$2&0245 & $-$0&1098 && $-$1&0281 & $-$0&3675 && 0&1146 & 0&9949 \\
    Kurt. & 10&8156 & 3&5817 && 42&0777 & 3&3413 && 7&9856 & 3&5763 && 47&3077 & 67&2765 \\
    Jarque-Bera & \multicolumn{2}{c}{29461$^{***}$} & \multicolumn{2}{c}{3267$^{***}$} && \multicolumn{2}{c}{449859$^{***}$} & \multicolumn{2}{c}{2823$^{***}$} && \multicolumn{2}{c}{17119$^{***}$} & \multicolumn{2}{c}{3356$^{***}$} && \multicolumn{2}{c}{563435$^{***}$} & \multicolumn{2}{c}{1140449$^{***}$} \\
    ADF & $-$15&6816$^{***}$ & $-$14&2483$^{***}$ && $-$77&5926$^{***}$ & $-$15&8689$^{***}$ && $-$24&8772$^{***}$ & $-$76&6446$^{***}$ && $-$22&9390$^{***}$ & $-$9&6883$^{***}$ \\
  \bottomrule
    \end{tabular}
    }%
  \begin{flushleft}
    \footnotesize
\justifying Note: This table reports the descriptive statistics of futures and spot return series for wheat, maize, soybean, and rice. The Jarque-Bera test is a normality test, where a statistic significantly greater than 0 implies that the series does not obey normality. The ADF test is a white noise test, in which a significant statistic indicates the stationarity of the series. *** denotes the significance at the 1\% level.
\end{flushleft} 
  \label{Tab:Agro_Stat_Test}%
\end{table}%

\section{Empirical analysis}
\label{S1:EmpAnal}

\subsection{Decomposition of grain return series}

\begin{figure}[!ht]
  \centering
  \includegraphics[width=0.985\linewidth]{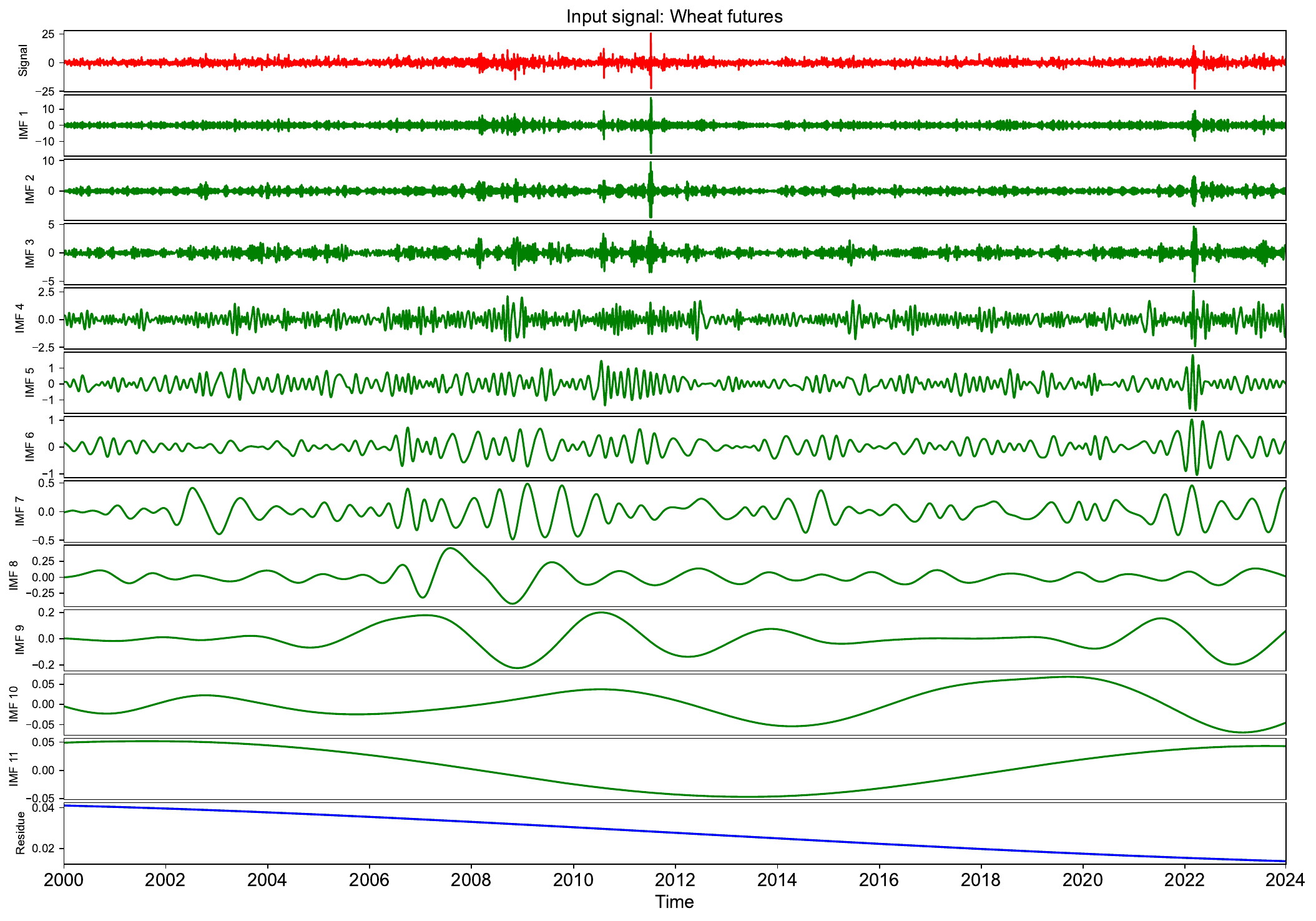}
  \caption{Decomposition results of wheat futures returns.} 
\label{Fig:AgroReturn_decomposition}
\end{figure}

We apply the ICEEMDAN method to decompose the futures and spot return series for the four staple crops. For clarity, the results for wheat futures returns are presented in Figure~\ref{Fig:AgroReturn_decomposition}, and the results for other return series are provided in the \ref{S:Appendix}, shown in Figures~\ref{Fig:AgroReturn_decomposition_WS}–\ref{Fig:AgroReturn_decomposition_RS}. All the IMFs are displayed in order from high to low frequency, followed by the residue. According to the model criterion, the return series are typically decomposed into 10 or 11 IMFs. These decomposed modes reflect the characteristics of their original return series, and different modes are driven by distinct underlying factors \citep{Huang-Shen-Long-Wu-Shih-Zheng-Yen-Tung-Liu-1998-PRSA}.

To analyze these decomposed modes, we compute their mean periods, correlations with the original return series, and their importance to the original series. The mean period, defined as the ratio of each IMF's total data points to its number of peaks, measures the timescale of each IMF. The Spearman rank correlation coefficient is adopted to evaluate the correlation between decomposed modes and original series, as this coefficient is distribution-independent and suitable for nonlinear data relationships. The importance of each decomposed mode to the original return series is assessed by calculating the ratio of its variance to the total variance. 

The results, presented in Table~\ref{Tab:Measures_of_IMFs}, reveal that the mean periods of IMFs for each return series increase gradually, corresponding to the order from short to long term. Additionally, IMF1s consistently have a mean period of about three working days across all return series, but as the order increases, the mean periods of the same-order IMFs show numerical differences. Furthermore, both correlation and importance coefficients decrease overall with increasing IMF order. These findings indicate that for each staple crop, IMF1 contributes the most to its market volatility, while the residue has the least impact. Notably, IMF1 explains 59.26\% of the fluctuations in wheat futures returns, followed by soybean returns. Even for rice spot, IMF1 accounts for around 37.80\% of the total volatility. In contrast, no significant correlation exists between residue and original series, and their variance proportions are also quite small.

\begin{table}[!ht]
  \belowrulesep=0pt 
  \aboverulesep=0pt
  \centering
  \setlength{\abovecaptionskip}{0pt}
  \setlength{\belowcaptionskip}{10pt}
  \caption{Measures of IMFs and residue for grain futures and spot returns}
  \setlength\tabcolsep{3pt}   \resizebox{\textwidth}{!}{ 
    \begin{tabular}{l r@{.}l r@{.}l r@{.}l cc | l r@{.}l r@{.}l r@{.}l cc}
    \toprule
         & \multicolumn{2}{c}{Mean period} & \multicolumn{2}{c}{Correlation} & \multicolumn{2}{c}{Importance} & Run-length & Group && \multicolumn{2}{c}{Mean period} & \multicolumn{2}{c}{Correlation} & \multicolumn{2}{c}{Importance} & Run-length & Group \\
    \midrule
    \multicolumn{9}{l|}{\textit{Panel A: Wheat futures}} & \multicolumn{9}{l}{\textit{Panel B: Wheat spot}} \\
    IMF1 & 2&76 & 0&7500$^{***}$ & 59&2598$\%$ & 4155 & STC & IMF1 & 2&78 & 0&6594$^{***}$ & 46&1774$\%$ & 4103 & STC \\
    IMF2 & 5&29 & 0&4612$^{***}$ & 18&2214$\%$ & 2173 & STC & IMF2 & 5&26 & 0&4582$^{***}$ & 16&7934$\%$ & 2151 & STC \\
    IMF3 & 10&00 & 0&3072$^{***}$ & 10&5273$\%$ & 1144 & MTC & IMF3 & 10&40 & 0&3807$^{***}$ & 13&3846$\%$ & 1112 & MTC \\
    IMF4 & 19&24 & 0&2180$^{***}$ & 6&0925$\%$ & 596 & MTC & IMF4 & 20&48 & 0&2815$^{***}$ & 9&6024$\%$ & 565 & MTC \\
    IMF5 & 40&55 & 0&1544$^{***}$ & 3&2528$\%$ & 283 & MTC & IMF5 & 44&10 & 0&2117$^{***}$ & 6&2241$\%$ & 260 & MTC \\
    IMF6 & 73&68 & 0&1067$^{***}$ & 1&4682$\%$ & 163 & LTC & IMF6 & 85&10 & 0&1595$^{***}$ & 3&4952$\%$ & 137 & LTC \\
    IMF7 & 140&51 & 0&0673$^{***}$ & 0&6468$\%$ & 86 & LTC & IMF7 & 201&40 & 0&0993$^{***}$ & 1&4580$\%$ & 58 & LTC \\
    IMF8 & 302&10 & 0&0332$^{***}$ & 0&2854$\%$ & 40 & LTC & IMF8 & 402&80 & 0&1072$^{***}$ & 1&6448$\%$ & 28 & LTC \\
    IMF9 & 755&25 & 0&0308$^{**}$ & 0&1871$\%$ & 17 & LTC & IMF9 & 1208&40 & 0&0863$^{***}$ & 1&0093$\%$ & 11 & LTC \\
    IMF10 & 2014&00 & 0&0129 & 0&0296$\%$ & 7 & LTC & IMF10 & 3021&00 & 0&0033 & 0&1586$\%$ & 5 & LTC \\
    IMF11 & 3021&00 & 0&0050 & 0&0275$\%$ & 3 & LTC & Residue & \multicolumn{2}{c}{} & 0&0138 & 0&0522$\%$ & \multicolumn{2}{c}{}  \\
    Residue & \multicolumn{2}{c}{} & 0&0128 & 0&0016$\%$ & \multicolumn{2}{c|}{} & \multicolumn{9}{c}{} \\
    \multicolumn{9}{c|}{\vspace{-1.2mm}} & \multicolumn{9}{c}{\vspace{-1.2mm}} \\

    \multicolumn{9}{l|}{\textit{Panel C: Maize futures}} & \multicolumn{9}{l}{\textit{Panel D: Maize spot}} \\
    IMF1 & 2&78 & 0&7300$^{***}$ & 54&1497$\%$ & 4151 & STC & IMF1 & 2&75 & 0&7027$^{***}$ & 54&1441$\%$ & 4142 & STC \\
    IMF2 & 5&27 & 0&4499$^{***}$ & 20&5695$\%$ & 2140 & STC & IMF2 & 5&27 & 0&4670$^{***}$ & 19&3971$\%$ & 2176 & STC \\
    IMF3 & 9&89 & 0&3052$^{***}$ & 11&8402$\%$ & 1179 & MTC & IMF3 & 9&76 & 0&3312$^{***}$ & 10&4270$\%$ & 1189 & MTC \\
    IMF4 & 18&14 & 0&1995$^{***}$ & 5&3490$\%$ & 651 & MTC & IMF4 & 18&71 & 0&2226$^{***}$ & 6&7541$\%$ & 618 & MTC \\
    IMF5 & 36&40 & 0&1533$^{***}$ & 4&2317$\%$ & 327 & MTC & IMF5 & 38&00 & 0&1663$^{***}$ & 4&6253$\%$ & 310 & MTC \\
    IMF6 & 78&47 & 0&1052$^{***}$ & 1&6997$\%$ & 150 & LTC & IMF6 & 79&50 & 0&1193$^{***}$ & 2&4460$\%$ & 145 & LTC \\
    IMF7 & 177&71 & 0&0793$^{***}$ & 1&2074$\%$ & 67 & LTC & IMF7 & 172&63 & 0&0967$^{***}$ & 1&1926$\%$ & 70 & LTC \\
    IMF8 & 402&80 & 0&0374$^{***}$ & 0&6363$\%$ & 31 & LTC & IMF8 & 402&80 & 0&0580$^{***}$ & 0&5984$\%$ & 30 & LTC \\
    IMF9 & 863&14 & 0&0098 & 0&1514$\%$ & 16 & LTC & IMF9 & 1208&40 & 0&0425$^{***}$ & 0&2608$\%$ & 11 & LTC \\
    IMF10 & 2014&00 & 0&0208 & 0&0983$\%$ & 6 & LTC & IMF10 & 3021&00 & 0&0287$^{**}$ & 0&1481$\%$ & 4 & LTC \\
    Residue & \multicolumn{2}{c}{} & $-$0&0035 & 0&0668$\%$ & \multicolumn{2}{c|}{} & Residue & \multicolumn{2}{c}{} & $-$0&0067 & 0&0065$\%$ & \multicolumn{2}{c}{} \\
    \multicolumn{9}{c|}{\vspace{-1.2mm}} & \multicolumn{9}{c}{\vspace{-1.2mm}} \\

    \multicolumn{9}{l|}{\textit{Panel E: Soybean futures}} & \multicolumn{9}{l}{\textit{Panel F: Soybean spot}} \\
    IMF1 & 2&71 & 0&7414$^{***}$ & 56&0677$\%$ & 4224 & STC & IMF1 & 2&75 & 0&7260$^{***}$ & 56&4743$\%$ & 4176 & STC \\
    IMF2 & 5&09 & 0&4391$^{***}$ & 17&3390$\%$ & 2246 & STC & IMF2 & 5&23 & 0&4434$^{***}$ & 18&6443$\%$ & 2172 & STC \\
    IMF3 & 9&70 & 0&3081$^{***}$ & 12&1844$\%$ & 1197 & MTC & IMF3 & 10&24 & 0&3162$^{***}$ & 11&4726$\%$ & 1121 & MTC \\
    IMF4 & 19&00 & 0&2107$^{***}$ & 6&6501$\%$ & 615 & MTC & IMF4 & 20&01 & 0&2355$^{***}$ & 5&9991$\%$ & 579 & MTC \\
    IMF5 & 37&53 & 0&1634$^{***}$ & 3&2406$\%$ & 309 & MTC & IMF5 & 40&01 & 0&1699$^{***}$ & 3&8043$\%$ & 286 & MTC \\
    IMF6 & 77&46 & 0&0878$^{***}$ & 1&6945$\%$ & 153 & LTC & IMF6 & 87&57 & 0&1015$^{***}$ & 1&8437$\%$ & 135 & LTC \\
    IMF7 & 172&63 & 0&0704$^{***}$ & 1&1454$\%$ & 66 & LTC & IMF7 & 208&34 & 0&0747$^{***}$ & 1&0628$\%$ & 58 & LTC \\
    IMF8 & 318&00 & 0&0536$^{***}$ & 0&8237$\%$ & 39 & LTC & IMF8 & 402&80 & 0&0448$^{***}$ & 0&3255$\%$ & 29 & LTC \\
    IMF9 & 755&25 & 0&0195 & 0&5315$\%$ & 19 & LTC & IMF9 & 1208&40 & 0&0490$^{***}$ & 0&3191$\%$ & 11 & LTC \\
    IMF10 & 1510&50 & 0&0077 & 0&3067$\%$ & 8 & LTC & IMF10 & 2014&00 & 0&0204 & 0&0399$\%$ & 5 & LTC \\
    IMF11 & 3021&00 & 0&0064 & 0&0148$\%$ & 3 & LTC & Residue & \multicolumn{2}{c}{} & 0&0194 & 0&0144$\%$ & \multicolumn{2}{c}{}  \\
    Residue & \multicolumn{2}{c}{} & $-$0&0128 & 0&0016$\%$ & \multicolumn{2}{c|}{} & \multicolumn{9}{c}{} \\
    \multicolumn{9}{c|}{\vspace{-1.2mm}} & \multicolumn{9}{c}{\vspace{-1.2mm}} \\

    \multicolumn{9}{l|}{\textit{Panel G: Rice futures}} & \multicolumn{9}{l}{\textit{Panel H: Rice spot}} \\
    IMF1 & 2&79 & 0&6914$^{***}$ & 49&3305$\%$ & 4055 & STC & IMF1 & 2&75 & 0&4779$^{***}$ & 37&7964$\%$ & 4162 & STC \\
    IMF2 & 5&25 & 0&4682$^{***}$ & 25&7247$\%$ & 2171 & STC & IMF2 & 4&81 & 0&3261$^{***}$ & 20&9636$\%$ & 2326 & STC \\
    IMF3 & 9&97 & 0&3045$^{***}$ & 12&9665$\%$ & 1152 & MTC & IMF3 & 8&89 & 0&2415$^{***}$ & 11&5784$\%$ & 1292 & MTC \\
    IMF4 & 19&43 & 0&2069$^{***}$ & 6&4227$\%$ & 599 & MTC & IMF4 & 16&16 & 0&1959$^{***}$ & 7&4560$\%$ & 726 & MTC \\
    IMF5 & 37&30 & 0&1699$^{***}$ & 2&8841$\%$ & 317 & MTC & IMF5 & 29&91 & 0&1880$^{***}$ & 4&8427$\%$ & 374 & MTC \\
    IMF6 & 78&47 & 0&0975$^{***}$ & 1&7369$\%$ & 154 & LTC & IMF6 & 57&00 & 0&2049$^{***}$ & 3&1308$\%$ & 208 & LTC \\
    IMF7 & 172&63 & 0&0569$^{***}$ & 0&4896$\%$ & 67 & LTC & IMF7 & 102&41 & 0&1559$^{***}$ & 3&8109$\%$ & 114 & LTC \\
    IMF8 & 355&41 & 0&0492$^{***}$ & 0&2994$\%$ & 34 & LTC & IMF8 & 208&34 & 0&1852$^{***}$ & 5&4092$\%$ & 58 & LTC \\
    IMF9 & 1007&00 & 0&0269$^{**}$ & 0&0792$\%$ & 12 & LTC & IMF9 & 464&77 & 0&1501$^{***}$ & 3&0737$\%$ & 28 & LTC \\
    IMF10 & 3021&00 & 0&0178 & 0&0614$\%$ & 5 & LTC & IMF10 & 1007&00 & 0&0908$^{***}$ & 1&2572$\%$ & 13 & LTC \\
    Residue & \multicolumn{2}{c}{} & $-$0&0084 & 0&0049$\%$ & \multicolumn{2}{c|}{} & IMF11 & 3021&00 & 0&1196$^{***}$ & 0&6774$\%$ & 4 & LTC \\
    \multicolumn{9}{c|}{} & Residue & \multicolumn{2}{c}{} & $-$0&0456 & 0&0037$\%$ & \multicolumn{2}{c}{} \\
  \bottomrule
    \end{tabular}
    }%
  \begin{flushleft}
    \footnotesize
    \justifying Note: This table gives the measures for the decomposed modes of futures and spot return series for wheat, maize, soybean, and rice. The abbreviations STC, MTC, and LTC denote the short-, medium-, and long-term components, respectively. ** and *** indicate significance at the 5\% and 1\% levels.
\end{flushleft} 
  \label{Tab:Measures_of_IMFs}%
\end{table}%

\subsection{Reconstruction of grain return series}

Next, we employ the proposed GMM-RLN method to reconstruct decomposed IMFs into short-term (high-frequency), medium-term (medium-frequency), and long-term (low-frequency) components. The last two columns in each panel of Table~\ref{Tab:Measures_of_IMFs} give the run-length number and cluster group for each decomposed IMF. For all return series, the run-length number decreases as the IMF order increases, implying greater volatility in higher-frequency IMFs with shorter timescales. This conclusion aligns with intuitive expectations. Moreover, the clustering results of decomposed IMFs show remarkable consistency across all return series. Specifically, IMF1 and IMF2 cluster into the short-term component (STC), IMF3, IMF4, and IMF5 cluster into the medium-term component (MTC), while IMF6 and the remaining longer-term IMFs cluster into the long-term component (LTC).

As can be seen in Table~\ref{Tab:Measures_of_IMFs}, each short-term component has an average period ceiling of about five working days, meaning that the timescale for short-term components falls within one week. Short-term components relate to short-run market fluctuations and are often driven by factors with short-term impacts, which tend to be intense, short-lived, and frequent. Furthermore, the mean period ceiling for the medium-term components is approximately 44 business days, indicating that the timescale associated with them generally spans up to two months. For the long-term components, the mean period ceiling is about 3,021 working days, which suggests that the timescale extends up to 12 years. Specially, the average period ceiling for maize futures and soybean spot is around 2,014 business days, or roughly eight years. The broad timescale of long-term components allows for diverse factors to exert influence. It is worth noting that some critical factors or major shocks are expected to simultaneously impact components at different timescales. In general, the grouping and mean periods of decomposed IMFs for the four main staple crops are largely consistent, which reflects the commonality of the international staple food markets to a certain extent.

\begin{figure}[!h]
  \centering
  \includegraphics[width=0.475\linewidth]{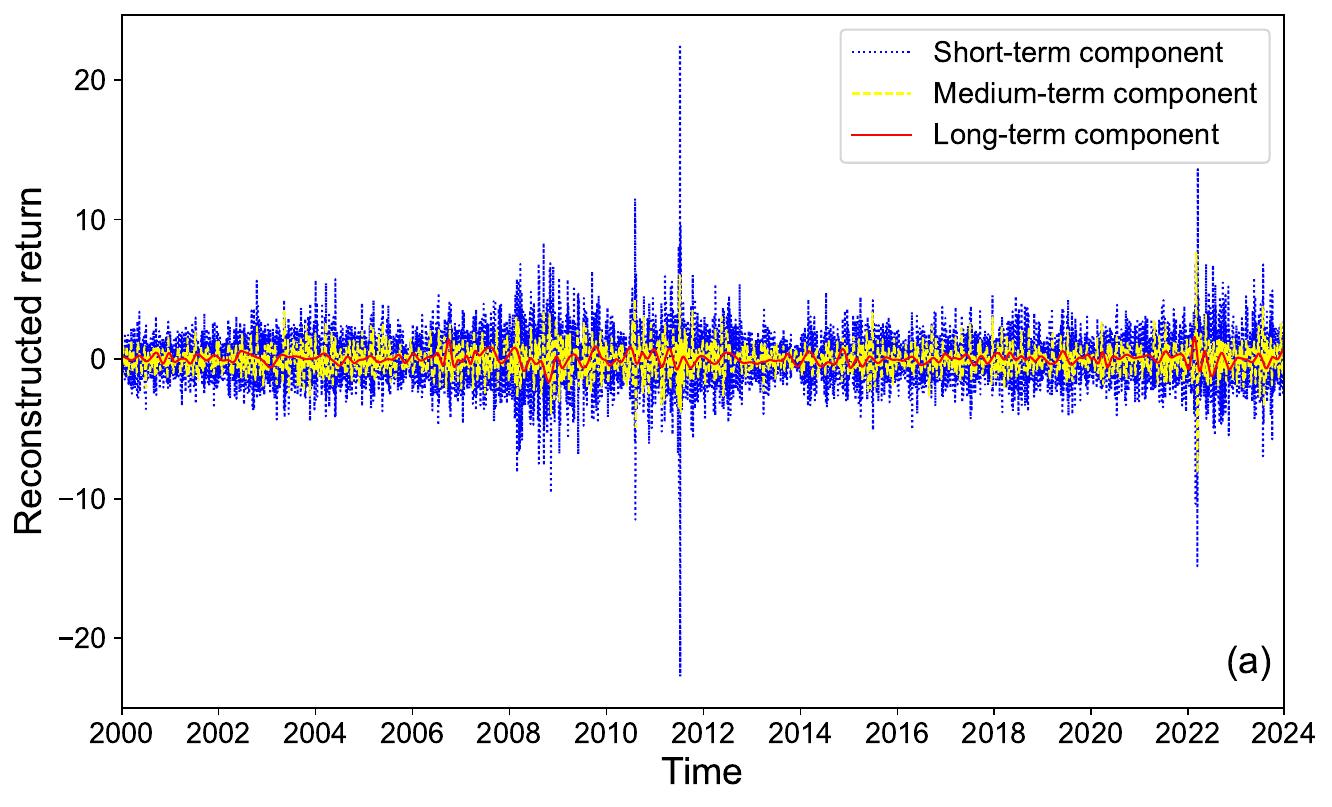}
  \includegraphics[width=0.475\linewidth]{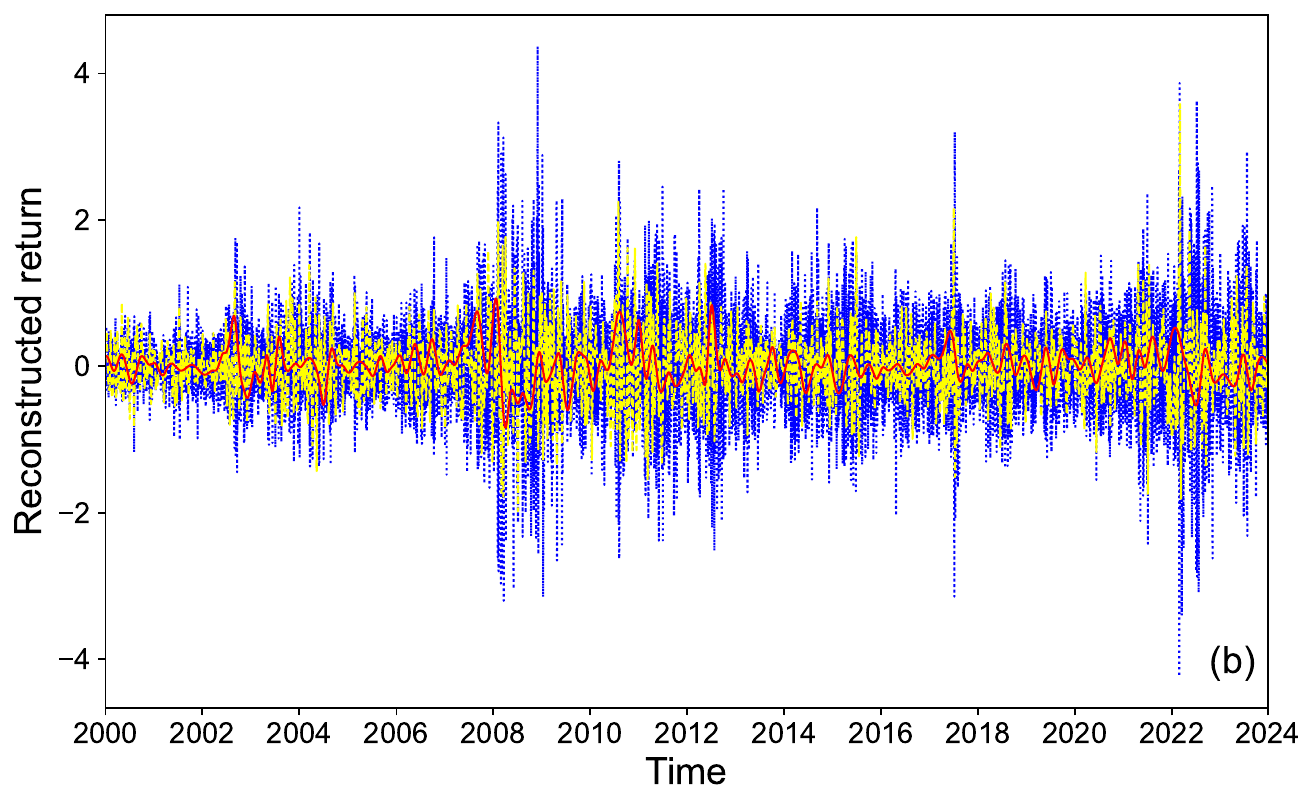}
  \includegraphics[width=0.475\linewidth]{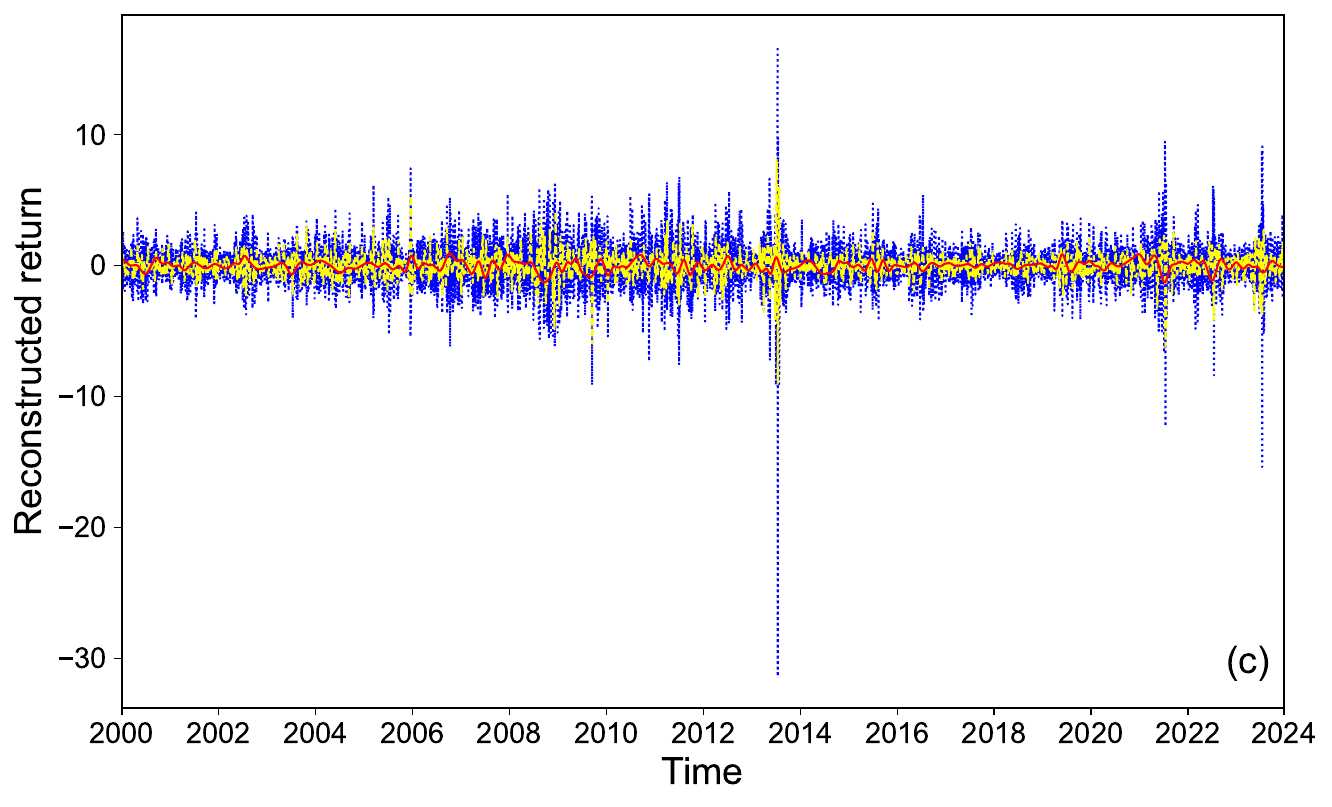}
  \includegraphics[width=0.475\linewidth]{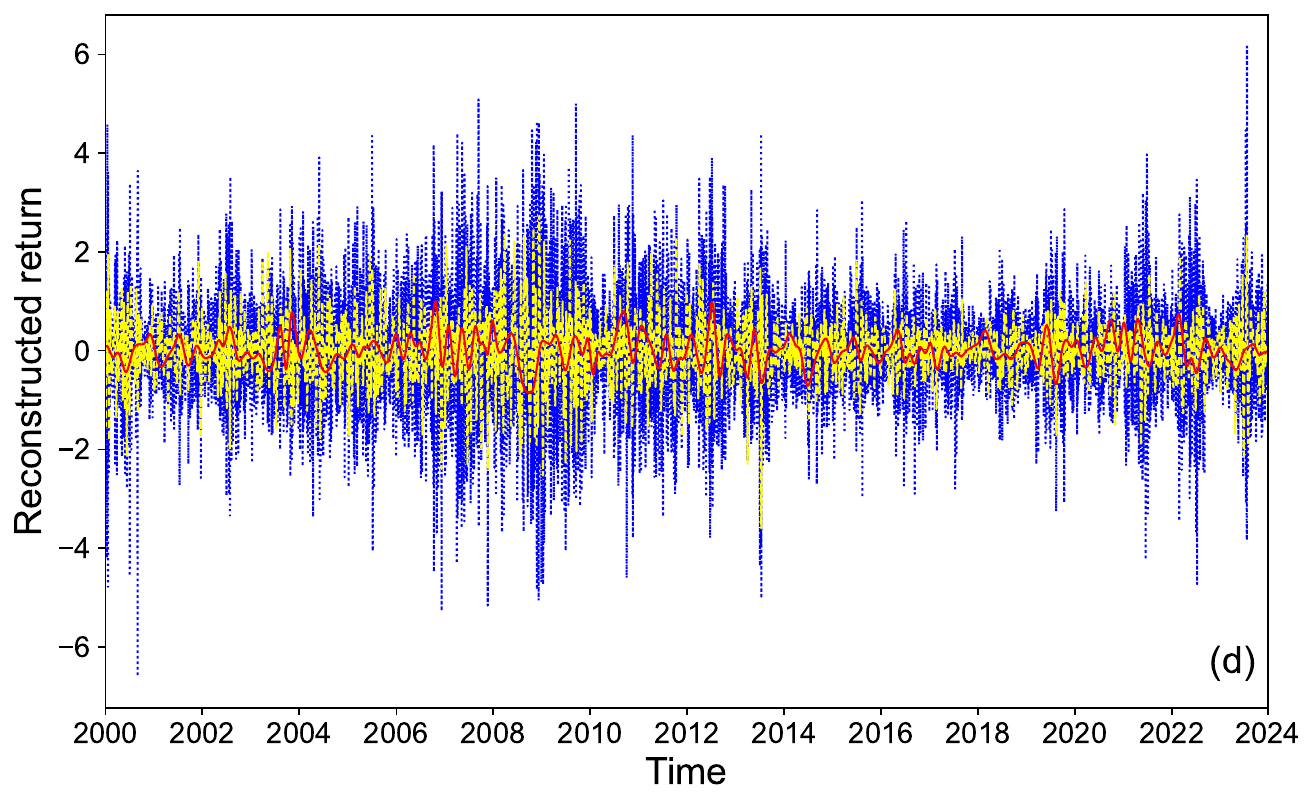}
  \includegraphics[width=0.475\linewidth]{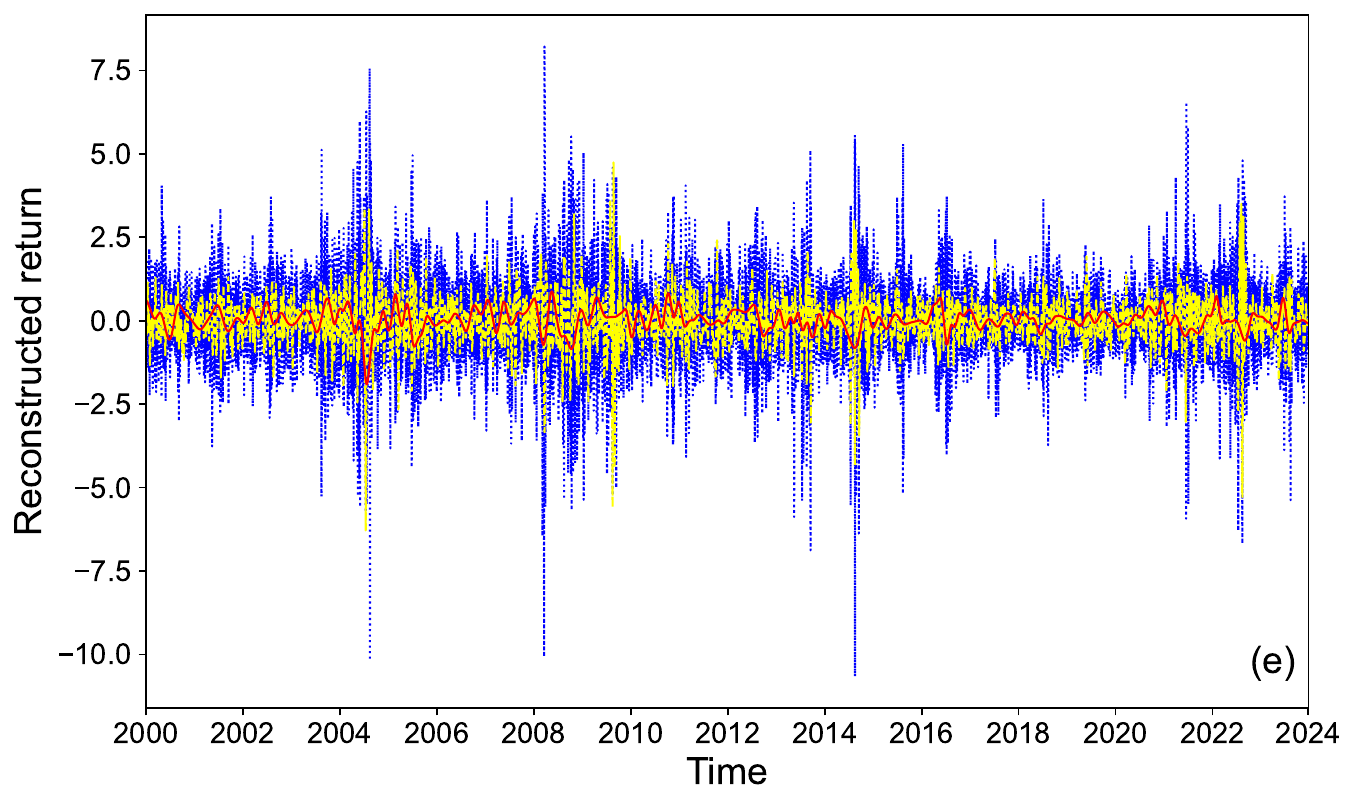}
  \includegraphics[width=0.475\linewidth]{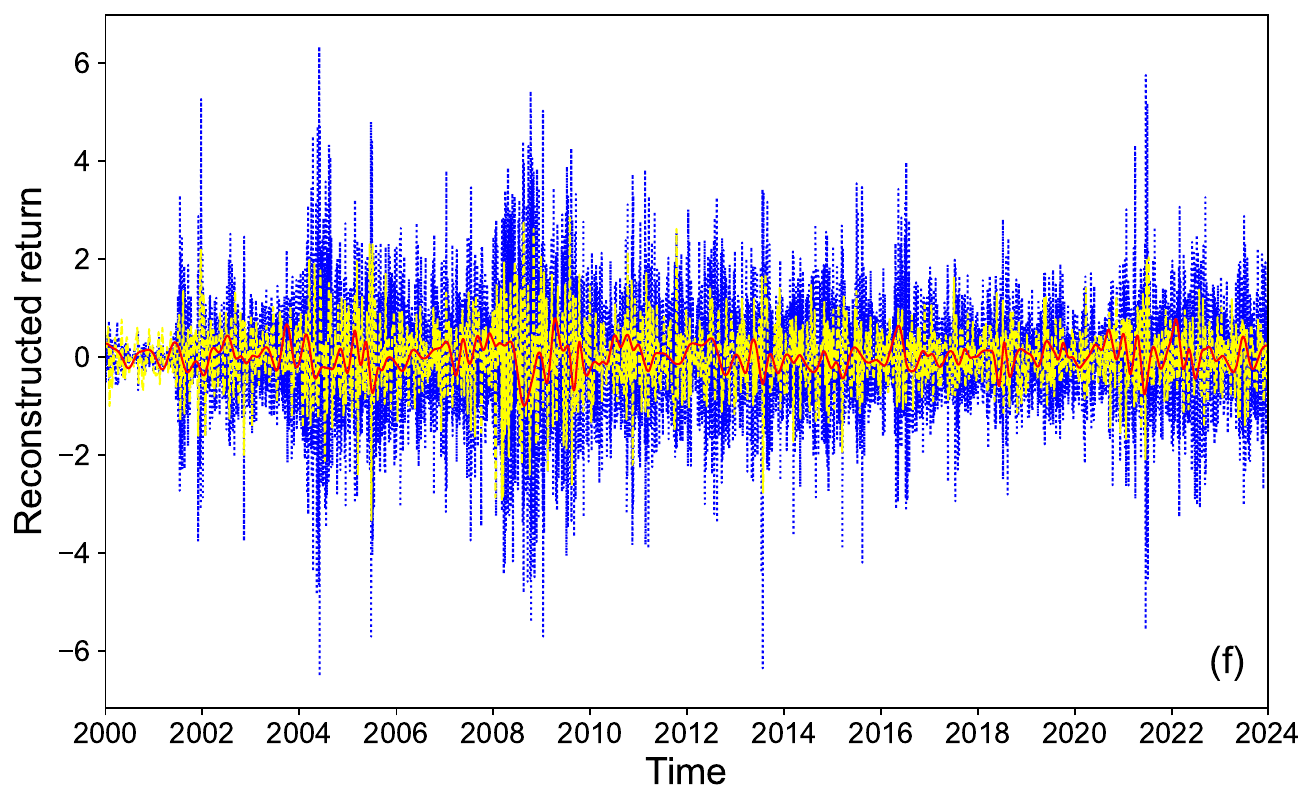}
  \includegraphics[width=0.475\linewidth]{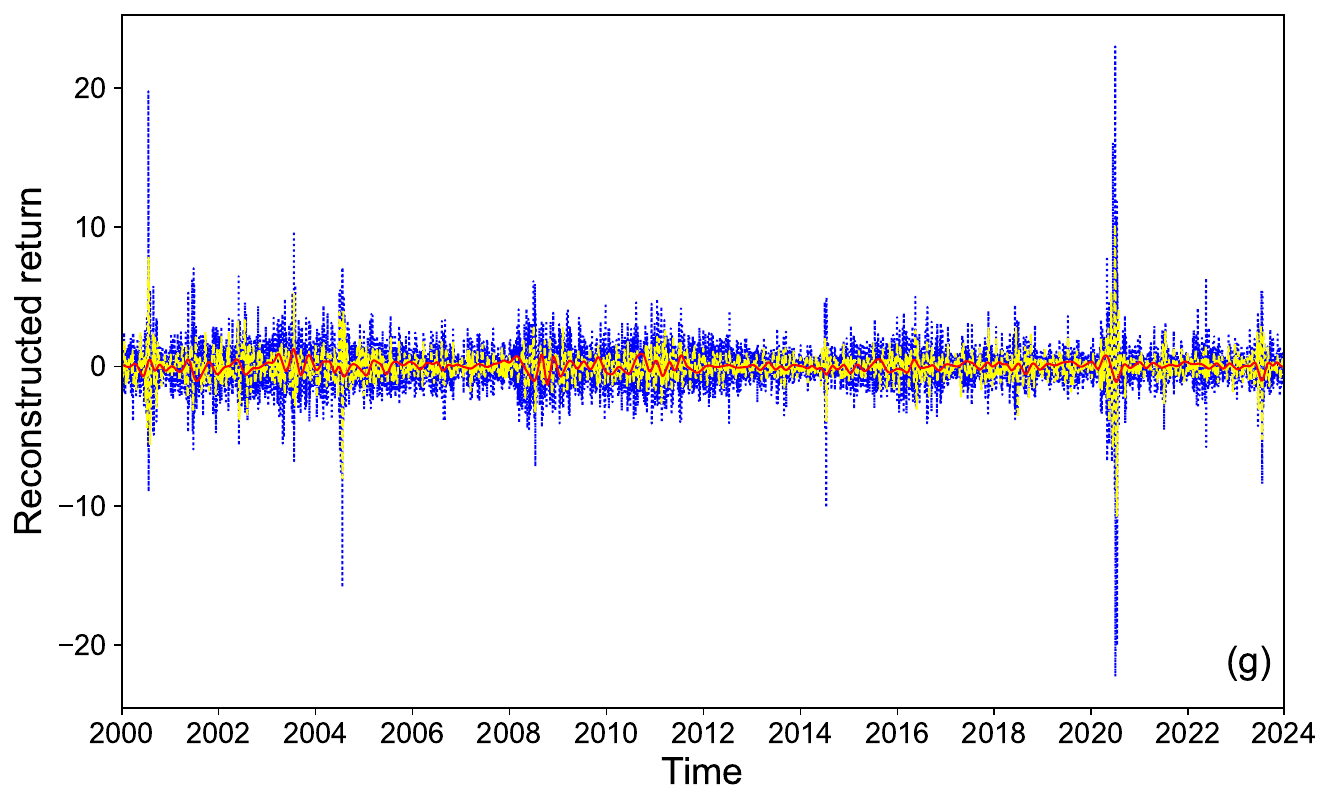}
  \includegraphics[width=0.475\linewidth]{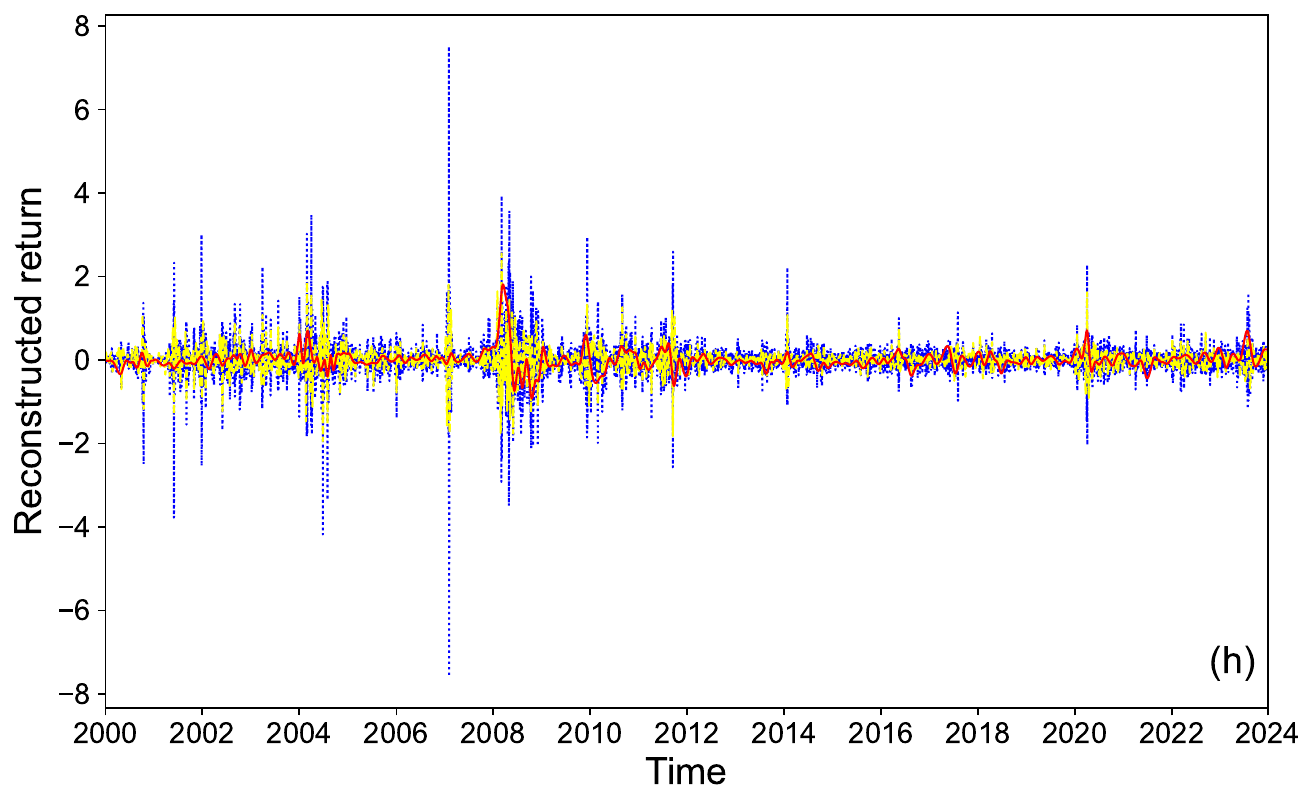}
  \caption{Evolution of reconstructed components for wheat futures (a), wheat spot (b), maize futures (c), maize spot (d), soybean futures (e), soybean spot (f), rice futures (g), and rice spot (h).}
\label{Fig:AgroReturn_reconstruction}
\end{figure}

The evolution of reconstructed components for grain futures and spot returns is presented in Figure~\ref{Fig:AgroReturn_reconstruction}. Short-term components show the most intense fluctuations, followed by medium-term components, with long-term components exhibiting the least volatility. This finding aligns with the run-length numbers in Table~\ref{Tab:Measures_of_IMFs}. Moreover, the short- and medium-term components generally oscillate around the long-term component. Despite differences in component trends across various futures and spots, all of the components show wild gyrations during common periods, including the 2006–2008 food crisis driven by industrial demand and trade protectionism, the 2008 global financial crisis, and the 2010–2012 food crisis fueled by abundant liquidity and abnormal climate, as well as the post-2020 food crisis triggered by the COVID-19 pandemic, extreme weather, and regional conflicts. Besides, maize return components experienced violent swings in 2013 due to global grain glut, while the Sino-US soybean trade dispute, along with changes in market supply and demand, led to spikes and drops of soybean return components in 2004 and 2013–2014.

Table~\ref{Tab:Reconstructed_series_Stat_Test} provides the descriptive statistics for all reconstructed components. Similar to the descriptive statistics for the original return series in Table~\ref{Tab:Agro_Stat_Test}, the volatility of short-, medium-, and long-term components of futures returns exceeds that of the corresponding components of spot returns, as shown in the maximum, minimum, and standard deviation. By component, consistent with Figure~\ref{Fig:AgroReturn_reconstruction}, the short-term component exhibits greater volatility than the medium- and long-term components, with the long-term component showing the least overall volatility. Particularly, the short-term component of wheat futures displays the highest volatility, followed by maize, rice, and soybean futures. In addition, none of these components follows a normal distribution, and all are stationary.

\begin{table}[!ht]
  \centering
  \setlength{\abovecaptionskip}{0pt}
  \setlength{\belowcaptionskip}{10pt}
  \caption{Descriptive statistics of reconstructed components for grain futures and spot returns}
  \setlength\tabcolsep{3pt}   \resizebox{\textwidth}{!}{ 
    \begin{tabular}{l r@{.}l r@{.}l r@{.}l r@{.}l r@{.}l r@{.}l r@{.}l r@{.}l}
    \toprule
         & \multicolumn{2}{c}{Max} & \multicolumn{2}{c}{Min} & \multicolumn{2}{c}{Mean} & \multicolumn{2}{c}{Std. Dev.} & \multicolumn{2}{c}{Skew.} & \multicolumn{2}{c}{Kurt.} & \multicolumn{2}{c}{Jarque-Bera} & \multicolumn{2}{c}{ADF} \\
    \midrule
    \multicolumn{17}{l}{\textit{Panel A: Reconstructed short-term components}} \\
    Wheat futures & 22&4061 & $-$22&7405 & $-$0&0095 & 1&8728 & $-$0&0254 & 9&4622 & \multicolumn{2}{r}{22541$^{***}$} & $-$13&5475$^{***}$ \\
    Wheat spot & 4&3667 & $-$4&2370 & 0&0058 & 0&6920 & 0&0530 & 2&4577 & \multicolumn{2}{r}{1523$^{***}$} & $-$17&9357$^{***}$ \\
    Maize futures & 16&7231 & $-$31&3706 & $-$0&0126 & 1&6750 & $-$1&0365 & 25&6041 & \multicolumn{2}{r}{166122$^{***}$} & $-$16&9757$^{***}$ \\
    Maize spot & 6&1595 & $-$6&6013 & $-$0&0089 & 1&1484 & $-$0&0413 & 1&8158 & \multicolumn{2}{r}{832$^{***}$} & $-$20&9247$^{***}$ \\
    Soybean futures & 8&2147 & $-$10&6594 & 0&0112 & 1&3667 & $-$0&1940 & 3&8182 & \multicolumn{2}{r}{3708$^{***}$} & $-$41&6938$^{***}$ \\
    Soybean spot & 6&3398 & $-$6&5199 & 0&0030 & 1&1714 & $-$0&0741 & 2&1657 & \multicolumn{2}{r}{1186$^{***}$} & $-$40&6646$^{***}$ \\
    Rice futures & 22&9851 & $-$22&2667 & $-$0&0013 & 1&6060 & 0&2449 & 26&9501 & \multicolumn{2}{r}{182908$^{***}$} & $-$14&5382$^{***}$ \\
    Rice spot & 7&5127 & $-$7&5856 & $-$0&0016 & 0&3940 & $-$0&0532 & 60&4812 & \multicolumn{2}{r}{920898$^{***}$} & $-$15&4142$^{***}$ \vspace{2mm}\\

    \multicolumn{17}{l}{\textit{Panel B: Reconstructed medium-term components}} \\
    Wheat futures & 7&6965 & $-$8&1793 & $-$0&0177 & 0&9641 & 0&2219 & 4&5807 & \multicolumn{2}{r}{5332$^{***}$} & $-$19&8646$^{***}$ \\
    Wheat spot & 3&5817 & $-$1&9797 & $-$0&0080 & 0&4640 & 0&3342 & 2&4102 & \multicolumn{2}{r}{1575$^{***}$} & $-$17&0459$^{***}$ \\
    Maize futures & 8&1562 & $-$8&9943 & 0&0064 & 0&9299 & $-$0&1772 & 9&6671 & \multicolumn{2}{r}{23559$^{***}$} & $-$19&5700$^{***}$ \\
    Maize spot & 2&7320 & $-$3&5977 & 0&0016 & 0&6287 & $-$0&0457 & 1&6728 & \multicolumn{2}{r}{707$^{***}$} & $-$17&5518$^{***}$ \\
    Soybean futures & 4&7404 & $-$6&2831 & $-$0&0049 & 0&7551 & $-$0&5431 & 6&1196 & \multicolumn{2}{r}{9725$^{***}$} & $-$18&7536$^{***}$ \\
    Soybean spot & 2&8535 & $-$3&3327 & 0&0094 & 0&6382 & $-$0&0700 & 1&7676 & \multicolumn{2}{r}{792$^{***}$} & $-$18&6482$^{***}$ \\
    Rice futures & 10&1721 & $-$10&7491 & $-$0&0120 & 1&0073 & $-$0&4368 & 18&2694 & \multicolumn{2}{r}{842189$^{***}$} & $-$21&4780$^{***}$ \\
    Rice spot & 2&5539 & $-$1&9477 & $-$0&0001 & 0&2811 & 0&0955 & 10&3078 & \multicolumn{2}{r}{26758$^{***}$} & $-$16&9298$^{***}$  \vspace{2mm}\\

    \multicolumn{17}{l}{\textit{Panel C: Reconstructed long-term components}} \\
    Wheat futures & 1&5916 & $-$1&6421 & 0&0150 & 0&3534 & $-$0&1409 & 2&3420 & \multicolumn{2}{r}{1401$^{***}$} & $-$6&9721$^{***}$ \\
    Wheat spot & 0&9262 & $-$0&8391 & 0&0135 & 0&2341 & 0&3944 & 1&7502 & \multicolumn{2}{r}{928$^{***}$} & $-$6&2368$^{***}$ \\
    Maize futures & 0&9410 & $-$1&2907 & $-$0&0117 & 0&3563 & $-$0&4489 & 1&0892 & \multicolumn{2}{r}{502$^{***}$} & $-$7&1483$^{***}$ \\
    Maize spot & 0&9998 & $-$0&8533 & 0&0113 & 0&2794 & 0&1554 & 0&8806 & \multicolumn{2}{r}{220$^{***}$} & $-$7&4820$^{***}$ \\
    Soybean futures & 0&8770 & $-$1&9077 & 0&0296 & 0&3157 & $-$0&9038 & 4&0693 & \multicolumn{2}{r}{4991$^{***}$} & $-$8&3336$^{***}$ \\
    Soybean spot & 0&7935 & $-$0&9941 & 0&0078 & 0&2556 & $-$0&2577 & 0&9651 & \multicolumn{2}{r}{301$^{***}$} & $-$7&5105$^{***}$ \\
    Rice futures & 1&2509 & $-$1&3200 & 0&0093 & 0&3327 & $-$0&2918 & 1&3893 & \multicolumn{2}{r}{572$^{***}$} & $-$7&2209$^{***}$ \\
    Rice spot & 1&8104 & $-$0&9233 & 0&0028 & 0&2321 & 2&2497 & 16&1078 & \multicolumn{2}{r}{70416$^{***}$} & $-$7&7542$^{***}$ \\    
  \bottomrule
    \end{tabular}
    }%
  \begin{flushleft}
    \footnotesize
    \justifying Note: This table reports the descriptive statistics for the reconstructed short-, medium-, and long-term components of futures and spot return series for wheat, maize, soybean, and rice. *** denotes the significance at the 1\% level.
\end{flushleft} 
  \label{Tab:Reconstructed_series_Stat_Test}%
\end{table}%

\subsection{Static analysis of risk spillovers}

After reconstructing the short-, medium-, and long-term components of futures and spot returns for each staple crop, we first adopt the $R^{2}$ decomposed connectedness approach to calculate risk connectedness measures between all components based on the full sample.

\begin{figure}[!h]
  \centering
  \includegraphics[width=0.75\linewidth]{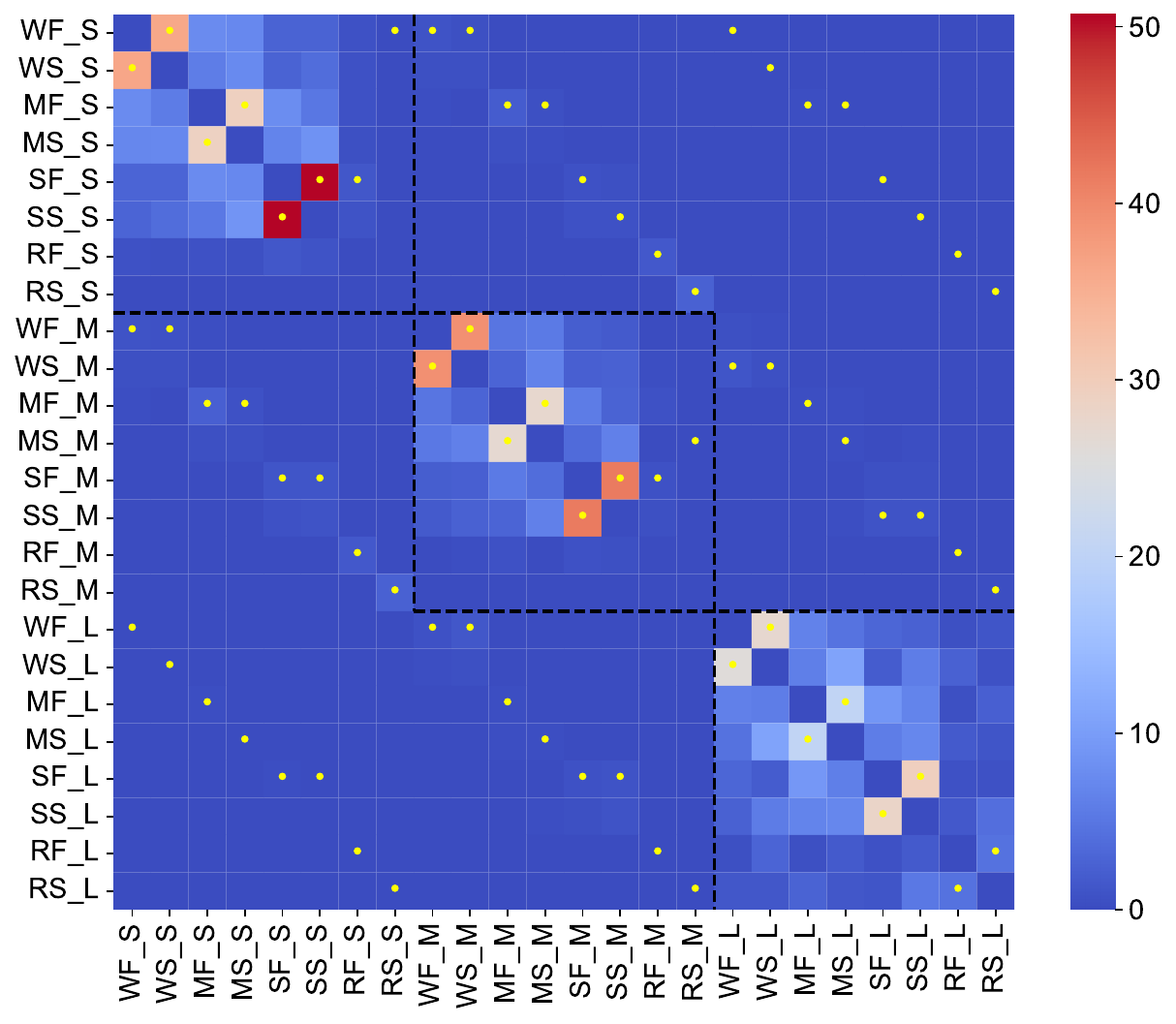}
  \caption{Heatmap of static risk spillovers across all components. Here, the first letters of the labels represent the grain type: W for wheat, M for maize, S for soybean, and R for rice. The second letters indicate the market type: F for futures and S for spot markets. The third letters correspond to the timescale: S for short-term, M for medium-term, and L for long-term components.}
\label{Fig:Heatmap_Static_AcrossComponents}
\end{figure}

Figure~\ref{Fig:Heatmap_Static_AcrossComponents} presents the static risk spillovers across all components. Each square in each row (column) represents the connectedness received (transmitted) by the corresponding component from (to) each of the other components, and the solid yellow circles indicate the highest risk connectedness levels within their respective ranges. In this figure, two major spillover patterns in international food markets stand out. First, there are stronger risk spillovers among components of the same timescale across different grains, which is evident in the three diagonal sub-blocks. Exceptions are observed with the short- and medium-term components of rice, but they exhibit small risk spillovers. Second, significant risk spillovers exist between components at different timescales within the same grain, as shown by the lines formed by solid yellow circles, although there are a few exceptions with low spillovers.

Therefore, in order to elucidate risk spillovers within the staple food markets in a more focused manner, we next explore risk spillovers between same-timescale components of different grains and between same-grain components of different timescales, respectively. This dual analysis offers a comprehensive view of risk transmission from both cross-grain and cross-timescale perspectives.

Table~\ref{Tab:Static_Connectedness_Commodity} reports the static risk spillover indices among same-timescale components of different grain futures and spot returns. It can be found that the TCI index in each panel is large, implying significant risk spillovers across diverse international food submarkets. For short-term components, the TCI reaches 47.19\%, with soybean contributing the most. The TO values associated with soybean futures and spot are 71.51\% and 71.20\%, respectively, followed by maize and wheat. In contrast, rice contributes very little to risk spillovers, with TO values of only 4.56\% and 0.05\% for futures and spot. The results for FROM are broadly similar to those for TO, with only minor numerical differences. Generally, staple foods with higher risk spillovers tend to receive greater risk spillovers. This finding holds true for both the medium- and long-term components as well. Overall, the soybean market plays an important role in transmitting and receiving risks in the global food markets, while rice exhibits few spillovers, which is related to its own particularities of low internationalization and high self-sufficiency in major consumers. Specifically, determined by climate conditions and dietary preferences, Asia serves as the main region for rice production and consumption, accounting for 88.91\% of global production and 83.90\% of consumption in 2023\footnote{\url{https://apps.fas.usda.gov/psdonline/app/index.html\#/app/downloads}}. Compared to the other three staple grains, rice is less internationalized, with limited futures trading volumes, low market liquidity, and relatively stable production and consumption structures.

The sign of NET reflects whether the component is a net risk transmitter or receiver. However, the NET values in each panel are quite small, indicating minimal differences between risk transmission and risk reception for each component. This also suggests that the futures and spot components of staple crops tend to be in a balanced state of cross-grain risk connectedness. Furthermore, except for rice, the risk spillovers between futures and spot components of the same grain are more significant than spillovers between futures and spot components of distinct grains. For example, the short-term component of wheat futures exhibits the spillover index of 57.20\% to others, with 36.00\% directed specifically towards wheat spot. Conversely, the FROM index corresponding to the short-term component of wheat futures is 57.01\%, of which the risk spillover from wheat spot accounts for 35.88\%.

\begin{table}[!ht]
  \centering
  \setlength{\abovecaptionskip}{0pt}
  \setlength{\belowcaptionskip}{10pt}
  \caption{Static connectedness indices between same-timescale components of different grains} 
  \setlength\tabcolsep{3pt}   \resizebox{\textwidth}{!}{ 
    \begin{tabular}{l r@{.}l r@{.}l r@{.}l r@{.}l r@{.}l r@{.}l r@{.}l r@{.}l r@{.}l}
    \toprule
         & \multicolumn{2}{c}{Wheat futures} & \multicolumn{2}{c}{Wheat spot} & \multicolumn{2}{c}{Maize futures} & \multicolumn{2}{c}{Maize spot} & \multicolumn{2}{c}{Soybean futures} & \multicolumn{2}{c}{Soybean spot} & \multicolumn{2}{c}{Rice futures} & \multicolumn{2}{c}{Rice spot} & \multicolumn{2}{c}{FROM}  \\
    \midrule
    \multicolumn{19}{l}{\textit{Panel A: Short-term components}} \\
    Wheat futures & 0&00 & 35&88 & 7&60 & 7&25 & 2&81 & 2&75 & 0&70 & 0&02 & 57&01 \\
    Wheat spot & 36&00 & 0&00 & 5&85 & 7&60 & 2&72 & 3&88 & 0&60 & 0&01 & 56&66 \\ 
    Maize futures & 7&60 & 5&77 & 0&00 & 29&32 & 7&77 & 5&02 & 0&66 & 0&02 & 56&15 \\ 
    Maize spot & 7&07 & 7&37 & 28&72 & 0&00 & 6&65 & 8&48 & 0&53 & 0&00 & 58&83 \\ 
    Soybean futures & 2&94 & 2&92 & 7&68 & 7&14 & 0&00 & 50&24 & 1&19 & 0&00 & 72&11 \\ 
    Soybean spot & 2&86 & 3&96 & 5&31 & 8&84 & 50&28 & 0&00 & 0&87 & 0&00 & 72&11 \\ 
    Rice futures & 0&71 & 0&58 & 0&66 & 0&50 & 1&27 & 0&82 & 0&00 & 0&00 & 4&55  \\
    Rice spot & 0&03 & 0&01 & 0&03 & 0&01 & 0&01 & 0&00 & 0&01 & 0&00 & 0&09  \\
    TO & 57&20 & 56&50 & 55&84 & 60&65 & 71&51 & 71&20 & 4&56 & 0&05 & \multicolumn{2}{c}{TCI}  \\
    NET & 0&19 & $-$0&16 & $-$0&30 & 1&82 & $-$0&61 & $-$0&92 & 0&01 & $-$0&03 & 47&19  \vspace{2mm}\\
    \multicolumn{19}{l}{\textit{Panel B: Medium-term components}} \\
    Wheat futures & 0&00 & 38&47 & 4&74 & 5&51 & 2&27 & 1&69 & 0&16 & 0&06 & 52&90 \\
    Wheat spot & 38&43 & 0&00 & 2&94 & 6&67 & 2&26 & 2&46 & 0&31 & 0&02 & 53&09 \\ 
    Maize futures & 4&87 & 2&90 & 0&00 & 27&16 & 5&62 & 2&77 & 0&65 & 0&03 & 44&00 \\ 
    Maize spot & 5&30 & 6&45 & 26&35 & 0&00 & 3&81 & 6&05 & 0&19 & 0&13 & 48&29 \\ 
    Soybean futures & 2&23 & 2&27 & 5&45 & 3&98 & 0&00 & 40&55 & 0&67 & 0&03 & 55&17 \\ 
    Soybean spot & 1&71 & 2&53 & 2&86 & 6&20 & 40&78 & 0&00 & 0&49 & 0&02 & 54&60 \\ 
    Rice futures & 0&15 & 0&34 & 0&74 & 0&18 & 0&73 & 0&49 & 0&00 & 0&01 & 2&63  \\
    Rice spot & 0&08 & 0&02 & 0&05 & 0&18 & 0&03 & 0&02 & 0&01 & 0&00 & 0&40  \\
    TO & 52&76 & 52&98 & 43&13 & 49&87 & 55&50 & 54&03 & 2&49 & 0&31 & \multicolumn{2}{c}{TCI}  \\
    NET & $-$0&14 & $-$0&11 & $-$0&86 & 1&59 & 0&33 & $-$0&57 & $-$0&14 & $-$0&09 & 38&88  \vspace{2mm}\\
    \multicolumn{19}{l}{\textit{Panel C: Long-term components}} \\
    Wheat futures & 0&00 & 26&88 & 6&55 & 4&61 & 3&04 & 2&50 & 0&40 & 1&02 & 45&00 \\
    Wheat spot & 25&38 & 0&00 & 6&09 & 10&89 & 1&87 & 5&86 & 2&49 & 0&70 & 53&28 \\ 
    Maize futures & 6&31 & 5&99 & 0&00 & 20&34 & 8&89 & 6&81 & 0&58 & 2&26 & 51&17 \\ 
    Maize spot & 4&42 & 10&96 & 20&35 & 0&00 & 6&01 & 6&97 & 1&67 & 1&04 & 51&42 \\ 
    Soybean futures & 3&13 & 1&95 & 9&20 & 6&28 & 0&00 & 28&96 & 0&72 & 0&66 & 50&89 \\ 
    Soybean spot & 2&43 & 5&68 & 6&88 & 6&95 & 27&86 & 0&00 & 1&57 & 4&25 & 55&62 \\ 
    Rice futures & 0&42 & 2&95 & 0&57 & 1&80 & 0&66 & 1&66 & 0&00 & 4&41 & 12&47  \\
    Rice spot & 1&37 & 1&23 & 2&69 & 1&27 & 1&02 & 5&46 & 4&34 & 0&00 & 17&38  \\
    TO & 43&46 & 55&64 & 52&31 & 52&15 & 49&33 & 58&22 & 11&77 & 14&33 & \multicolumn{2}{c}{TCI}  \\
    NET & $-$1&53 & 2&36 & 1&14 & 0&72 & $-$1&55 & 2&61 & $-$0&70 & $-$3&05 & 42&15 \\
  \bottomrule
    \end{tabular}
    }%
   \begin{flushleft}
     \footnotesize
 \justifying Note: This table presents static risk spillover indices between same-timescale components across different grains, where Panels A, B, and C correspond to the results for short-, medium-, and long-term components, respectively.
 \end{flushleft} 
  \label{Tab:Static_Connectedness_Commodity}%
\end{table}%

Table~\ref{Tab:Static_Connectedness_Timescale} presents the static risk spillover indices across components at different timescales for each grain. As can be seen, the TCI is highest for the soybean market at 59.30\%, followed by wheat and maize, and finally rice. This result reaffirms the limited risk connectedness within the rice market compared to the other three staple crops. A comparison of TO in all panels shows that, for wheat, maize, and soybean, short-term components contribute the most risk spillover to other components, followed by medium- and long-term components. For instance, in wheat futures and spot markets, the TO values for short-term components are 55.95\% and 55.47\%, for medium-term components are 53.98\% and 54.47\%, and for long-term components are 42.97\% and 42.04\%, respectively. In the rice market, however, long-term components exhibit the highest risk spillovers, with TO values of 5.95\% and 6.14\% for futures and spot, followed by medium-term components (1.42\% and 2.84\%) and short-term components (1.39\% and 2.66\%). The results of FROM align closely with the TO results.

In addition, the Net values are all below 2\%, which means that all of the components are in equilibrium of cross-timescale risk connectedness. Besides, it can be observed from Table~\ref{Tab:Static_Connectedness_Timescale} that, with the exception of rice, the risk spillovers between futures and spot components of the same timescale are more prominent than spillovers between futures and spot components of distinct timescales. For example, the TO index of the short-term component of maize futures is 56.19\%, of which the risk spillover to the short-term component of maize spot accounts for 52.75\%. Similarly, the short-term component of maize futures has a FROM index of 55.14\%, with 52.17\% originating from the short-term component of maize spot.

\begin{table}[!ht]
  \centering
  \setlength{\abovecaptionskip}{0pt}
  \setlength{\belowcaptionskip}{10pt}
  \caption{Static connectedness indices between same-grain components of different timescales}
  \setlength\tabcolsep{3pt}   \resizebox{\textwidth}{!}{ 
    \begin{tabular}{l r@{.}l r@{.}l r@{.}l r@{.}l r@{.}l r@{.}l r@{.}l}
    \toprule
         & \multicolumn{2}{c}{Futures STC} & \multicolumn{2}{c}{Spot STC} & \multicolumn{2}{c}{Futures MTC} & \multicolumn{2}{c}{Spot MTC} & \multicolumn{2}{c}{Futures LTC} & \multicolumn{2}{c}{Spot LTC} & \multicolumn{2}{c}{FROM}  \\
    \midrule
    \multicolumn{15}{l}{\textit{Panel A: Wheat}} \\
    Futures STC & 0&00 & 54&05 & 0&81 & 0&67 & 0&14 & 0&06 & 55&74 \\
    Spot STC & 54&20 & 0&00 & 0&74 & 0&42 & 0&05 & 0&07 & 55&47 \\ 
    Futures MTC & 0&85 & 0&82 & 0&00 & 51&26 & 0&47 & 0&51 & 53&91 \\ 
    Spot MTC & 0&65 & 0&43 & 51&17 & 0&00 & 1&26 & 0&57 & 54&07 \\ 
    Futures LTC & 0&17 & 0&07 & 0&64 & 1&45 & 0&00 & 40&83 & 43&17 \\ 
    Spot LTC & 0&08 & 0&09 & 0&63 & 0&67 & 41&05 & 0&00 & 42&52 \\ 
    TO & 55&95 & 55&47 & 53&98 & 54&47 & 42&97 & 42&04 & \multicolumn{2}{c}{TCI}  \\
    NET & 0&21 & 0&00 & 0&07 & 0&40 & $-$0&20 & $-$0&48 & 50&81  \vspace{2mm}\\
    \multicolumn{15}{l}{\textit{Panel B: Maize}} \\
    Futures STC & 0&00 & 52&17 & 1&74 & 0&89 & 0&21 & 0&12 & 55&14 \\
    Spot STC & 52&75 & 0&00 & 0&77 & 0&39 & 0&07 & 0&09 & 54&07 \\ 
    Futures MTC & 2&04 & 1&06 & 0&00 & 40&72 & 0&22 & 0&28 & 44&32 \\ 
    Spot MTC & 1&00 & 0&52 & 41&06 & 0&00 & 0&38 & 0&33 & 43&29 \\ 
    Futures LTC & 0&25 & 0&10 & 0&23 & 0&37 & 0&00 & 42&29 & 43&23 \\ 
    Spot LTC & 0&15 & 0&12 & 0&29 & 0&32 & 42&31 & 0&00 & 43&18 \\ 
    TO & 56&19 & 53&97 & 44&09 & 42&69 & 43&19 & 43&10 & \multicolumn{2}{c}{TCI}  \\
    NET & 1&05 & $-$0&10 & $-$0&23 & $-$0&60 & $-$0&04 & $-$0&07 & 47&21  \vspace{2mm}\\
    \multicolumn{15}{l}{\textit{Panel C: Soybean}} \\
    Futures STC & 0&00 & 70&87 & 0&77 & 0&61 & 0&19 & 0&07 & 72&51 \\
    Spot STC & 70&84 & 0&00 & 0&82 & 0&67 & 0&09 & 0&12 & 72&54 \\ 
    Futures MTC & 1&26 & 1&30 & 0&00 & 53&37 & 0&55 & 0&70 & 57&18 \\ 
    Spot MTC & 0&97 & 1&03 & 53&30 & 0&00 & 1&06 & 0&90 & 57&26 \\ 
    Futures LTC & 0&34 & 0&21 & 0&70 & 1&20 & 0&00 & 45&78 & 48&24 \\ 
    Spot LTC & 0&15 & 0&21 & 0&87 & 1&04 & 45&83 & 0&00 & 48&10 \\ 
    TO & 73&55 & 73&63 & 56&46 & 56&89 & 47&73 & 47&57 & \multicolumn{2}{c}{TCI}  \\
    NET & 1&04 & 1&09 & $-$0&71 & $-$0&37 & $-$0&51 & $-$0&54 & 59&30  \vspace{2mm}\\
    \multicolumn{15}{l}{\textit{Panel D: Rice}} \\
    Futures STC & 0&00 & 0&00 & 1&31 & 0&02 & 0&05 & 0&01 & 1&39 \\
    Spot STC & 0&00 & 0&00 & 0&00 & 2&56 & 0&00 & 0&10 & 2&67 \\ 
    Futures MTC & 1&31 & 0&00 & 0&00 & 0&01 & 0&10 & 0&01 & 1&43 \\ 
    Spot MTC & 0&02 & 2&56 & 0&01 & 0&00 & 0&03 & 0&23 & 2&84 \\ 
    Futures LTC & 0&04 & 0&00 & 0&10 & 0&02 & 0&00 & 5&78 & 5&96 \\ 
    Spot LTC & 0&01 & 0&10 & 0&01 & 0&23 & 5&78 & 0&00 & 6&12 \\ 
    TO & 1&39 & 2&66 & 1&42 & 2&84 & 5&95 & 6&14 & \multicolumn{2}{c}{TCI}  \\
    NET & 0&00 & 0&00 & 0&00 & 0&00 & 0&00 & 0&01 & 3&40 \\
  \bottomrule
    \end{tabular}
    }%
   \begin{flushleft}
     \footnotesize
 \justifying Note: This table reports static risk spillover indices among same-grain components of different timescales, in which Panels A, B, C, and D contain results for wheat, maize, soybean, and rice, respectively. STC, MTC, and LTC denote the short-, medium-, and long-term components.
 \end{flushleft} 
  \label{Tab:Static_Connectedness_Timescale}%
\end{table}%

\subsection{Dynamic analysis of risk spillovers}

Following the static full-sample analysis, we further employ a one-year rolling window to examine dynamic risk spillovers within the global futures and spot markets of diverse staple crops. 

Table~\ref{Tab:Dynamic_Average_Connectedness_Commodity} reports the dynamic average connectedness measures for the same-timescale components across various grains. The dynamic average TCI in each panel is significantly higher than the corresponding static TCI in Table~\ref{Tab:Static_Connectedness_Commodity}, implying that the cross-grain risk spillovers in the dynamic analysis are more pronounced than those observed in the static analysis. For the long-term components, in particular, the dynamic average TCI reaches 81.30\%. This high level may be attributed to the long duration of risk spillovers between long-term components, resulting in persistently high spillovers during the rolling window. Additionally, soybean and maize contribute the most to the risk spillovers, with their futures and spot components having TO values exceeding 89\%. Rice, with a relatively lower contribution, still shows a TO index of more than 53\%. The FROM values demonstrate that staple food markets are exposed to large risk spillovers alongside substantial risk spill-ins, consistent with the findings from the static analysis.

Combining the results of short-, medium-, and long-term components, soybean emerges as the primary risk transmitter and receiver in the global markets of staple crops, followed by maize and wheat, with rice showing the least effect. Furthermore, the NET values in Panels A and B are within 3\%, suggesting that the short- and medium-term components of each grain's futures and spot returns generally maintain a balanced level of risk connectedness. However, the NET results in Panel C reveal that, in the dynamic analysis, the long-term components of both soybean and maize are net risk transmitters, while the long-term components of rice futures and spot are net risk receivers.

\begin{table}[!ht]
  \centering
  \setlength{\abovecaptionskip}{0pt}
  \setlength{\belowcaptionskip}{10pt}
  \caption{Average connectedness measures between same-timescale components of different grains} 
  \setlength\tabcolsep{3pt}   \resizebox{\textwidth}{!}{ 
    \begin{tabular}{l r@{.}l r@{.}l r@{.}l r@{.}l r@{.}l r@{.}l r@{.}l r@{.}l r@{.}l}
    \toprule
         & \multicolumn{2}{c}{Wheat futures} & \multicolumn{2}{c}{Wheat spot} & \multicolumn{2}{c}{Maize futures} & \multicolumn{2}{c}{Maize spot} & \multicolumn{2}{c}{Soybean futures} & \multicolumn{2}{c}{Soybean spot} & \multicolumn{2}{c}{Rice futures} & \multicolumn{2}{c}{Rice spot} & \multicolumn{2}{c}{FROM}  \\
    \midrule
    \multicolumn{19}{l}{\textit{Panel A: Short-term components}} \\
    Wheat futures & 0&00 & 38&12 & 10&46 & 7&54 & 3&42 & 2&90 & 1&30 & 0&34 & 64&09 \\
    Wheat spot & 38&60 & 0&00 & 6&96 & 8&32 & 3&31 & 3&90 & 1&23 & 0&27 & 62&59 \\ 
    Maize futures & 10&29 & 6&91 & 0&00 & 35&35 & 8&88 & 6&69 & 1&37 & 0&26 & 69&74 \\ 
    Maize spot & 7&53 & 8&24 & 35&82 & 0&00 & 7&25 & 8&12 & 1&16 & 0&30 & 68&42 \\ 
    Soybean futures & 3&44 & 3&49 & 8&87 & 7&54 & 0&00 & 50&73 & 1&79 & 0&35 & 76&20 \\ 
    Soybean spot & 2&95 & 4&05 & 6&83 & 8&28 & 50&77 & 0&00 & 1&48 & 0&39 & 74&74 \\ 
    Rice futures & 1&47 & 1&37 & 1&65 & 1&35 & 2&14 & 1&71 & 0&00 & 0&58 & 10&29  \\
    Rice spot & 0&51 & 0&42 & 0&43 & 0&48 & 0&58 & 0&61 & 0&61 & 0&00 & 3&65  \\
    TO & 64&80 & 62&61 & 71&02 & 68&85 & 76&36 & 74&66 & 8&93 & 2&49 & \multicolumn{2}{c}{TCI}  \\
    NET & 0&71 & 0&01 & 1&28 & 0&43 & 0&15 & $-$0&08 & $-$1&36 & $-$1&16 & 53&72  \vspace{2mm}\\
    \multicolumn{19}{l}{\textit{Panel B: Medium-term components}} \\
    Wheat futures & 0&00 & 35&18 & 8&88 & 6&56 & 3&49 & 2&98 & 1&84 & 1&06 & 59&99 \\
    Wheat spot & 35&51 & 0&00 & 5&04 & 7&87 & 3&92 & 3&85 & 2&08 & 0&83 & 59&11 \\ 
    Maize futures & 8&75 & 5&04 & 0&00 & 29&72 & 8&18 & 6&00 & 2&31 & 0&56 & 60&55 \\ 
    Maize spot & 6&61 & 7&91 & 30&53 & 0&00 & 5&22 & 6&91 & 1&61 & 0&98 & 59&77 \\ 
    Soybean futures & 3&44 & 3&95 & 8&03 & 5&28 & 0&00 & 42&45 & 1&75 & 0&73 & 65&65 \\ 
    Soybean spot & 3&02 & 4&01 & 6&22 & 7&06 & 42&88 & 0&00 & 1&76 & 0&79 & 65&74 \\ 
    Rice futures & 2&23 & 2&51 & 2&88 & 2&07 & 2&13 & 2&16 & 0&00 & 1&65 & 15&63  \\
    Rice spot & 1&49 & 1&19 & 0&88 & 1&44 & 1&14 & 1&22 & 1&74 & 0&00 & 9&11  \\
    TO & 61&05 & 59&80 & 62&47 & 60&01 & 66&96 & 65&56 & 13&08 & 6&59 & \multicolumn{2}{c}{TCI}  \\
    NET & 1&06 & 0&69 & 1&92 & 0&24 & 1&32 & $-$0&18 & $-$2&55 & $-$2&51 & 49&44  \vspace{2mm}\\
    \multicolumn{19}{l}{\textit{Panel C: Long-term components}} \\
    Wheat futures & 0&00 & 21&37 & 14&86 & 9&63 & 10&40 & 8&34 & 7&22 & 8&76 & 80&59 \\
    Wheat spot & 20&84 & 0&00 & 12&22 & 13&80 & 8&15 & 10&98 & 10&77 & 6&25 & 83&00 \\ 
    Maize futures & 13&62 & 11&80 & 0&00 & 20&26 & 14&26 & 11&56 & 6&90 & 8&54 & 86&94 \\ 
    Maize spot & 8&71 & 13&65 & 21&38 & 0&00 & 11&76 & 13&76 & 9&25 & 6&72 & 85&23 \\ 
    Soybean futures & 9&34 & 7&75 & 14&51 & 11&64 & 0&00 & 29&54 & 6&17 & 7&45 & 86&40 \\ 
    Soybean spot & 6&94 & 10&12 & 11&43 & 13&19 & 29&00 & 0&00 & 9&06 & 6&76 & 86&51 \\ 
    Rice futures & 8&89 & 12&63 & 9&00 & 11&38 & 8&14 & 11&94 & 0&00 & 9&33 & 71&30  \\
    Rice spot & 10&13 & 8&10 & 12&68 & 9&18 & 10&58 & 9&99 & 9&78 & 0&00 & 70&44  \\
    TO & 78&47 & 85&42 & 96&07 & 89&07 & 92&28 & 96&10 & 59&16 & 53&82 & \multicolumn{2}{c}{TCI}  \\
    NET & $-$2&12 & 2&42 & 9&13 & 3&85 & 5&89 & 9&60 & $-$12&14 & $-$16&63 & 81&30 \\
  \bottomrule
    \end{tabular}
    }%
   \begin{flushleft}
     \footnotesize
 \justifying Note: This table shows the average connectedness measures for same-timescale components across different staple foods computed by the one-year rolling window, where Panels A, B, and C refer to the results for short-, medium-, and long-term components, respectively.
 \end{flushleft} 
  \label{Tab:Dynamic_Average_Connectedness_Commodity}%
\end{table}%

\begin{figure}[!h]
  \centering
  \includegraphics[width=0.65\linewidth]{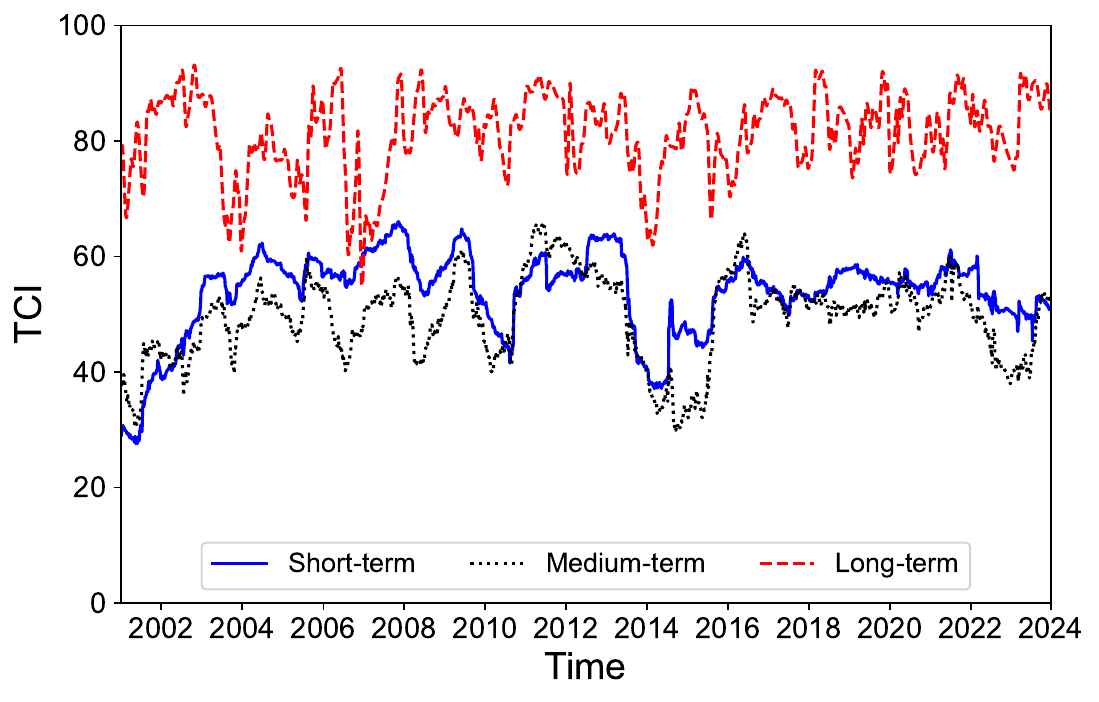}
  \caption{Evolution of the total risk spillover index TCI corresponding to the short-term components, the medium-term components, and the long-term components of the futures and spot returns for wheat, maize, soybean, and rice.} 
\label{Fig:TCI:CrossGrain}
\end{figure}

To observe changes in TCI over time more intuitively, Figure~\ref{Fig:TCI:CrossGrain} illustrates the dynamic evolution of the TCI index among same-timescale components of different grains. Corresponding to the average results in Table~\ref{Tab:Dynamic_Average_Connectedness_Commodity}, the risk spillovers between long-term components consistently remain high, followed by the short- and medium-term components. Notably, the TCI index for components at each timescale shows a significant decrease around 2014. In recent years, however, the risk spillovers in the medium- and long-term components have shown a clear upward trend. Moreover, each TCI index shows substantial fluctuations over time, and all of them experienced elevated risk spillovers during the 2006 global food crisis, the 2008 global financial crisis, the 2012 global food crisis, and the more recent food crisis since 2020. These findings reinforce the evidence of significant dynamic risk transmission among same-timescale components across different staple foods, highlighting periods of heightened interconnectedness that correlate with major global disruptions.

\begin{figure}[!h]
  \centering
  \includegraphics[width=0.43\linewidth]{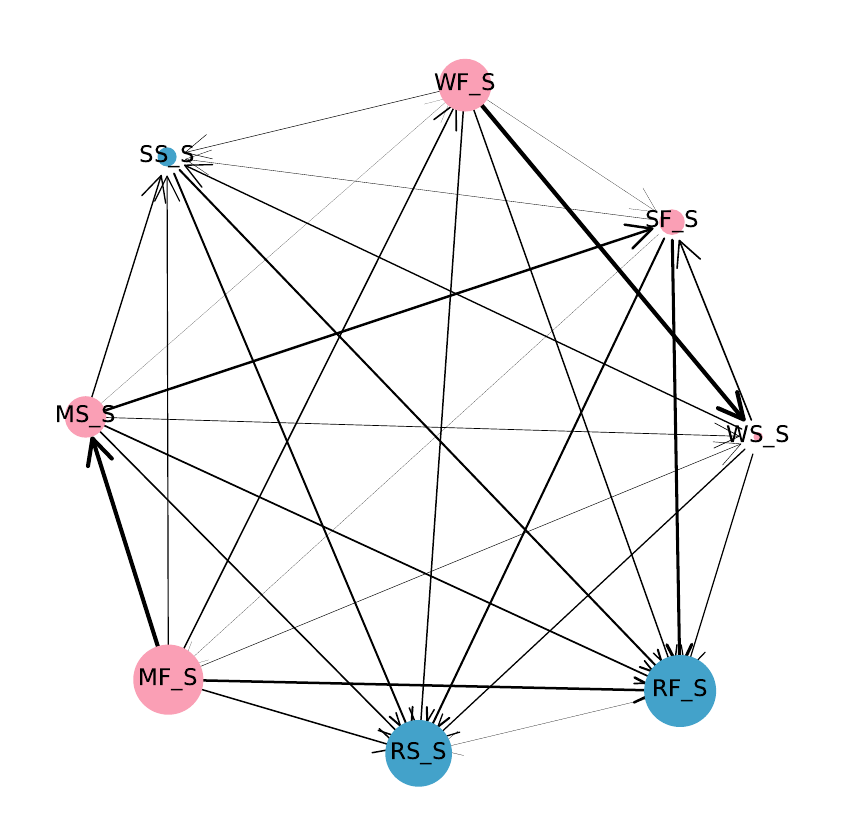}
  \includegraphics[width=0.43\linewidth]{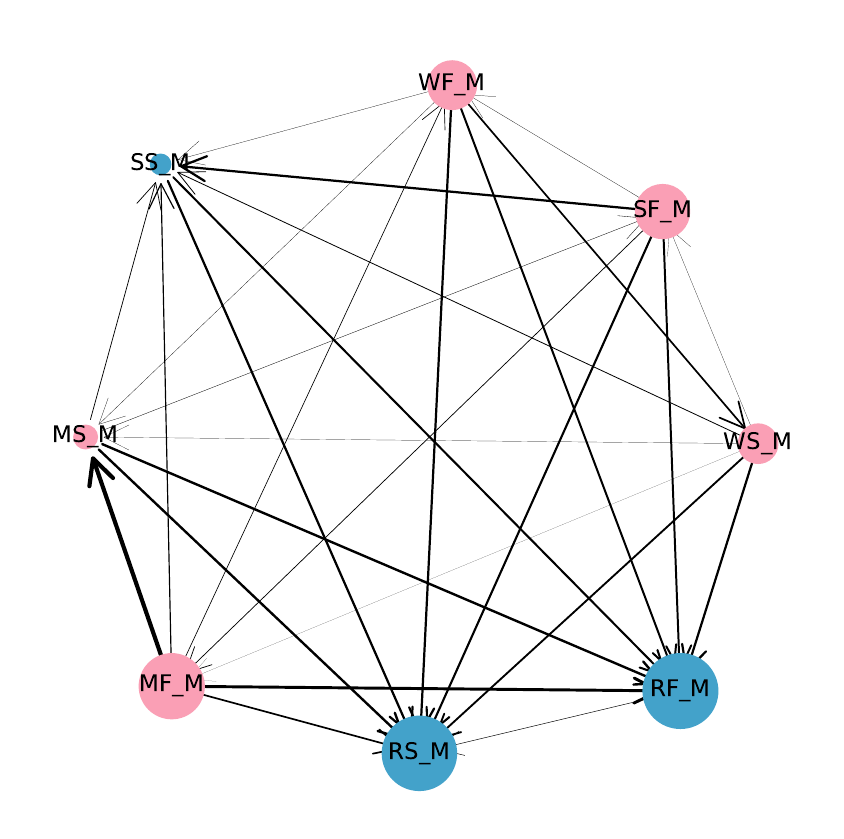}
  \includegraphics[width=0.43\linewidth]{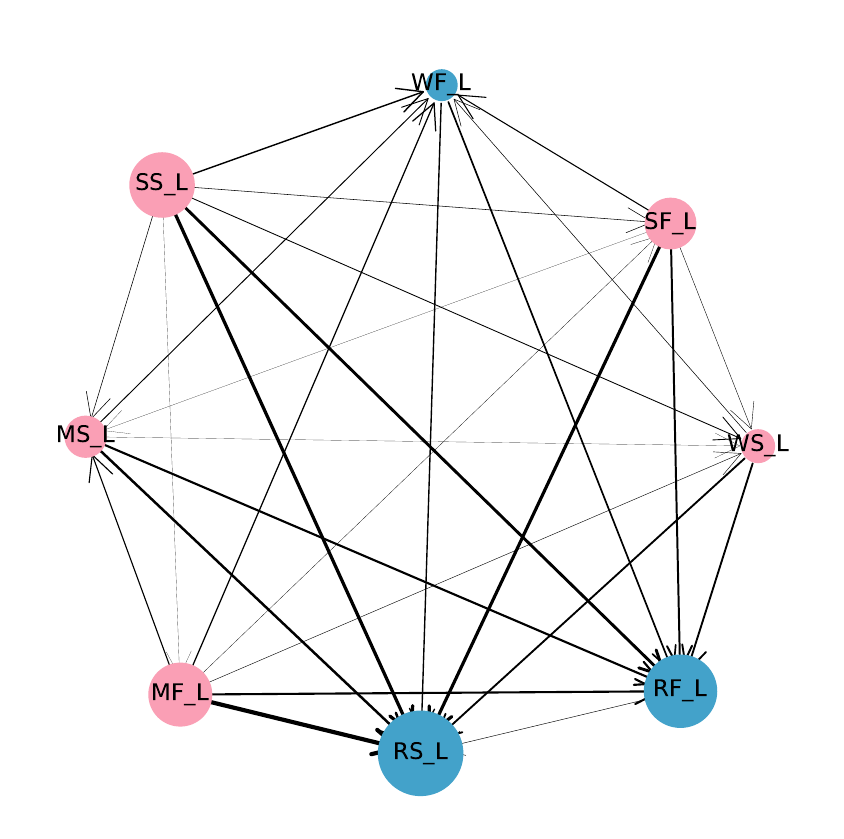}
  \caption{Net risk spillover networks between same-timescale components of futures and spot returns for different staple grains.} 
\label{Fig:Network:CrossGrain}
\end{figure}

The net risk spillover network is further constructed to visualize the risk spillovers between diverse components. Figure~\ref{Fig:Network:CrossGrain} presents the net risk spillover networks among same-timescale components of different grain futures and spot returns, where the first, second, and third subfigures represent networks for the short-, medium-, and long-term components, respectively. These networks are weighted directed networks, with the net pairwise directional connectedness index NPDC serving as the weight of each edge. Consequently, the thickness of the edges reflects the magnitude of the net risk spillover effect between any two nodes. The size of each node is determined by its out-strength, and larger nodes indicate greater risk spillovers. The node color refers to the role of each node within the net risk spillover network, in which pink denotes positive net risk spillovers (risk transmitter), while blue represents negative net risk spillovers (risk receiver).

As shown in Figure~\ref{Fig:Network:CrossGrain}, the components of maize futures and wheat futures are the main risk transmitters in the net risk spillover network of short-term components. The paths of the largest NPDC values are from wheat futures to wheat spot and from maize futures to maize spot, proving the existence of substantial risk spillover effects from agricultural futures markets to their corresponding spot markets. In addition, there are large risk spillovers between the short-term components of maize and soybean, likely due to their substitutive relationship in use. Wheat futures also show evident risk spillovers to rice futures in short-term components, possibly owing to their shared role as major food rations. The net risk spillover network of medium-term components is broadly similar to that of short-term components, though in the medium-term network, soybean futures and wheat spot also act as primary risk transmitters. In the net risk spillover network of long-term components, the dominant risk spillover paths lead from futures and spots of maize and soybean to those of rice. Furthermore, across the short-, medium-, and long-term networks, rice futures and spot consistently play the role of main risk receivers.

Turning to cross-timescale risk spillovers, Table~\ref{Tab:Dynamic_Average_Connectedness_Timescale} gives the dynamic average connectedness measures for the same-grain components across different timescales. The TCI results align with the static full-sample analysis in Table~\ref{Tab:Static_Connectedness_Timescale}, which indicates that the internal spillover risks corresponding to soybean, maize, and wheat are high, whereas the spillover risk for rice remains low. The TO and FROM values show that, for soybean, maize, and wheat, the short-term component plays a critical role in dynamic risk spillovers within their individual markets, followed by the medium- and long-term components. This may be linked to the higher volatility observed in the short-term components. For rice, however, the long-term component acts as the primary risk transmitter and receiver, despite the low levels of spillovers within the rice market. Furthermore, the NET values in Table~\ref{Tab:Dynamic_Average_Connectedness_Timescale} are all quite small, confirming that these same-grain components of different timescales are in equilibrium of cross-timescale risk connectedness.

\begin{table}[!ht]
  \centering
  \setlength{\abovecaptionskip}{0pt}
  \setlength{\belowcaptionskip}{10pt}
  \caption{Average connectedness measures between same-grain components of different timescales}  
  \setlength\tabcolsep{3pt}   \resizebox{\textwidth}{!}{ 
    \begin{tabular}{l r@{.}l r@{.}l r@{.}l r@{.}l r@{.}l r@{.}l r@{.}l}
    \toprule
         & \multicolumn{2}{c}{Futures STC} & \multicolumn{2}{c}{Spot STC} & \multicolumn{2}{c}{Futures MTC} & \multicolumn{2}{c}{Spot MTC} & \multicolumn{2}{c}{Futures LTC} & \multicolumn{2}{c}{Spot LTC} & \multicolumn{2}{c}{FROM}  \\
    \midrule
    \multicolumn{15}{l}{\textit{Panel A: Wheat}} \\
    Futures STC & 0&00 & 58&26 & 0&78 & 0&74 & 0&14 & 0&10 & 60&02 \\
    Spot STC & 58&38 & 0&00 & 0&75 & 0&49 & 0&08 & 0&09 & 59&79 \\ 
    Futures MTC & 0&89 & 0&88 & 0&00 & 51&88 & 1&01 & 1&10 & 55&74 \\ 
    Spot MTC & 0&77 & 0&52 & 51&65 & 0&00 & 2&11 & 1&10 & 56&15 \\ 
    Futures LTC & 0&18 & 0&12 & 1&47 & 2&84 & 0&00 & 38&23 & 42&83 \\ 
    Spot LTC & 0&14 & 0&14 & 1&54 & 1&52 & 38&33 & 0&00 & 41&68 \\ 
    TO & 60&35 & 59&92 & 56&20 & 57&46 & 41&66 & 40&61 & \multicolumn{2}{c}{TCI}  \\
    NET & 0&33 & 0&14 & 0&46 & 1&31 & $-$1&18 & $-$1&06 & 52&70  \vspace{2mm}\\
    \multicolumn{15}{l}{\textit{Panel B: Maize}} \\
    Futures STC & 0&00 & 62&17 & 1&33 & 0&78 & 0&22 & 0&15 & 64&66 \\
    Spot STC & 62&39 & 0&00 & 0&80 & 0&46 & 0&08 & 0&09 & 63&82 \\ 
    Futures MTC & 1&67 & 1&13 & 0&00 & 48&28 & 0&60 & 1&02 & 52&70 \\ 
    Spot MTC & 0&98 & 0&62 & 48&40 & 0&00 & 0&95 & 1&08 & 52&03 \\ 
    Futures LTC & 0&27 & 0&12 & 1&03 & 1&47 & 0&00 & 45&54 & 48&43 \\ 
    Spot LTC & 0&20 & 0&14 & 1&79 & 1&76 & 45&44 & 0&00 & 49&33 \\ 
    TO & 65&52 & 64&17 & 53&37 & 52&75 & 47&29 & 47&87 & \multicolumn{2}{c}{TCI}  \\
    NET & 0&86 & 0&35 & 0&67 & 0&72 & $-$1&14 & $-$1&46 & 55&16  \vspace{2mm}\\
    \multicolumn{15}{l}{\textit{Panel C: Soybean}} \\
    Futures STC & 0&00 & 72&26 & 0&78 & 0&64 & 0&16 & 0&08 & 73&92 \\
    Spot STC & 72&06 & 0&00 & 1&09 & 0&87 & 0&10 & 0&12 & 74&23 \\ 
    Futures MTC & 1&31 & 1&58 & 0&00 & 59&35 & 0&88 & 1&03 & 64&14 \\ 
    Spot MTC & 1&08 & 1&25 & 59&30 & 0&00 & 1&44 & 1&17 & 64&25 \\ 
    Futures LTC & 0&30 & 0&23 & 1&39 & 1&95 & 0&00 & 52&17 & 56&04 \\ 
    Spot LTC & 0&18 & 0&23 & 1&55 & 1&62 & 52&32 & 0&00 & 55&90 \\ 
    TO & 74&94 & 75&55 & 64&11 & 64&42 & 54&89 & 54&58 & \multicolumn{2}{c}{TCI}  \\
    NET & 1&02 & 1&32 & $-$0&03 & 0&17 & $-$1&16 & $-$1&32 & 64&75  \vspace{2mm}\\
    \multicolumn{15}{l}{\textit{Panel D: Rice}} \\
    Futures STC & 0&00 & 0&57 & 0&76 & 0&24 & 0&09 & 0&05 & 1&71 \\
    Spot STC & 0&57 & 0&00 & 0&07 & 2&07 & 0&04 & 0&13 & 2&88 \\ 
    Futures MTC & 0&76 & 0&07 & 0&00 & 1&70 & 0&54 & 0&49 & 3&56 \\ 
    Spot MTC & 0&23 & 2&06 & 1&69 & 0&00 & 0&49 & 0&56 & 5&03 \\ 
    Futures LTC & 0&08 & 0&04 & 0&52 & 0&47 & 0&00 & 13&62 & 14&73 \\ 
    Spot LTC & 0&05 & 0&12 & 0&46 & 0&52 & 13&60 & 0&00 & 14&75 \\ 
    TO & 1&69 & 2&85 & 3&51 & 5&00 & 14&76 & 14&84 & \multicolumn{2}{c}{TCI}  \\
    NET & $-$0&02 & $-$0&02 & $-$0&05 & $-$0&03 & 0&03 & 0&10 & 7&11 \\
  \bottomrule
    \end{tabular}
    }%
   \begin{flushleft}
     \footnotesize
 \justifying Note: This table presents the average connectedness measures for same-grain components at different timescales calculated based on the one-year rolling window, with Panels A, B, C, and D corresponding to wheat, maize, soybean, and rice, respectively.
 \end{flushleft} 
  \label{Tab:Dynamic_Average_Connectedness_Timescale}%
\end{table}%

\begin{figure}[!h]
  \centering
  \includegraphics[width=0.65\linewidth]{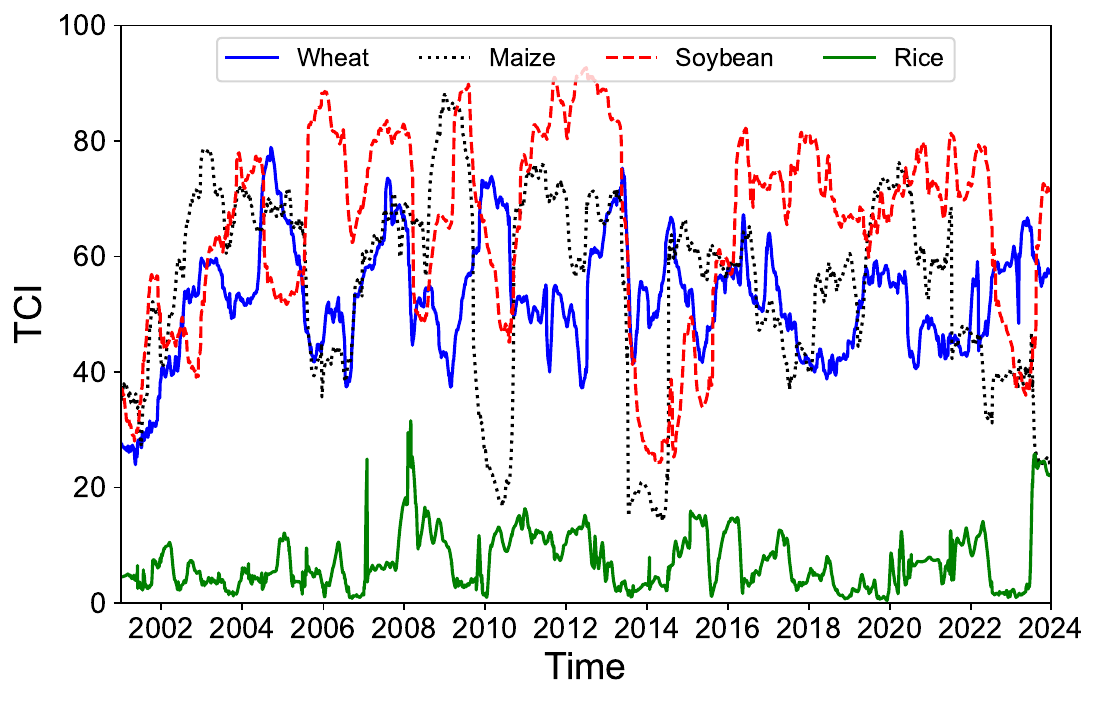}
  \caption{Evolution of the total risk spillover index TCI between different timescale components of futures and spot returns for wheat, maize, soybean, and rice.} 
\label{Fig:TCI:CrossTimescale}
\end{figure}

The dynamic evolution of the TCI index for same-grain components at different timescales is depicted in Figure~\ref{Fig:TCI:CrossTimescale}. Consistent with previous findings, the risk transmission within the rice market is markedly weaker than those observed in the other three staple food markets. Each TCI index varies very dramatically over time, and notable peaks have been identified in grain submarkets during critical periods, such as the 2008 global financial crisis and the food crises of 2006–2008, 2010–2012, and since 2020. It is particularly noteworthy that risk spillovers within the wheat, soybean, and rice markets have been intensifying in recent years, as evidenced by higher TCI levels. Another striking observation is the common sharp decline in TCI values for all grains prior to 2015, followed by a rapid increase. This synchronized shift may be related to the commodity market downturn around 2015, triggered by global economic slowdown and abundant food supply. Overall, these results demonstrate the significant dynamic risk transmission between components of different timescales within the same grain market.

\begin{figure}[!h]
  \centering
  \includegraphics[width=0.43\linewidth]{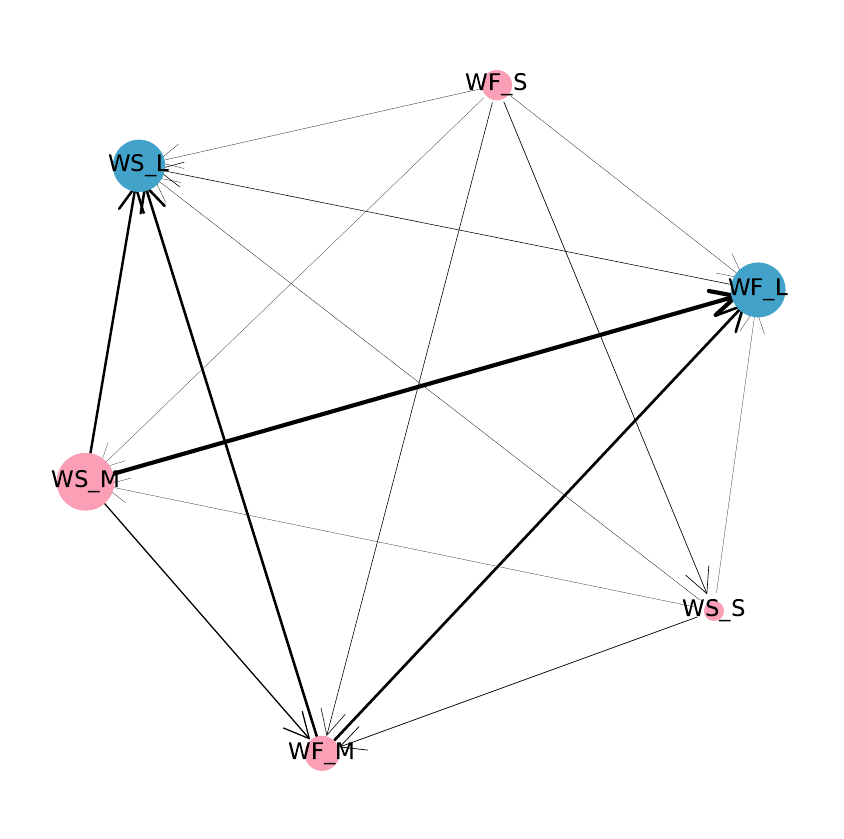}
  \includegraphics[width=0.43\linewidth]{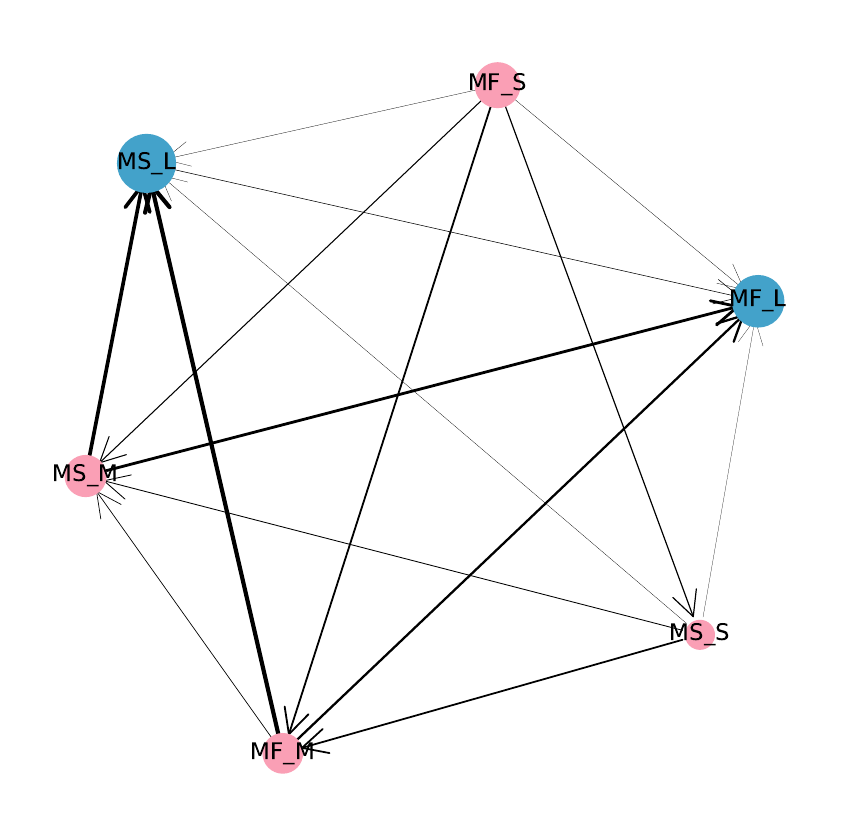}
  \includegraphics[width=0.43\linewidth]{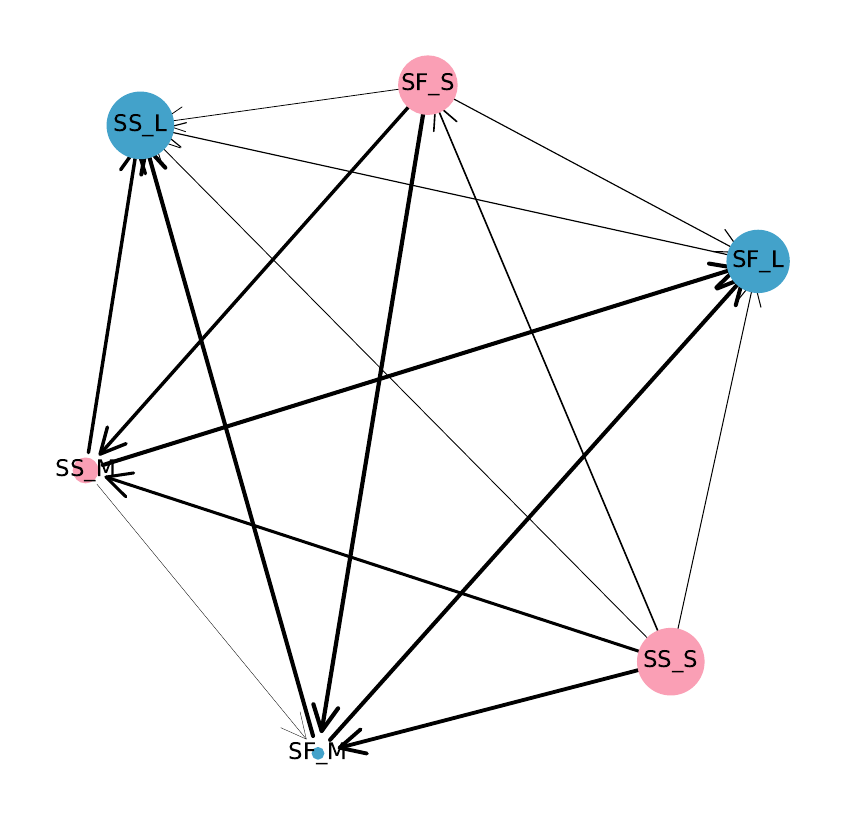}
  \includegraphics[width=0.43\linewidth]{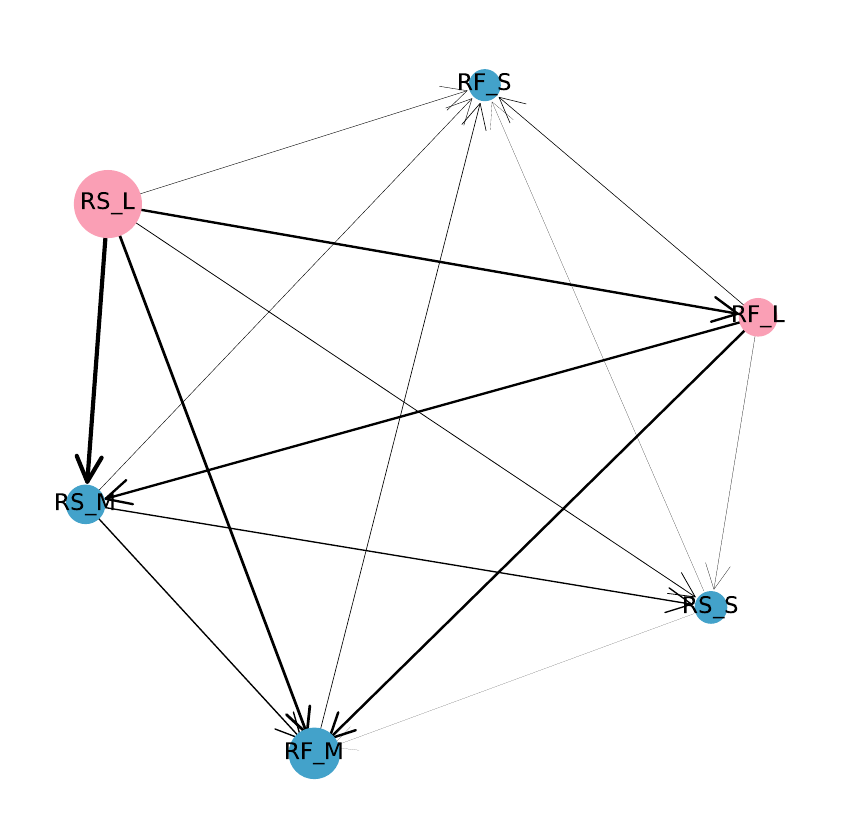}
  \caption{Net risk spillover networks between same-grain components of futures and spot returns at different timescales.} 
\label{Fig:Network:CrossTimescale}
\end{figure}

The net risk spillover networks for same-grain components of diverse timescales are illustrated in Figure~\ref{Fig:Network:CrossTimescale}, with the first, second, third, and fourth networks corresponding to wheat, maize, soybean, and rice, respectively. For wheat, maize, and soybean, the long-term components of futures and spot returns are risk receivers within their respective networks, while the short- and medium-term components serve as risk transmitters, except for the medium-term component of soybean futures, the node of which is quite small. The opposite is true for rice, again highlighting the particularity of the rice market. Besides, in the wheat network, the main net risk spillover paths are directed from the medium-term components toward the long-term components, a channel similarly observed in the maize market. For soybean, the short-term components display stronger risk spillovers than the medium-term ones, resulting in key net spillover paths from short to medium term and from medium to long term. In the rice network, the long-term components exhibit significant net risk spillovers toward the medium-term components, which suggests that the risk transmission pattern within the rice market is distinct from that within the other three staple crop markets.

\subsection{Factors influencing risk spillovers}

After identifying the significant cross-grain and cross-timescale risk spillover effects in global staple food futures and spot markets, we aim to explore the underlying factors influencing these spillovers. Compared to traditional econometric models, machine learning approaches demonstrate significantly enhanced predictive capabilities \citep{Gu-Kelly-Xiu-2020-RevFinancStud}. Among these, the random forest model is particularly well-suited due to its robustness, stability, and generalization capacity. It effectively addresses overfitting, multicollinearity, and nonlinear relationships, making it an ideal choice for further analysis. 

Following the prior studies \citep{Bakas-Triantafyllou-2018-JIntMoneyFinan, Lin-Chang-Hsiao-2019-TranspResPte-LogistTranspRev, Tiwari-Abakah-Adewuyi-Lee-2022-EnergyEcon, Wei-Gephart-Iizumi-Ramankutty-Davis-2023-NatSustain}, we select 12 potential explanatory variables, namely ARMI, FPI, EPI, BDI, GPR, EPU, CPU, GND, PROD, IMP, CONS, and ES. These factors encompass crucial aspects of food systems, including supply, demand, production, and transportation, while also concerning external uncertainties closely related to food security. By investigating their explanatory power for the TCIs corresponding to short-, medium-, and long-term components, as well as wheat, maize, soybean, and rice, we provide a comprehensive understanding of how these factors influence various risk spillovers in staple food markets.

\begin{figure}[!ht]
  \centering
  \includegraphics[width=0.97\linewidth]{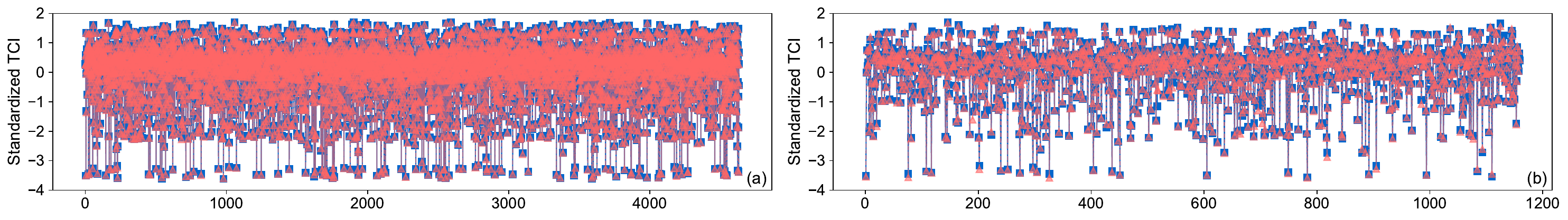}
  \includegraphics[width=0.97\linewidth]{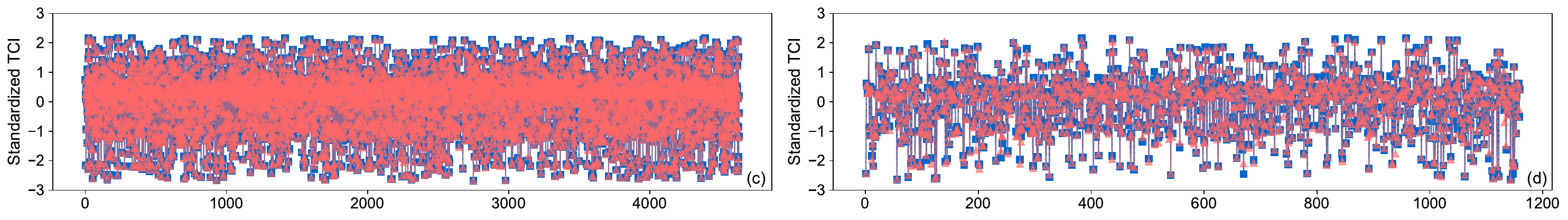}
  \includegraphics[width=0.97\linewidth]{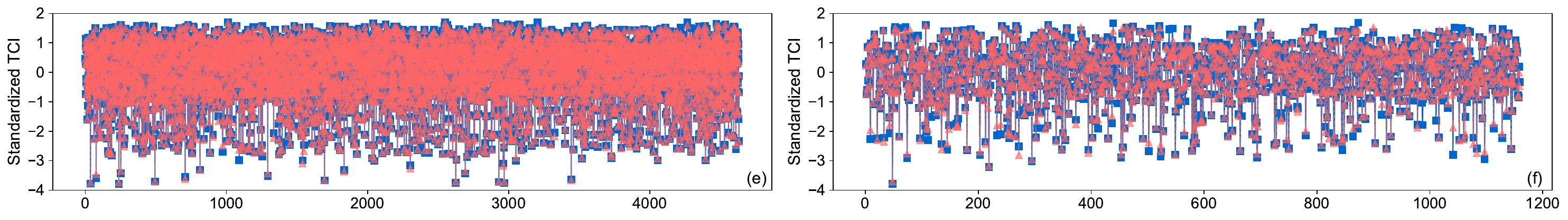}
  \includegraphics[width=0.97\linewidth]{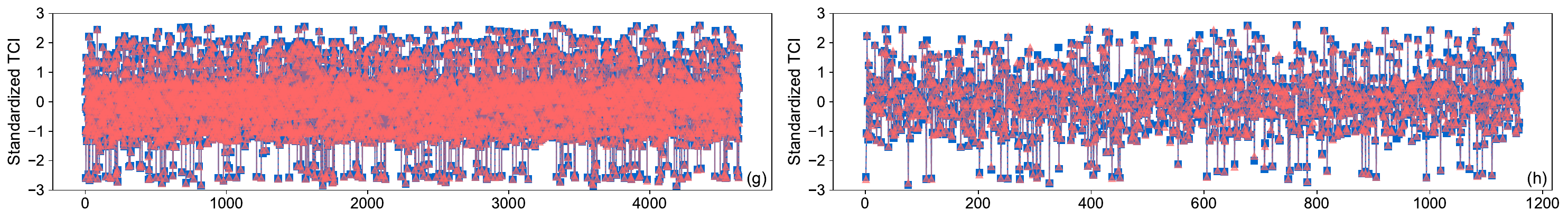}
  \includegraphics[width=0.97\linewidth]{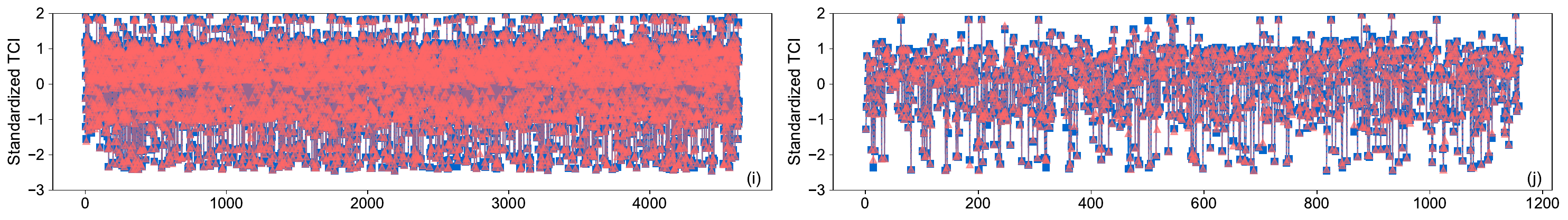}
  \includegraphics[width=0.97\linewidth]{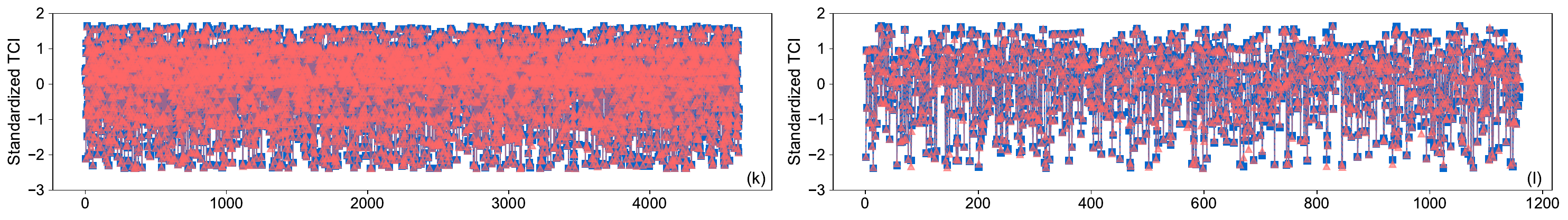}
  \includegraphics[width=0.97\linewidth]{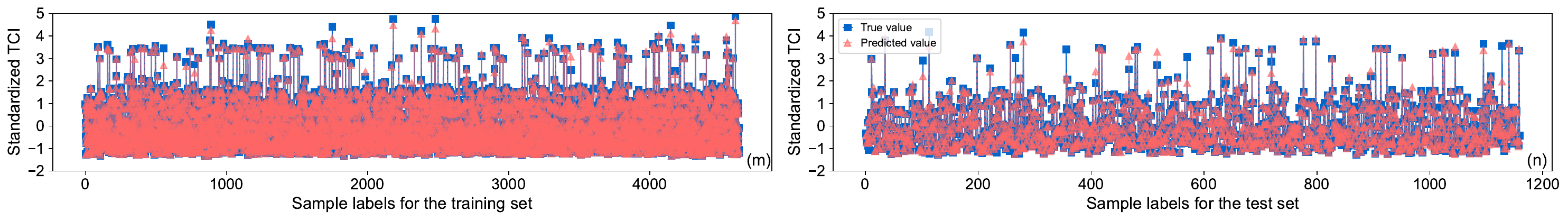}
  \caption{Learning results of the random forest model for training and test sets of TCIs corresponding to short term (a–b), medium term (c–d), long term (e–f), wheat (g–h), maize (i–j), soybean (k–l), and rice (m–n).} 
\label{Fig:RF_TruePredValue}
\end{figure}

Considering the varying dimensions of different data, both the explanatory variables and TCIs are standardized before applying the random forest machine learning algorithm. Referring to \cite{Wei-Gephart-Iizumi-Ramankutty-Davis-2023-NatSustain}, we split the dataset into 80\% training and 20\% test sets. Figure~\ref{Fig:RF_TruePredValue} presents the learning results of the random forest model, with the left and right columns corresponding to the training and test sets, respectively. It can be seen that the random forest model achieves an excellent fit for each set. To assess the performance of the model, we compute MAE and MSE as evaluation metrics. As shown in Table~\ref{Tab:RF_Metrics}, although the MAE and MSE for test sets are slightly higher than those for training sets, their values remain very small, which suggests that the 12 selected explanatory variables effectively explain the variance of each TCI index. In summary, these evaluation metrics confirm the good performance of the random forest model we constructed. Moreover, these selected driving factors significantly impact the risk spillover effects in the grain markets, making them viable early-warning indicators for systemic risks within the global food system.

\begin{table}[!ht]
  \centering
  \setlength{\abovecaptionskip}{0pt}
  \setlength{\belowcaptionskip}{10pt}
  \caption{Metrics for the random forest model}
  \setlength\tabcolsep{3pt} 
  \resizebox{\textwidth}{!}{ 
    \begin{tabular}{l cc c cc c cc c cc c cc c cc c cc}
    \toprule
         & \multicolumn{2}{c}{STC} && \multicolumn{2}{c}{MTC} && \multicolumn{2}{c}{LTC} && \multicolumn{2}{c}{Wheat} && \multicolumn{2}{c}{Maize} && \multicolumn{2}{c}{Soybean} && \multicolumn{2}{c}{Rice} \\
    \cline{2-3} \cline{5-6} \cline{8-9} \cline{11-12} \cline{14-15} \cline{17-18} \cline{20-21} 
         & Training & Test && Training & Test && Training & Test && Training & Test && Training & Test && Training & Test && Training & Test  \\
    \midrule
    MAE & 0.0142 & 0.0423 && 0.0212 & 0.0577 && 0.0350 & 0.0885 && 0.0209 & 0.0573 && 0.0151 & 0.0394 && 0.0135 & 0.0389 && 0.0287 & 0.0795 \\
    MSE & 0.0007 & 0.0061 && 0.0012 & 0.0082 && 0.0039 & 0.0234 && 0.0015 & 0.0121 && 0.0016 & 0.0097 && 0.0008 & 0.0063 && 0.0035 & 0.0232 \\
    \bottomrule
    \end{tabular}
  }%
  \begin{flushleft}
    \footnotesize
    \justifying 
    Note: This table reports the MAE and MSE of random forest models for the training and test sets. The abbreviations STC, MTC, and LTC denote the short-, medium-, and long-term components, respectively.
  \end{flushleft} 
  \label{Tab:RF_Metrics}%
\end{table}%

\begin{figure}[!ht]
  \centering
  \includegraphics[width=0.475\linewidth]{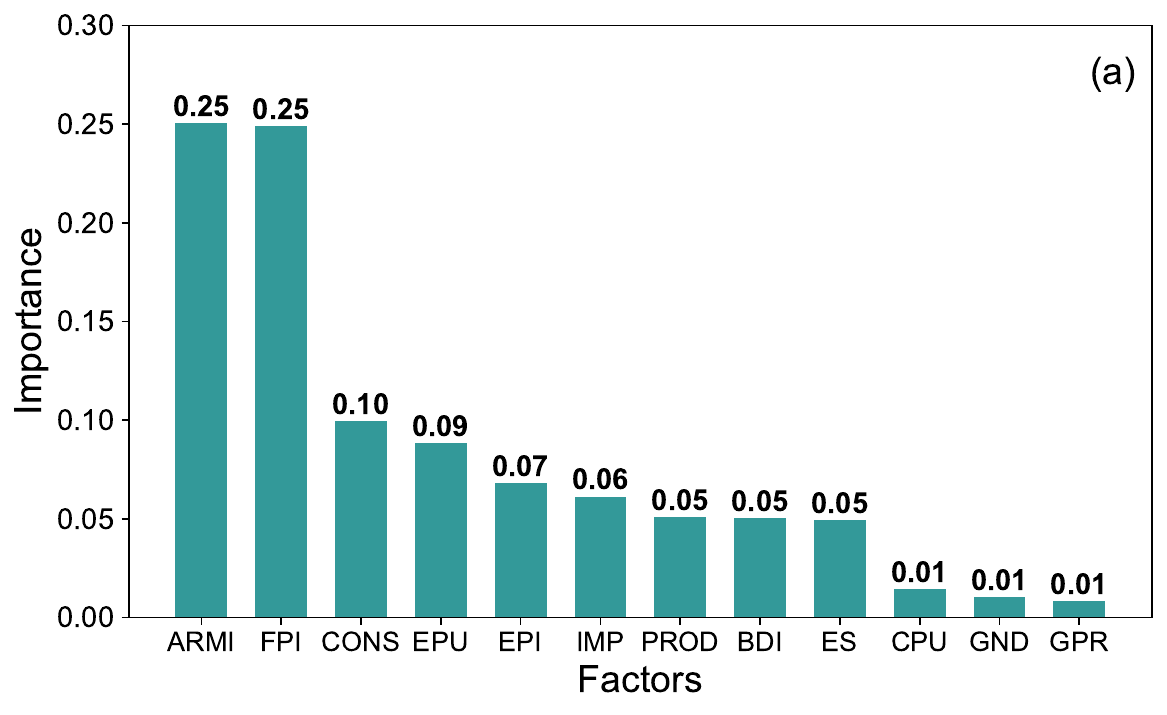}
  \includegraphics[width=0.475\linewidth]{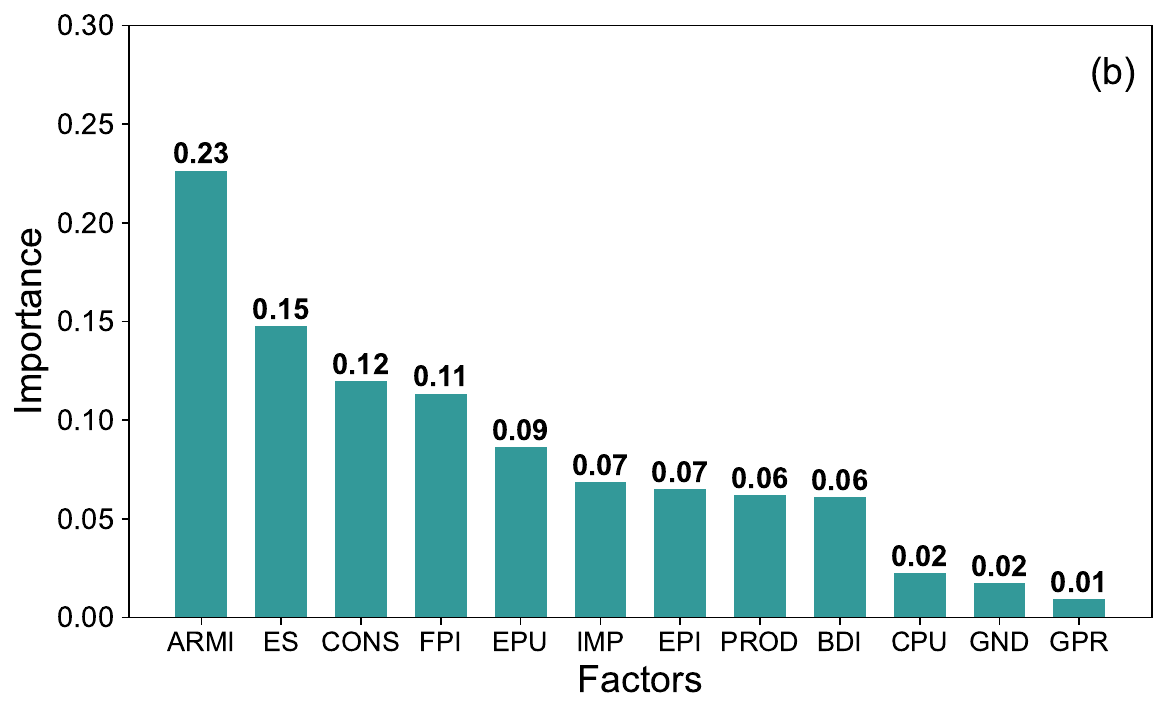}\\
  \includegraphics[width=0.475\linewidth]{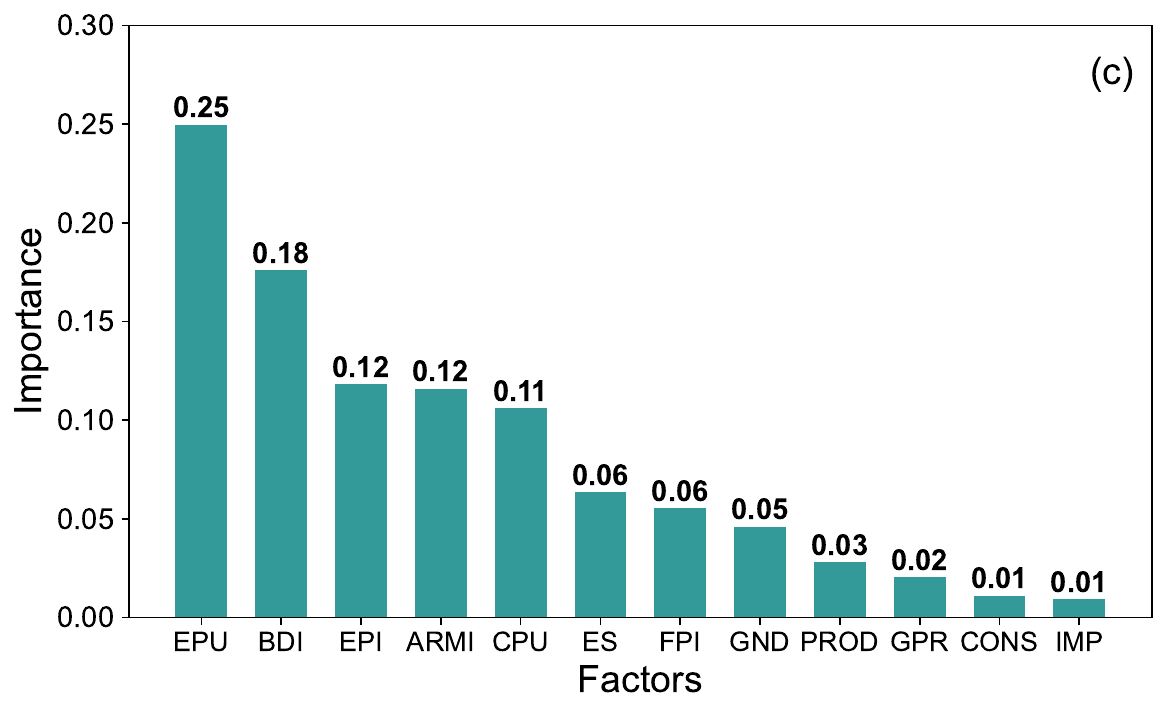}
  \caption{Importance scores of different factors in explaining TCIs for short-term (a), medium-term (b), and long-term (c) components based on the random forest regression.}
\label{Fig:RF_Factors_TCI_Timescale}
\end{figure}

With the model accuracy validated, we calculate the importance of each variable in each regression to evaluate their respective explanatory power for various types of risk spillovers. Figure~\ref{Fig:RF_Factors_TCI_Timescale} illustrates the importance of each factor in explaining the cross-grain TCIs for short-, medium-, and long-term components.

For short-term components, ARMI and FPI emerge as the two most important explanatory variables, highlighting that changes in upstream production costs and price volatility at the consumer level have substantial impacts on the risk spillovers between short-term components. ARMI, which includes the prices of fertilizers, feed, and other agricultural inputs, indirectly affects planting costs and market supply capacity. FPI, on the other hand, is a crucial indicator of the global balance between supply and demand in the food market, directly reflecting trends in food prices, inflationary effects, and food security. Together, these two factors capture market dynamics from the pricing dimension, with ARMI focusing on the production side and FPI on the consumption side.

For medium-term components, the variable importance rankings reveal that, in addition to ARMI and FPI, ES and CONS also significantly influence the risk spillovers between medium-term components. ES acts as a buffer in the global food supply chain, reflecting supply security and reserve capacity. CONS, representing actual grain usage, is a key measure of market demand. These two factors primarily reflect supply-demand dynamics from the quantitative perspective, with ES emphasizing reserves and CONS focusing on demand. Thus, market changes and supply-demand dynamics can be considered as the main drivers of risk spillovers among medium-term components across different grain submarkets. In contrast, variables associated with longer timescales, such as CPU, GND, and GPR, exert limited effect on short- and medium-term spillovers.

For long-term components, a diverse set of factors plays a critical role in explaining their risk spillovers, including EPU, BDI, EPI, ARMI, and CPU. EPU denotes uncertainties in economic policies, such as tariffs and trade agreements, which can disrupt grain production, trade, and pricing, thereby affecting long-term risk spillovers in the food market. CPU, closely tied to climate policies such as carbon taxes and environmental regulations, has long-run impacts on grain production and supply. BDI, a critical benchmark for global shipping costs, significantly influences international food prices. EPI, which captures energy prices including oil and natural gas, directly affects the costs of farm machinery, fertilizer production, and food transportation, ultimately influencing grain prices. As discussed earlier, the broad time horizon of long-term components means that their influencing factors span a wide range, including policy uncertainties, production costs, and transportation costs.

Moreover, as noted earlier, certain critical factors or major shocks may simultaneously impact risk spillovers at different timescales. For instance, ARMI consistently plays an important role in driving risk spillovers for the short-, medium-, and long-term components, which underscores that production costs are a fundamental source of cross-grain risk spillovers.

\begin{figure}[!ht]
  \centering
  \includegraphics[width=0.475\linewidth]{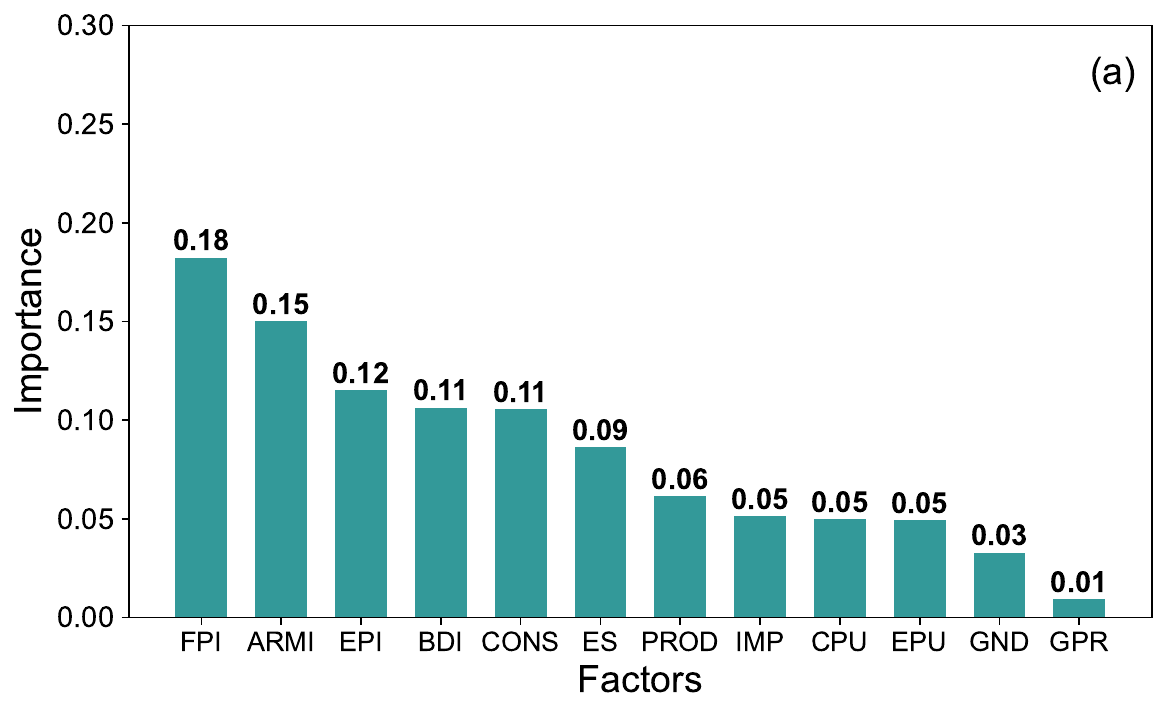}
  \includegraphics[width=0.475\linewidth]{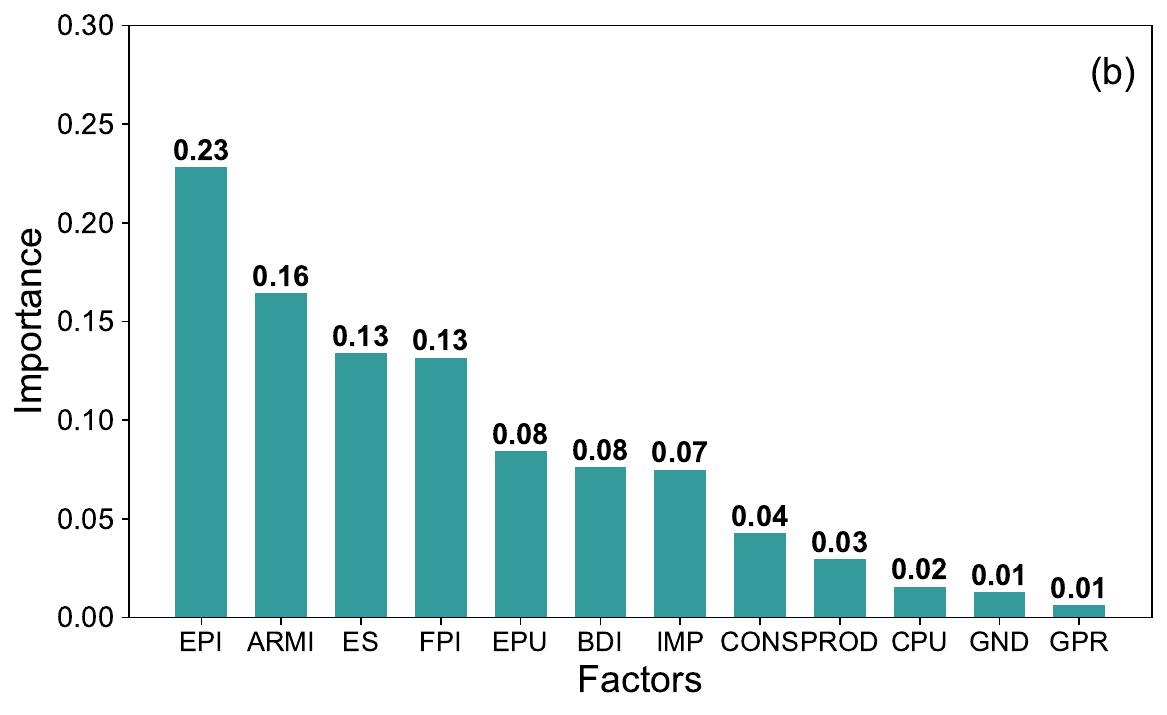}\\
  \includegraphics[width=0.475\linewidth]{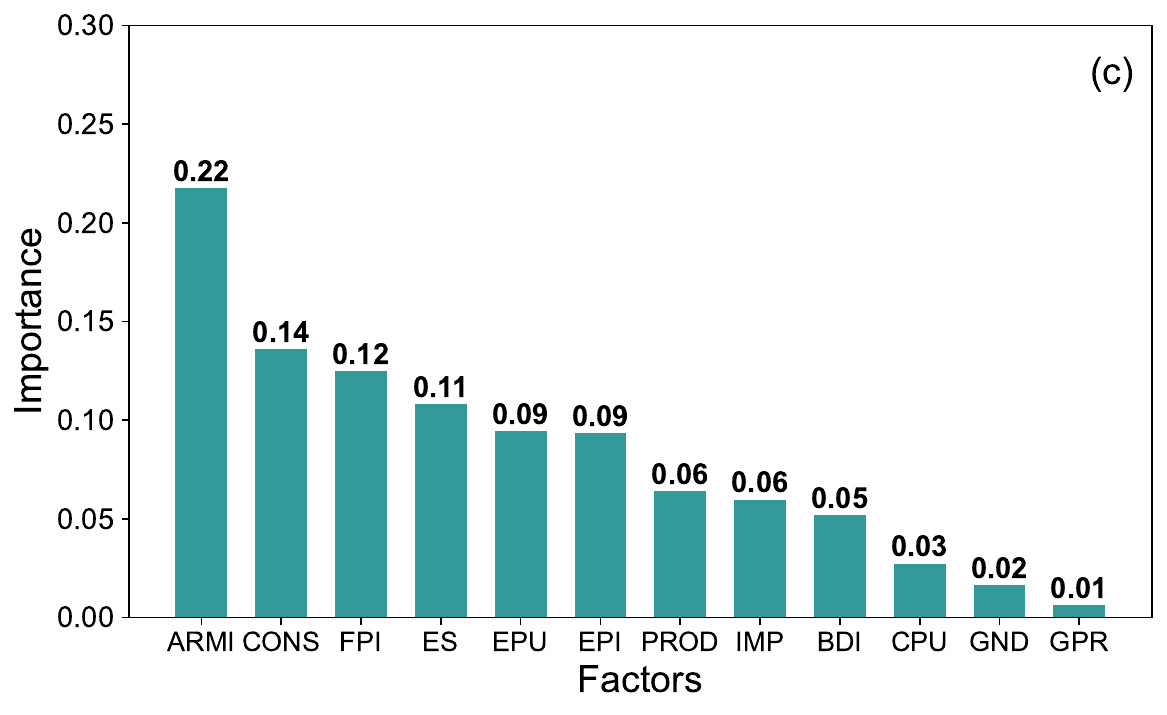}
  \includegraphics[width=0.475\linewidth]{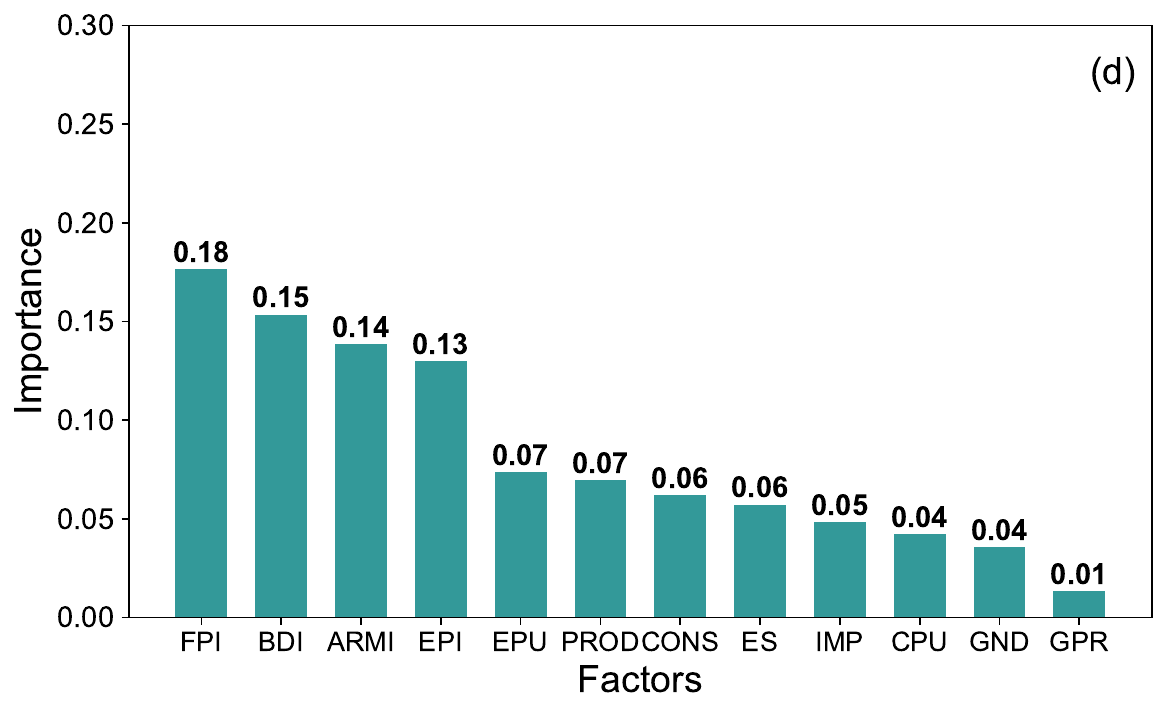}
  \caption{Importance scores of different factors in explaining TCIs for wheat (a), maize (b), soybean (c), and rice (d) based on the random forest regression.}
\label{Fig:RF_Factors_TCI_Commodity}
\end{figure}

Figure~\ref{Fig:RF_Factors_TCI_Commodity} depicts the importance of each factor in explaining the cross-timescale TCIs for wheat, maize, soybean, and rice. As can be seen, FPI, ARMI, and EPI always exhibit high importance for risk spillovers within the four staple food submarkets. FPI and ARMI, which measure the respective price levels of global food products and agricultural raw materials, respectively, serve as essential indicators of market price dynamics and agricultural production costs. Accordingly, fluctuations in these factors directly influence the risk spillover effect within each staple crop market. In addition, the close ties between these grains and the energy market underscore the importance of EPI. Specifically, wheat and maize are primary feedstocks for biofuels like ethanol, so energy prices, especially oil prices, significantly affect ethanol demand and thus the consumption structure and market prices of wheat and maize. Similarly, soybean oil, a key input for biodiesel, is directly impacted by energy price fluctuations. Besides, rice production relies heavily on energy-intensive processes, such as irrigation and mechanized operations, making energy price volatility an important driver of planting and harvesting costs. Furthermore, energy prices determine fuel costs for agricultural machinery, as well as transportation and processing expenses, which ultimately influence production conditions, trade prices, and market risks for all staple grains.

ES also exhibits substantial influence on risk spillovers within wheat, maize, and soybean markets, but its impact on the rice market is comparatively limited. This discrepancy is largely attributable to the distinct characteristics of these crops. Wheat, maize, and soybean are widely used, highly internationalized, and traded on a large scale, so stock changes can quickly affect their consumption and supply-demand balances through price mechanisms. Moreover, the production and trade of these crops rely on specific major exporting economies, making their supply more susceptible to external disruptions. ES, as a key buffer of supply, plays a significant role in mitigating price volatility during periods of tight supply and demand. Hence, the high degree of international dependence positions ending stocks as a critical factor in risk spillovers for these three food submarkets. In contrast, rice is mainly used as an edible ration, with production and consumption concentrated regionally, especially in Asian countries. The low proportion of international trade to total global production means that the majority of rice is consumed domestically and its market risk is less affected by international stocks. Furthermore, many major rice producers implement government interventions to stabilize domestic rice supplies and prices, such as minimum procurement prices and export controls. These policies often maintain high national stock levels, as exemplified by China and India's reserve policies, which further reduce the impact of international stock changes on rice market risk.

Additionally, BDI plays an important role in explaining risk spillovers in wheat and rice markets. Wheat is one of the most extensively traded staple grains globally, with production concentrated in a few major exporters, such as Russia and the United States, while demand is distributed worldwide. Thus, wheat prices are highly sensitive to international transportation costs. Although rice accounts for a smaller share of global trade, the cost of transportation significantly influences prices in both export-oriented countries (e.g., Thailand and Vietnam) and import-dependent regions (e.g., parts of Africa). Besides, wheat and rice are primarily transported as bulk commodities in international trade, whereas maize and soybean often utilize alternative modes such as land transport. As foundational staples with inelastic demand, wheat and rice require stable supplies in international markets, which indicates high real-time responsiveness but low flexibility of their supply chains. Maize and soybean, on the other hand, are versatile crops with elastic demand, so the impact of transportation costs on their market risks can be mitigated by adjusting their end-use or seeking substitutes. Furthermore, these two crops benefit from robust storage infrastructure and seasonal redistribution mechanisms, enabling importing economies to alleviate short-term fluctuations in transportation costs by pre-importing or reducing exports. As a result, maize and soybean exhibit lower sensitivity to changes in spot transportation costs compared to wheat and rice.

\subsection{Robustness checks}

\begin{figure}[!ht]
  \centering
  \includegraphics[width=0.985\linewidth]{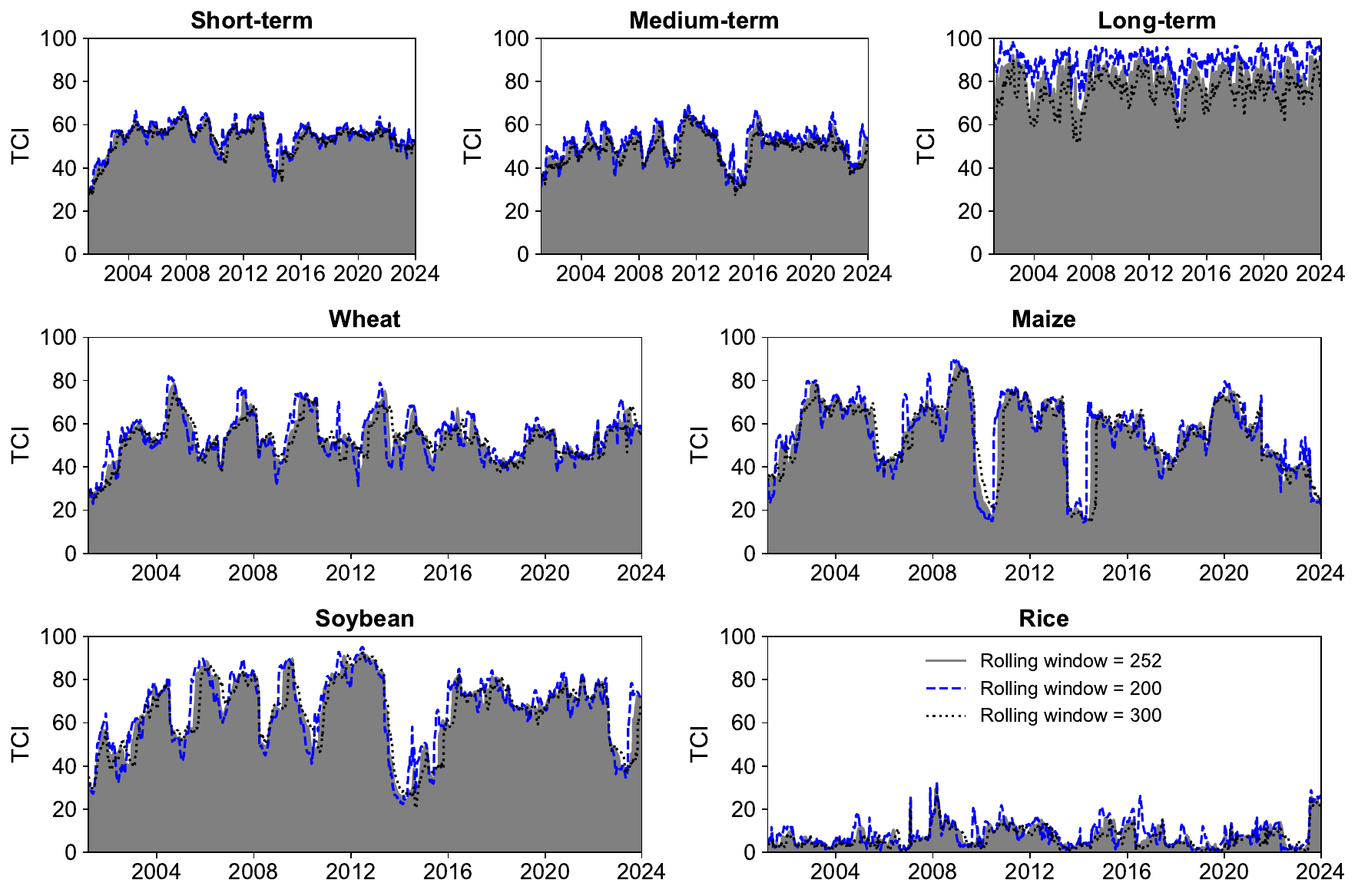}
  \caption{Window-size robustness check for dynamic total connectedness.}
\label{Fig:RobustnessCheck_WindowSize}
\end{figure}

In the dynamic analysis, we set the rolling window length to one year (252 trading days). To ensure the reliability of the empirical results, we conduct robustness analysis using alternative rolling window lengths of 200 and 300 trading days. The results, as shown in Figure~\ref{Fig:RobustnessCheck_WindowSize}, indicate only minor numerical differences in the TCIs calculated with different rolling windows. The overall trends remain highly consistent, with pronounced volatility observed during global crises. This demonstrates that the TCIs are not sensitive to the choice of rolling window length, affirming the robustness of our analysis of dynamic risk spillovers across grains and timescales in global staple food markets.

We also perform robustness checks for the empirical results of the random forest model. The entire dataset is split into 70\% for training and 30\% for testing, and the optimal parameters are re-determined by the grid search method. Table~\ref{Tab:RF_Metrics_RobustnessCheck} reports the evaluation metrics of the random forest model. Comparing these results with those in Table~\ref{Tab:RF_Metrics}, we find consistent outcomes with only minor numerical differences. These findings reaffirm the robust performance of random forest models we constructed, highlighting the significant influence of our selected price drivers, external uncertainties, and core supply-demand indicators on risk spillovers of various dimensions in the grain market, with notable explanatory power.

\begin{figure}[!ht]
  \centering
  \includegraphics[width=0.985\linewidth]{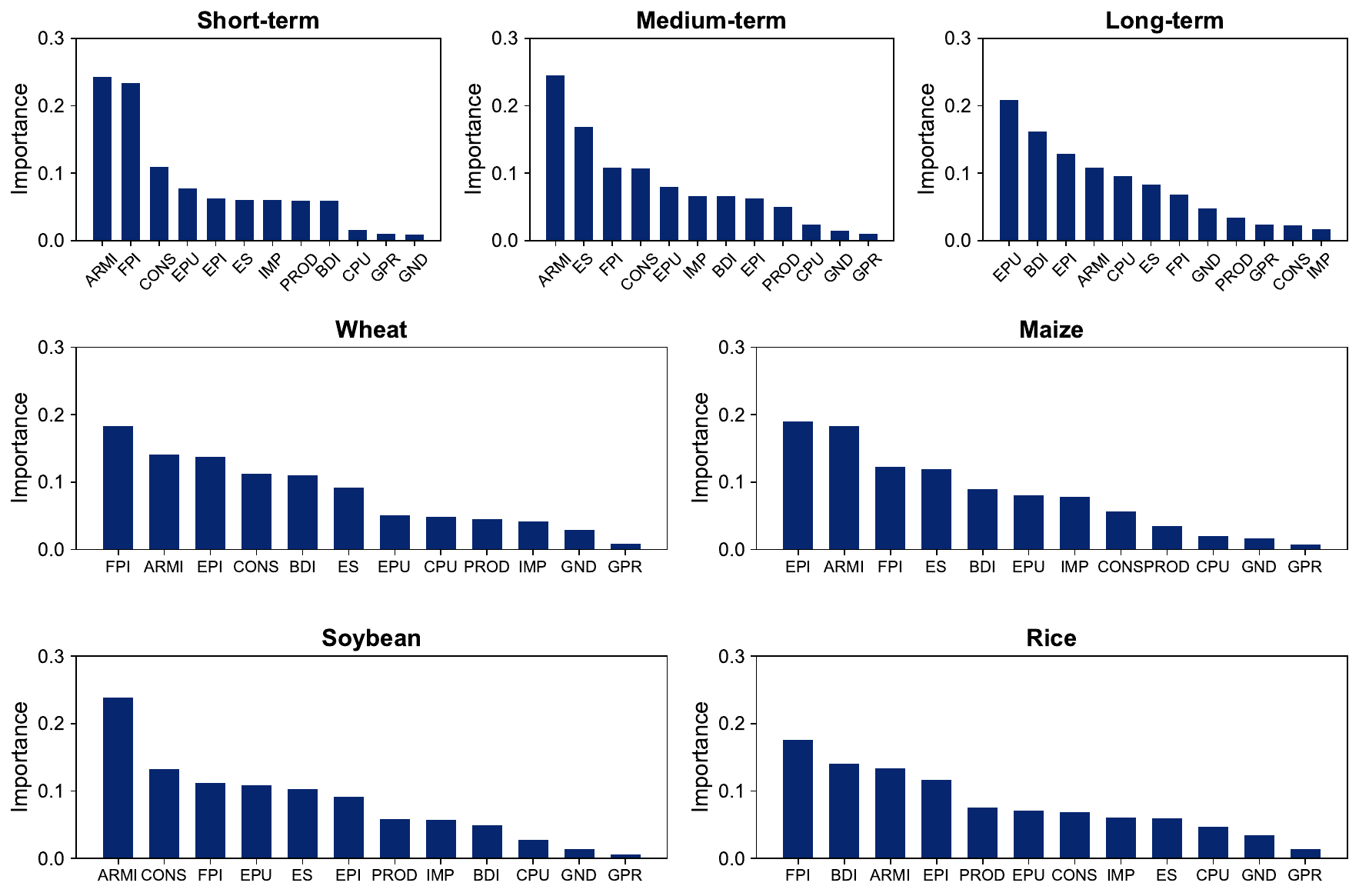}
  \caption{Factor importance from random forest regression with adjusted parameters for robustness check.}
\label{Fig:RF_Factors_TCI__RobustnessCheck}
\end{figure}

Figure~\ref{Fig:RF_Factors_TCI__RobustnessCheck} further illustrates the importance of individual factors in explaining various risk spillovers. The results show that the key determinants driving cross-grain and cross-timescale risk spillovers remain largely consistent with the previous analysis, with only minor reordering among a few closely ranked factors. For short-term spillovers, price-related market dynamics, including production- and consumption-side variations, are particularly important. Both price dynamics and supply-demand quantities, such as reserves and demand, play crucial roles in medium-term spillovers. Due to the broader time horizon, long-term spillovers are shaped by multifaceted factors, including policy uncertainties, energy price fluctuations, and production and transportation costs. Notably, external uncertainties, such as economic and climate policy uncertainties, warrant greater attention for their significant impact on long-term spillovers. Moreover, the main factors driving spillovers within grain submarkets exhibit both commonalities and specificities across different staple foods. For instance, market dynamics and energy prices are critical for all four staple grains; global ending stocks significantly influence spillovers in wheat, maize, and soybean markets but have limited impact on the rice market; the Baltic dry index plays an important role in explaining spillovers in wheat and rice markets. These results, in line with our empirical analysis, validate the robustness of the random forest model under different specifications and underscore the reliability of our analysis of risk spillover drivers.

\begin{table}[!ht]
  \centering
  \setlength{\abovecaptionskip}{0pt}
  \setlength{\belowcaptionskip}{10pt}
  \caption{Metrics for the random forest model with adjusted parameters}
  \setlength\tabcolsep{3pt} 
  \resizebox{\textwidth}{!}{ 
    \begin{tabular}{l cc c cc c cc c cc c cc c cc c cc}
    \toprule
         & \multicolumn{2}{c}{STC} && \multicolumn{2}{c}{MTC} && \multicolumn{2}{c}{LTC} && \multicolumn{2}{c}{Wheat} && \multicolumn{2}{c}{Maize} && \multicolumn{2}{c}{Soybean} && \multicolumn{2}{c}{Rice} \\
    \cline{2-3} \cline{5-6} \cline{8-9} \cline{11-12} \cline{14-15} \cline{17-18} \cline{20-21} 
         & Training & Test && Training & Test && Training & Test && Training & Test && Training & Test && Training & Test && Training & Test  \\
    \midrule
    MAE & 0.0147 & 0.0427 && 0.0218 & 0.0597 && 0.0370 & 0.0986 && 0.0218 & 0.0616 && 0.0160 & 0.0410 && 0.0138 & 0.0404 && 0.0298 & 0.0822 \\
    MSE & 0.0008 & 0.0060 && 0.0013 & 0.0086 && 0.0040 & 0.0287 && 0.0016 & 0.0133 && 0.0017 & 0.0099 && 0.0008 & 0.0081 && 0.0038 & 0.0259 \\
    \bottomrule
    \end{tabular}
  }%
  \begin{flushleft}
    \footnotesize
    \justifying 
    Note: This table gives the MAE and MSE of random forest models with adjusted parameters for robustness check.
  \end{flushleft} 
  \label{Tab:RF_Metrics_RobustnessCheck}%
\end{table}%

\section{Conclusions}
\label{S1:Conclude}

Multiple unfavorable factors, including cost disturbances, supply-demand imbalances, and external shocks, are posing escalating threats to global food security. Coping with these growing challenges requires a comprehensive and nuanced understanding of risk contagion within the global food market. To this end, this study adopts a dual perspective—across grains and across timescales—to clarify the risk spillover effects in the international markets of staple foods. Furthermore, this paper provides novel insights into how these spillovers are influenced by price-driven factors, external uncertainties, and core supply-demand indicators.

The decomposition-reconstruction method is first employed to extract timescale-specific components of grain futures and spot returns. The ICEEMDAN decomposition results reveal that the mean periods of intrinsic modes increase sequentially, corresponding to short-, medium-, and long-term scales, with decreasing correlation and importance to the original series. The GMM-RLN reconstruction results show that decomposed modes with shorter timescales and higher frequencies exhibit greater volatility, and futures components tend to fluctuate more than spot components. Meanwhile, the striking similarity in mode groupings for the four grains reflects, to some extent, the commonality of international staple food markets. Additionally, short-term components are the most volatile, followed by medium- and long-term components, with short- and medium-term movements generally oscillating around the long-term trend.

Applying the $R^{2}$ decomposed connectedness approach to all components uncovers two main spillover patterns: the first occurs between same-timescale components of different grains, and the second occurs between same-grain components of different timescales. Both the full-sample static and rolling-window dynamic analyses consistently confirm the existence of significant risk transmission within the global food market, with dynamic spillovers being more pronounced and closely tied to major global crises. Besides, components with larger spillovers as transmitters also experience greater spillovers as receivers, resulting in most components being in equilibrium of risk connectedness. For cross-grain spillovers, soybean, maize, and wheat emerge as important contributors and recipients of risk, whereas rice exhibits minimal spillovers due to its lower internationalization and high self-sufficiency of major consumers. For cross-timescale spillovers, short-term components dominate internal risk transmission for soybean, maize, and wheat markets, while long-term components play a similar role in the rice market, albeit with limited overall spillovers. Our constructed net risk spillover networks intuitively capture the inter-component spillover relationships and transmission pathways. Moreover, the spillover effects between futures and spot components for the same grain and at the same timescale are particularly prominent. Notably, the level of risk spillovers within the international grain market has shown a clear upward trend in recent years.

The twelve potential factors, encompassing food supply, demand, cost, and external uncertainties, are further examined to identify the determinants of cross-grain and cross-timescale risk spillovers. The robust performance of the random forest model demonstrates the substantial impact of these drivers on risk contagion, suggesting their promise as early-warning indicators for systemic risks in the global food system. Variable importance analysis reveals distinct drivers for different spillovers. For short-term spillovers, ARMI and FPI are the most critical, reflecting market changes primarily from production cost and consumer price dimensions, respectively. In medium-term spillovers, ES and CONS are also important, highlighting supply-demand dynamics from the perspectives of reserve levels and market demand. Long-term spillovers, associated with broader temporal horizons, are notably influenced by diverse factors, including policy uncertainties and cost-related drivers. Furthermore, FPI, ARMI, and EPI consistently play significant roles in risk transmission within each of the four food submarkets. ES strongly impacts spillovers within wheat, maize, and soybean markets but has limited influence on rice, owing to the distinct characteristics of different crops. Besides, BDI is instrumental in explaining spillovers in wheat and rice markets due to their reliance on dry bulk shipping, unlike maize and soybean, which benefit from greater transport flexibility.

The global food security situation remains precarious, with significant risk contagion within staple food markets serving as a major impediment to progress. Addressing excessive food price volatility requires collaborative efforts among international stakeholders, alongside a vigilant focus on the multidimensional risk transmission dynamics. Clear evidence of notable risk spillover effects, as provided by this study, highlights the vital need for targeted monitoring of cross-grain and cross-timescale risk transmission. For one thing, economies are expected to bolster their risk tolerance, actively engage in initiatives aimed at mitigating risk contagion in staple crop markets, and implement government intervention during periods of high spillovers, particularly for essential commodities like rice and wheat. For another, stakeholders should establish scientifically sound early-warning systems for food market risks, which are capable of capturing shifts in internal drivers and external shocks, as well as emerging risk events, in a timely and comprehensive manner. Moreover, it is imperative to accelerate the transformation of the global agrifood system and to enhance its resilience against critical challenges and potential structural vulnerabilities. Policymakers need to adopt context-specific and adaptive strategies to effectively manage complex spillover risks, tackle their underlying determinants, and alleviate their cascading effects on global and local food security.

Given the constraints of data availability, this study primarily examines the impact of price drivers, external uncertainties, and core supply-demand indicators on risk spillovers in global staple food markets. Future research could expand upon this by integrating market-specific and socioeconomic factors that may influence spillover effects, such as supply chain constraints or varying levels of market maturity. In addition, the data employed in this study may have inherent limitations, including potential biases, which are expected to be addressed in the future by incorporating a broader range of supplementary data to enrich the explanatory analysis of risk spillovers.

\section*{Acknowledgment}

This work was supported by the National Natural Science Foundation of China (Grant Numbers: 72201099, 72171083), the China Scholarship Council, the Fundamental Research Funds for the Central Universities, and the Central Universities' Program for Building World-Class Universities (Disciplines) and Special Development Guidance--Cultural Heritage and Innovation.

\section*{Data availability}

The price data for staple foods used in this study are obtained from the \href{https://www.wind.com.cn/}{Wind Database} and the \href{https://www.igc.int/}{International Grains Council Website}, and the data for influencing factors are sourced from the \href{http://www.policyuncertainty.com/}{Policy Uncertainty Website}, the \href{https://www.emdat.be/}{International Disaster Database}, the \href{https://www.imf.org/}{International Monetary Fund Website}, and the \href{https://www.fao.org/faostat/}{Food and Agriculture Organization of the United Nations Statistical Database}.


\begin{thebibliography}{63}
\expandafter\ifx\csname natexlab\endcsname\relax\def\natexlab#1{#1}\fi
\providecommand{\url}[1]{\texttt{#1}}
\providecommand{\href}[2]{#2}
\providecommand{\path}[1]{#1}
\providecommand{\DOIprefix}{doi:}
\providecommand{\ArXivprefix}{arXiv:}
\providecommand{\URLprefix}{URL: }
\providecommand{\Pubmedprefix}{pmid:}
\providecommand{\doi}[1]{\href{http://dx.doi.org/#1}{\path{#1}}}
\providecommand{\Pubmed}[1]{\href{pmid:#1}{\path{#1}}}
\providecommand{\bibinfo}[2]{#2}
\ifx\xfnm\relax \def\xfnm[#1]{\unskip,\space#1}\fi
\bibitem[{Abuzayed et~al.(2021)Abuzayed, Bouri, Al-Fayoumi and
  Jalkh}]{Abuzayed-Bouri-AlFayoumi-Jalkh-2021-EconAnalPolicy}
\bibinfo{author}{Abuzayed, B.}, \bibinfo{author}{Bouri, E.},
  \bibinfo{author}{Al-Fayoumi, N.}, \bibinfo{author}{Jalkh, N.},
  \bibinfo{year}{2021}.
\newblock \bibinfo{title}{Systemic risk spillover across global and country
  stock markets during the {COVID}-19 pandemic}.
\newblock \bibinfo{journal}{Econ. Anal. Policy} \bibinfo{volume}{71},
  \bibinfo{pages}{180--197}.
\newblock \DOIprefix\doi{10.1016/j.eap.2021.04.010}.
\bibitem[{Adeleke and Awodumi(2022)}]{Adeleke-Awodumi-2022-JApplEcon}
\bibinfo{author}{Adeleke, M.A.}, \bibinfo{author}{Awodumi, O.B.},
  \bibinfo{year}{2022}.
\newblock \bibinfo{title}{Modelling time and frequency connectedness among
  energy, agricultural raw materials and food markets}.
\newblock \bibinfo{journal}{J. Appl. Econ} \bibinfo{volume}{25},
  \bibinfo{pages}{644--662}.
\newblock \DOIprefix\doi{10.1080/15140326.2022.2056300}.
\bibitem[{Adrian and Brunnermeier(2016)}]{Adrian-Brunnermeier-2016-AmEconRev}
\bibinfo{author}{Adrian, T.}, \bibinfo{author}{Brunnermeier, M.K.},
  \bibinfo{year}{2016}.
\newblock \bibinfo{title}{Co{V}a{R}}.
\newblock \bibinfo{journal}{Am. Econ. Rev.} \bibinfo{volume}{106},
  \bibinfo{pages}{1705--1741}.
\newblock \DOIprefix\doi{10.1257/aer.20120555}.
\bibitem[{Arzandeh and Frank(2019)}]{Arzandeh-Frank-2019-AmJAgrEcon}
\bibinfo{author}{Arzandeh, M.}, \bibinfo{author}{Frank, J.},
  \bibinfo{year}{2019}.
\newblock \bibinfo{title}{Price discovery in agricultural futures markets:
  {S}hould we look beyond the best bid-ask spread?}
\newblock \bibinfo{journal}{Am. J. Agr. Econ.} \bibinfo{volume}{101},
  \bibinfo{pages}{1482--1498}.
\newblock \DOIprefix\doi{10.1093/ajae/aaz001}.
\bibitem[{Athey et~al.(2019)Athey, Tibshirani and
  Wager}]{Athey-Tibshirani-Wager-2019-AnnStat}
\bibinfo{author}{Athey, S.}, \bibinfo{author}{Tibshirani, J.},
  \bibinfo{author}{Wager, S.}, \bibinfo{year}{2019}.
\newblock \bibinfo{title}{Generalized random forests}.
\newblock \bibinfo{journal}{Ann. Stat.} \bibinfo{volume}{47},
  \bibinfo{pages}{1148--1178}.
\newblock \DOIprefix\doi{10.1214/18-AOS1709}.
\bibitem[{Bakas and
  Triantafyllou(2018)}]{Bakas-Triantafyllou-2018-JIntMoneyFinan}
\bibinfo{author}{Bakas, D.}, \bibinfo{author}{Triantafyllou, A.},
  \bibinfo{year}{2018}.
\newblock \bibinfo{title}{The impact of uncertainty shocks on the volatility of
  commodity prices}.
\newblock \bibinfo{journal}{J. Int. Money Finan.} \bibinfo{volume}{87},
  \bibinfo{pages}{96--111}.
\newblock \DOIprefix\doi{10.1016/j.jimonfin.2018.06.001}.
\bibitem[{Baker et~al.(2016)Baker, Bloom and
  Davis}]{Baker-Bloom-Davis-2016-QJEcon}
\bibinfo{author}{Baker, S.R.}, \bibinfo{author}{Bloom, N.},
  \bibinfo{author}{Davis, S.J.}, \bibinfo{year}{2016}.
\newblock \bibinfo{title}{{Measuring economic policy uncertainty}}.
\newblock \bibinfo{journal}{Quart. J. Econ.} \bibinfo{volume}{131},
  \bibinfo{pages}{1593--1636}.
\newblock \DOIprefix\doi{10.1093/qje/qjw024}.
\bibitem[{Balli et~al.(2023)Balli, Balli, Dang and
  Gabauer}]{Balli-Balli-Dang-Gabauer-2023-FinancResLett}
\bibinfo{author}{Balli, F.}, \bibinfo{author}{Balli, H.O.},
  \bibinfo{author}{Dang, T.H.N.}, \bibinfo{author}{Gabauer, D.},
  \bibinfo{year}{2023}.
\newblock \bibinfo{title}{Contemporaneous and lagged {$R^2$} decomposed
  connectedness approach: New evidence from the energy futures market}.
\newblock \bibinfo{journal}{Financ. Res. Lett.} \bibinfo{volume}{57},
  \bibinfo{pages}{104168}.
\newblock \DOIprefix\doi{10.1016/j.frl.2023.104168}.
\bibitem[{Barunik and Krehlik(2018)}]{Barunik-Krehlik-2018-JFinancEconom}
\bibinfo{author}{Barunik, J.}, \bibinfo{author}{Krehlik, T.},
  \bibinfo{year}{2018}.
\newblock \bibinfo{title}{Measuring the frequency dynamics of financial
  connectedness and systemic risk}.
\newblock \bibinfo{journal}{J. Financ. Econom.} \bibinfo{volume}{16},
  \bibinfo{pages}{271--296}.
\newblock \DOIprefix\doi{10.1093/jjfinec/nby001}.
\bibitem[{Basak and Pavlova(2016)}]{Basak-Pavlova-2016-JFinanc}
\bibinfo{author}{Basak, S.}, \bibinfo{author}{Pavlova, A.},
  \bibinfo{year}{2016}.
\newblock \bibinfo{title}{A model of financialization of commodities}.
\newblock \bibinfo{journal}{J. Financ.} \bibinfo{volume}{71},
  \bibinfo{pages}{1511--1556}.
\newblock \DOIprefix\doi{10.1111/jofi.12408}.
\bibitem[{Bekaert et~al.(2005)Bekaert, Harvey and
  Ng}]{Bekaert-Harvey-Ng-2005-JBus}
\bibinfo{author}{Bekaert, G.}, \bibinfo{author}{Harvey, C.},
  \bibinfo{author}{Ng, A.}, \bibinfo{year}{2005}.
\newblock \bibinfo{title}{Market integration and contagion}.
\newblock \bibinfo{journal}{J. Bus.} \bibinfo{volume}{78},
  \bibinfo{pages}{39--69}.
\newblock \DOIprefix\doi{10.1086/426519}.
\bibitem[{Ben~Ameur et~al.(2024)Ben~Ameur, Ftiti and
  Louhichi}]{BenAmeur-Ftiti-Louhichi-2024-AnnOperRes}
\bibinfo{author}{Ben~Ameur, H.}, \bibinfo{author}{Ftiti, Z.},
  \bibinfo{author}{Louhichi, W.}, \bibinfo{year}{2024}.
\newblock \bibinfo{title}{Interconnectedness of cryptocurrency markets: {A}n
  intraday analysis of volatility spillovers based on realized volatility
  decomposition}.
\newblock \bibinfo{journal}{Ann. Oper. Res.} \bibinfo{volume}{341},
  \bibinfo{pages}{757--779}.
\newblock \DOIprefix\doi{10.1007/s10479-023-05757-w}.
\bibitem[{Bergstra and Bengio(2012)}]{Bergstra-Bengio-2012-JMachLearnRes}
\bibinfo{author}{Bergstra, J.}, \bibinfo{author}{Bengio, Y.},
  \bibinfo{year}{2012}.
\newblock \bibinfo{title}{Random search for hyper-parameter optimization}.
\newblock \bibinfo{journal}{J. Mach. Learn. Res.} \bibinfo{volume}{13},
  \bibinfo{pages}{281–305}.
\newblock \DOIprefix\doi{10.5555/2188385.2188395}.
\bibitem[{Bianchi et~al.(2020)Bianchi, Fan and
  Todorova}]{Bianchi-Fan-Todorova-2020-IntRevFinancAnal}
\bibinfo{author}{Bianchi, R.J.}, \bibinfo{author}{Fan, J.H.},
  \bibinfo{author}{Todorova, N.}, \bibinfo{year}{2020}.
\newblock \bibinfo{title}{Financialization and de-financialization of commodity
  futures: {A} quantile regression approach}.
\newblock \bibinfo{journal}{Int. Rev. Financ. Anal.} \bibinfo{volume}{68}.
\newblock \DOIprefix\doi{10.1016/j.irfa.2019.101451}.
\bibitem[{Breiman(2001)}]{Breiman-2001-MachLearn}
\bibinfo{author}{Breiman, L.}, \bibinfo{year}{2001}.
\newblock \bibinfo{title}{Random forests}.
\newblock \bibinfo{journal}{Mach. Learn.} \bibinfo{volume}{45},
  \bibinfo{pages}{5--32}.
\newblock \DOIprefix\doi{10.1023/A:1010933404324}.
\bibitem[{Caldara and Iacoviello(2022)}]{Caldara-Iacoviello-2022-AmEconRev}
\bibinfo{author}{Caldara, D.}, \bibinfo{author}{Iacoviello, M.},
  \bibinfo{year}{2022}.
\newblock \bibinfo{title}{Measuring geopolitical risk}.
\newblock \bibinfo{journal}{Am. Econ. Rev.} \bibinfo{volume}{112},
  \bibinfo{pages}{1194--1225}.
\newblock \DOIprefix\doi{10.1257/aer.20191823}.
\bibitem[{Cao and Cheng(2021)}]{Cao-Cheng-2021-ResourPolicy}
\bibinfo{author}{Cao, Y.}, \bibinfo{author}{Cheng, S.}, \bibinfo{year}{2021}.
\newblock \bibinfo{title}{Impact of {COVID}-19 outbreak on multi-scale
  asymmetric spillovers between food and oil prices}.
\newblock \bibinfo{journal}{Resour. Policy} \bibinfo{volume}{74},
  \bibinfo{pages}{102364}.
\newblock \DOIprefix\doi{10.1016/j.resourpol.2021.102364}.
\bibitem[{Chen et~al.(2022)Chen, Zheng and Hao}]{Chen-Zheng-Hao-2022-JBusRes}
\bibinfo{author}{Chen, Z.}, \bibinfo{author}{Zheng, C.}, \bibinfo{author}{Hao,
  X.}, \bibinfo{year}{2022}.
\newblock \bibinfo{title}{Volatility spillover effect between internet finance
  and banks}.
\newblock \bibinfo{journal}{J. Bus. Res.} \bibinfo{volume}{141},
  \bibinfo{pages}{512--519}.
\newblock \DOIprefix\doi{10.1016/j.jbusres.2021.11.049}.
\bibitem[{Cheng et~al.(2006)Cheng, Yu and
  Yang}]{Cheng-Yu-Yang-2006-MechSystSignalProc}
\bibinfo{author}{Cheng, J.S.}, \bibinfo{author}{Yu, D.J.},
  \bibinfo{author}{Yang, Y.}, \bibinfo{year}{2006}.
\newblock \bibinfo{title}{Research on the intrinsic mode function ({IMF})
  criterion in {EMD} method}.
\newblock \bibinfo{journal}{Mech. Syst. Signal Proc.} \bibinfo{volume}{20},
  \bibinfo{pages}{817--822}.
\newblock \DOIprefix\doi{10.1016/j.ymssp.2005.09.011}.
\bibitem[{Cheng et~al.(2024)Cheng, Li, Cui, Wei, Wang and
  Hong}]{Cheng-Li-Cui-Wei-Wang-Hong-2024-IntRevFinancAnal}
\bibinfo{author}{Cheng, Z.}, \bibinfo{author}{Li, M.}, \bibinfo{author}{Cui,
  R.}, \bibinfo{author}{Wei, Y.}, \bibinfo{author}{Wang, S.},
  \bibinfo{author}{Hong, Y.}, \bibinfo{year}{2024}.
\newblock \bibinfo{title}{The impact of {COVID}-19 on global financial markets:
  {A} multiscale volatility spillover analysis}.
\newblock \bibinfo{journal}{Int. Rev. Financ. Anal.} \bibinfo{volume}{95},
  \bibinfo{pages}{103454}.
\newblock \DOIprefix\doi{10.1016/j.irfa.2024.103454}.
\bibitem[{Colominas et~al.(2014)Colominas, Schlotthauer and
  Torres}]{Colominas-Schlotthauer-Torres-2014-BiomedSignalProcessControl}
\bibinfo{author}{Colominas, M.A.}, \bibinfo{author}{Schlotthauer, G.},
  \bibinfo{author}{Torres, M.E.}, \bibinfo{year}{2014}.
\newblock \bibinfo{title}{Improved complete ensemble {EMD}: {A} suitable tool
  for biomedical signal processing}.
\newblock \bibinfo{journal}{Biomed. Signal Process. Control}
  \bibinfo{volume}{14}, \bibinfo{pages}{19--29}.
\newblock \DOIprefix\doi{10.1016/j.bspc.2014.06.009}.
\bibitem[{Dai et~al.(2023)Dai, Dai and
  Zhou}]{Dai-Dai-Zhou-2023-JIntFinancMarkInstMoney}
\bibinfo{author}{Dai, Y.S.}, \bibinfo{author}{Dai, P.F.},
  \bibinfo{author}{Zhou, W.X.}, \bibinfo{year}{2023}.
\newblock \bibinfo{title}{Tail dependence structure and extreme risk spillover
  effects between the international agricultural futures and spot markets}.
\newblock \bibinfo{journal}{J. Int. Financ. Mark. Inst. Money}
  \bibinfo{volume}{88}, \bibinfo{pages}{101820}.
\newblock \DOIprefix\doi{10.1016/j.intfin.2023.101820}.
\bibitem[{De~Jong et~al.(2022)De~Jong, Sonnemans and
  Tuinstra}]{DeJong-Sonnemans-Tuinstra-2022-JEconBehavOrgan}
\bibinfo{author}{De~Jong, J.}, \bibinfo{author}{Sonnemans, J.},
  \bibinfo{author}{Tuinstra, J.}, \bibinfo{year}{2022}.
\newblock \bibinfo{title}{The effect of futures markets on the stability of
  commodity prices}.
\newblock \bibinfo{journal}{J. Econ. Behav. Organ.} \bibinfo{volume}{198},
  \bibinfo{pages}{176--211}.
\newblock \DOIprefix\doi{10.1016/j.jebo.2022.03.025}.
\bibitem[{Diebold and Yilmaz(2009)}]{Diebold-Yilmaz-2009-EJ}
\bibinfo{author}{Diebold, F.X.}, \bibinfo{author}{Yilmaz, K.},
  \bibinfo{year}{2009}.
\newblock \bibinfo{title}{{Measuring financial asset return and volatility
  spillovers, with application to global equity markets}}.
\newblock \bibinfo{journal}{Econ. J.} \bibinfo{volume}{119},
  \bibinfo{pages}{158--171}.
\newblock \DOIprefix\doi{10.1111/j.1468-0297.2008.02208.x}.
\bibitem[{Diebold and Yilmaz(2012)}]{Diebold-Yilmaz-2012-IntJForecast}
\bibinfo{author}{Diebold, F.X.}, \bibinfo{author}{Yilmaz, K.},
  \bibinfo{year}{2012}.
\newblock \bibinfo{title}{Better to give than to receive: {P}redictive
  directional measurement of volatility spillovers}.
\newblock \bibinfo{journal}{Int. J. Forecast.} \bibinfo{volume}{28},
  \bibinfo{pages}{57--66}.
\newblock \DOIprefix\doi{10.1016/j.ijforecast.2011.02.006}.
\bibitem[{Diebold and Yilmaz(2014)}]{Diebold-Yilmaz-2014-JEconom}
\bibinfo{author}{Diebold, F.X.}, \bibinfo{author}{Yilmaz, K.},
  \bibinfo{year}{2014}.
\newblock \bibinfo{title}{On the network topology of variance decompositions:
  {M}easuring the connectedness of financial firms}.
\newblock \bibinfo{journal}{J. Econom.} \bibinfo{volume}{182},
  \bibinfo{pages}{119--134}.
\newblock \DOIprefix\doi{10.1016/j.jeconom.2014.04.012}.
\bibitem[{Diebold and Yilmaz(2023)}]{Diebold-Yilmaz-2023-JEconom}
\bibinfo{author}{Diebold, F.X.}, \bibinfo{author}{Yilmaz, K.},
  \bibinfo{year}{2023}.
\newblock \bibinfo{title}{On the past, present, and future of the
  {D}iebold-{Y}ilmaz approach to dynamic network connectedness}.
\newblock \bibinfo{journal}{J. Econom.} \bibinfo{volume}{234},
  \bibinfo{pages}{115--120}.
\newblock \DOIprefix\doi{10.1016/j.jeconom.2023.01.021}.
\bibitem[{Ding et~al.(2022)Ding, Chen, Zhou and
  Wang}]{Ding-Chen-Zhou-Wang-2022-ExpertSystAppl}
\bibinfo{author}{Ding, Z.}, \bibinfo{author}{Chen, H.}, \bibinfo{author}{Zhou,
  L.}, \bibinfo{author}{Wang, Z.}, \bibinfo{year}{2022}.
\newblock \bibinfo{title}{A forecasting system for deterministic and uncertain
  prediction of air pollution data}.
\newblock \bibinfo{journal}{Expert Syst. Appl.} \bibinfo{volume}{208},
  \bibinfo{pages}{118123}.
\newblock \DOIprefix\doi{10.1016/j.eswa.2022.118123}.
\bibitem[{Engle et~al.(1990)Engle, Ito and
  Lin}]{Engle-Ito-Lin-1990-Econometrica}
\bibinfo{author}{Engle, R.F.}, \bibinfo{author}{Ito, T.}, \bibinfo{author}{Lin,
  W.L.}, \bibinfo{year}{1990}.
\newblock \bibinfo{title}{Meteor showers or heat waves? {H}eteroskedastic
  intra-daily volatility in the foreign exchange market}.
\newblock \bibinfo{journal}{Econometrica} \bibinfo{volume}{58},
  \bibinfo{pages}{525–542}.
\newblock \DOIprefix\doi{10.3386/w2609}.
\bibitem[{Finta and Aboura(2020)}]{Finta-Aboura-2020-JFinancMark}
\bibinfo{author}{Finta, M.A.}, \bibinfo{author}{Aboura, S.},
  \bibinfo{year}{2020}.
\newblock \bibinfo{title}{Risk premium spillovers among stock markets: evidence
  from higher-order moments}.
\newblock \bibinfo{journal}{J. Financ. Mark.} \bibinfo{volume}{49},
  \bibinfo{pages}{100533}.
\newblock \DOIprefix\doi{10.1016/j.finmar.2020.100533}.
\bibitem[{Gavriilidis(2021)}]{Gavriilidis-2021-SSRN}
\bibinfo{author}{Gavriilidis, K.}, \bibinfo{year}{2021}.
\newblock \bibinfo{title}{Measuring climate policy uncertainty}.
\newblock \DOIprefix\doi{10.2139/ssrn.3847388}.
\bibitem[{Giglio et~al.(2016)Giglio, Kelly and
  Pruitt}]{Giglio-Kelly-Pruitt-2016-JFinancEcon}
\bibinfo{author}{Giglio, S.}, \bibinfo{author}{Kelly, B.},
  \bibinfo{author}{Pruitt, S.}, \bibinfo{year}{2016}.
\newblock \bibinfo{title}{Systemic risk and the macroeconomy: {A}n empirical
  evaluation}.
\newblock \bibinfo{journal}{J. Financ. Econ.} \bibinfo{volume}{119},
  \bibinfo{pages}{457--471}.
\newblock \DOIprefix\doi{10.1016/j.jfineco.2016.01.010}.
\bibitem[{Gilbert and
  Morgan(2010)}]{Gilbert-Morgan-2010-PhilosTransRSocB-BiolSci}
\bibinfo{author}{Gilbert, C.L.}, \bibinfo{author}{Morgan, C.W.},
  \bibinfo{year}{2010}.
\newblock \bibinfo{title}{Food price volatility}.
\newblock \bibinfo{journal}{Philos. Trans. R. Soc. B-Biol. Sci.}
  \bibinfo{volume}{365}, \bibinfo{pages}{3023--3034}.
\newblock \DOIprefix\doi{10.1098/rstb.2010.0139}.
\bibitem[{Gu et~al.(2020)Gu, Kelly and Xiu}]{Gu-Kelly-Xiu-2020-RevFinancStud}
\bibinfo{author}{Gu, S.}, \bibinfo{author}{Kelly, B.}, \bibinfo{author}{Xiu,
  D.}, \bibinfo{year}{2020}.
\newblock \bibinfo{title}{Empirical asset pricing via machine learning}.
\newblock \bibinfo{journal}{Rev. Financ. Stud.} \bibinfo{volume}{33},
  \bibinfo{pages}{2223--2273}.
\newblock \DOIprefix\doi{10.1093/rfs/hhaa009}.
\bibitem[{Hanif et~al.(2021)Hanif, Hernandez, Shahzad and
  Yoon}]{Hanif-Hernandez-Shahzad-Yoon-2021-QRevEconFinanc}
\bibinfo{author}{Hanif, W.}, \bibinfo{author}{Hernandez, J.A.},
  \bibinfo{author}{Shahzad, S.J.H.}, \bibinfo{author}{Yoon, S.M.},
  \bibinfo{year}{2021}.
\newblock \bibinfo{title}{Tail dependence risk and spillovers between oil and
  food prices}.
\newblock \bibinfo{journal}{Q. Rev. Econ. Financ.} \bibinfo{volume}{80},
  \bibinfo{pages}{195--209}.
\newblock \DOIprefix\doi{10.1016/j.qref.2021.01.019}.
\bibitem[{Headey and Fan(2008)}]{Headey-Fan-2008-AgricEcon}
\bibinfo{author}{Headey, D.}, \bibinfo{author}{Fan, S.}, \bibinfo{year}{2008}.
\newblock \bibinfo{title}{Anatomy of a crisis: {T}he causes and consequences of
  surging food prices}.
\newblock \bibinfo{journal}{Agric. Econ.} \bibinfo{volume}{39},
  \bibinfo{pages}{375--391}.
\newblock \DOIprefix\doi{10.1111/j.1574-0862.2008.00345.x}.
\bibitem[{Hu et~al.(2012)Hu, Peng and
  Hwang}]{Hu-Peng-Hwang-2012-IEEETransSignalProcess}
\bibinfo{author}{Hu, X.}, \bibinfo{author}{Peng, S.}, \bibinfo{author}{Hwang,
  W.L.}, \bibinfo{year}{2012}.
\newblock \bibinfo{title}{{EMD} revisited: {A} new understanding of the
  envelope and resolving the mode-mixing problem in {AM-FM} signals}.
\newblock \bibinfo{journal}{IEEE Trans. Signal Process.} \bibinfo{volume}{60},
  \bibinfo{pages}{1075--1086}.
\newblock \DOIprefix\doi{10.1109/TSP.2011.2179650}.
\bibitem[{Huang et~al.(1998)Huang, Shen, Long, Wu, Shih, Zheng, Yen, Tung and
  Liu}]{Huang-Shen-Long-Wu-Shih-Zheng-Yen-Tung-Liu-1998-PRSA}
\bibinfo{author}{Huang, N.E.}, \bibinfo{author}{Shen, Z.},
  \bibinfo{author}{Long, S.R.}, \bibinfo{author}{Wu, M.C.},
  \bibinfo{author}{Shih, H.H.}, \bibinfo{author}{Zheng, Q.},
  \bibinfo{author}{Yen, N.C.}, \bibinfo{author}{Tung, C.C.},
  \bibinfo{author}{Liu, H.H.}, \bibinfo{year}{1998}.
\newblock \bibinfo{title}{{The empirical mode decomposition and the Hilbert
  spectrum for nonlinear and non-stationary time series analysis}}.
\newblock \bibinfo{journal}{Proc. R. Soc. Lond. A} \bibinfo{volume}{454},
  \bibinfo{pages}{903--995}.
\newblock \DOIprefix\doi{10.1098/rspa.1998.0193}.
\bibitem[{Huynh et~al.(2020)Huynh, Burggraf and
  Nasir}]{Huynh-Burggraf-Nasir-2020-ResourPolicy}
\bibinfo{author}{Huynh, T.L.D.}, \bibinfo{author}{Burggraf, T.},
  \bibinfo{author}{Nasir, M.A.}, \bibinfo{year}{2020}.
\newblock \bibinfo{title}{Financialisation of natural resources \& instability
  caused by risk transfer in commodity markets}.
\newblock \bibinfo{journal}{Resour. Policy} \bibinfo{volume}{66},
  \bibinfo{pages}{101620}.
\newblock \DOIprefix\doi{10.1016/j.resourpol.2020.101620}.
\bibitem[{Iqbal et~al.(2023)Iqbal, Bouri, Grebinevych and
  Roubaud}]{Iqbal-Bouri-Grebinevych-Roubaud-2023-AnnOperRes}
\bibinfo{author}{Iqbal, N.}, \bibinfo{author}{Bouri, E.},
  \bibinfo{author}{Grebinevych, O.}, \bibinfo{author}{Roubaud, D.},
  \bibinfo{year}{2023}.
\newblock \bibinfo{title}{Modelling extreme risk spillovers in the commodity
  markets around crisis periods including {COVID}19}.
\newblock \bibinfo{journal}{Ann. Oper. Res.} \bibinfo{volume}{330},
  \bibinfo{pages}{305--334}.
\newblock \DOIprefix\doi{10.1007/s10479-022-04522-9}.
\bibitem[{Ji et~al.(2018)Ji, Bouri, Roubaud and
  Shahzad}]{Ji-Bouri-Roubaud-Shahzad-2018-EnergyEcon}
\bibinfo{author}{Ji, Q.}, \bibinfo{author}{Bouri, E.},
  \bibinfo{author}{Roubaud, D.}, \bibinfo{author}{Shahzad, S.J.H.},
  \bibinfo{year}{2018}.
\newblock \bibinfo{title}{Risk spillover between energy and agricultural
  commodity markets: {A} dependence-switching {CoVaR}-copula model}.
\newblock \bibinfo{journal}{Energy Econ.} \bibinfo{volume}{75},
  \bibinfo{pages}{14--27}.
\newblock \DOIprefix\doi{10.1016/j.eneco.2018.08.015}.
\bibitem[{Ji et~al.(2020)Ji, Liu, Zhao and
  Fan}]{Ji-Liu-Zhao-Fan-2020-IntRevFinancAnal}
\bibinfo{author}{Ji, Q.}, \bibinfo{author}{Liu, B.Y.}, \bibinfo{author}{Zhao,
  W.L.}, \bibinfo{author}{Fan, Y.}, \bibinfo{year}{2020}.
\newblock \bibinfo{title}{Modelling dynamic dependence and risk spillover
  between all oil price shocks and stock market returns in the {BRICS}}.
\newblock \bibinfo{journal}{Int. Rev. Financ. Anal.} \bibinfo{volume}{68},
  \bibinfo{pages}{101238}.
\newblock \DOIprefix\doi{10.1016/j.irfa.2018.08.002}.
\bibitem[{Kumar et~al.(2021)Kumar, Tiwari, Raheem and
  Hille}]{Kumar-Tiwari-Raheem-Hille-2021-ResourPolicy}
\bibinfo{author}{Kumar, S.}, \bibinfo{author}{Tiwari, A.K.},
  \bibinfo{author}{Raheem, I.D.}, \bibinfo{author}{Hille, E.},
  \bibinfo{year}{2021}.
\newblock \bibinfo{title}{Time-varying dependence structure between oil and
  agricultural commodity markets: {A} dependence-switching {C}o{V}a{R} copula
  approach}.
\newblock \bibinfo{journal}{Resour. Policy} \bibinfo{volume}{72},
  \bibinfo{pages}{102049}.
\newblock \DOIprefix\doi{10.1016/j.resourpol.2021.102049}.
\bibitem[{Li and Chavas(2023)}]{Li-Chavas-2023-AmJAgrEcon}
\bibinfo{author}{Li, J.}, \bibinfo{author}{Chavas, J.P.}, \bibinfo{year}{2023}.
\newblock \bibinfo{title}{A dynamic analysis of the distribution of commodity
  futures and spot prices}.
\newblock \bibinfo{journal}{Am. J. Agr. Econ.} \bibinfo{volume}{105},
  \bibinfo{pages}{122--143}.
\newblock \DOIprefix\doi{10.1111/ajae.12309}.
\bibitem[{Li and Wei(2018)}]{Li-Wei-2018-EnergyEcon}
\bibinfo{author}{Li, X.}, \bibinfo{author}{Wei, Y.}, \bibinfo{year}{2018}.
\newblock \bibinfo{title}{The dependence and risk spillover between crude oil
  market and {C}hina stock market: {N}ew evidence from a variational mode
  decomposition-based copula method}.
\newblock \bibinfo{journal}{Energy Econ.} \bibinfo{volume}{74},
  \bibinfo{pages}{565--581}.
\newblock \DOIprefix\doi{10.1016/j.eneco.2018.07.011}.
\bibitem[{Lin et~al.(2019)Lin, Chang and
  Hsiao}]{Lin-Chang-Hsiao-2019-TranspResPte-LogistTranspRev}
\bibinfo{author}{Lin, A.J.}, \bibinfo{author}{Chang, H.Y.},
  \bibinfo{author}{Hsiao, J.L.}, \bibinfo{year}{2019}.
\newblock \bibinfo{title}{Does the {B}altic {D}ry {I}ndex drive volatility
  spillovers in the commodities, currency, or stock markets?}
\newblock \bibinfo{journal}{Transp. Res. Part E: Logist. Transp. Rev.}
  \bibinfo{volume}{127}, \bibinfo{pages}{265--283}.
\newblock \DOIprefix\doi{10.1016/j.tre.2019.05.013}.
\bibitem[{Lin and Sim(2013)}]{Lin-Sim-2013-EurEconRev}
\bibinfo{author}{Lin, F.}, \bibinfo{author}{Sim, N.C.S.}, \bibinfo{year}{2013}.
\newblock \bibinfo{title}{Trade, income and the {B}altic {D}ry {I}ndex}.
\newblock \bibinfo{journal}{Eur. Econ. Rev.} \bibinfo{volume}{59},
  \bibinfo{pages}{1--18}.
\newblock \DOIprefix\doi{10.1016/j.euroecorev.2012.12.004}.
\bibitem[{Lundberg et~al.(2020)Lundberg, Erion, Chen, DeGrave, Prutkin, Nair,
  Katz, Himmelfarb, Bansal and
  Lee}]{Lundberg-Erion-Chen-DeGrave-Prutkin-Nair-Katz-Himmelfarb-Bansal-Lee-2020-NatMachIntell}
\bibinfo{author}{Lundberg, S.M.}, \bibinfo{author}{Erion, G.},
  \bibinfo{author}{Chen, H.}, \bibinfo{author}{DeGrave, A.},
  \bibinfo{author}{Prutkin, J.M.}, \bibinfo{author}{Nair, B.},
  \bibinfo{author}{Katz, R.}, \bibinfo{author}{Himmelfarb, J.},
  \bibinfo{author}{Bansal, N.}, \bibinfo{author}{Lee, S.I.},
  \bibinfo{year}{2020}.
\newblock \bibinfo{title}{From local explanations to global understanding with
  explainable {AI} for trees}.
\newblock \bibinfo{journal}{Nat. Mach. Intell.} \bibinfo{volume}{2},
  \bibinfo{pages}{56--67}.
\newblock \DOIprefix\doi{10.1038/s42256-019-0138-9}.
\bibitem[{Luo et~al.(2021)Luo, Liu and Wang}]{Luo-Liu-Wang-2021-NAmEconFinanc}
\bibinfo{author}{Luo, C.}, \bibinfo{author}{Liu, L.}, \bibinfo{author}{Wang,
  D.}, \bibinfo{year}{2021}.
\newblock \bibinfo{title}{Multiscale financial risk contagion between
  international stock markets: {E}vidence from {EMD}-{C}opula-{C}o{V}a{R}
  analysis}.
\newblock \bibinfo{journal}{N. Am. Econ. Financ.} \bibinfo{volume}{58},
  \bibinfo{pages}{101512}.
\newblock \DOIprefix\doi{10.1016/j.najef.2021.101512}.
\bibitem[{Naeem et~al.(2024)Naeem, Chatziantoniou, Gabauer and
  Karim}]{Naeem-Chatziantoniou-Gabauer-Karim-2024-IntRevFinancAnal}
\bibinfo{author}{Naeem, M.A.}, \bibinfo{author}{Chatziantoniou, I.},
  \bibinfo{author}{Gabauer, D.}, \bibinfo{author}{Karim, S.},
  \bibinfo{year}{2024}.
\newblock \bibinfo{title}{Measuring the {G}20 stock market return transmission
  mechanism: Evidence from the {$R^2$} connectedness approach}.
\newblock \bibinfo{journal}{Int. Rev. Financ. Anal.} \bibinfo{volume}{91},
  \bibinfo{pages}{102986}.
\newblock \DOIprefix\doi{10.1016/j.irfa.2023.102986}.
\bibitem[{Noussair et~al.(2016)Noussair, Tucker and
  Xu}]{Noussair-Tucker-Xu-2016-JEconBehavOrgan}
\bibinfo{author}{Noussair, C.N.}, \bibinfo{author}{Tucker, S.},
  \bibinfo{author}{Xu, Y.}, \bibinfo{year}{2016}.
\newblock \bibinfo{title}{Futures markets, cognitive ability, and mispricing in
  experimental asset markets}.
\newblock \bibinfo{journal}{J. Econ. Behav. Organ.} \bibinfo{volume}{130},
  \bibinfo{pages}{166--179}.
\newblock \DOIprefix\doi{10.1016/j.jebo.2016.07.010}.
\bibitem[{Podgorski and Berg(2020)}]{Podgorski-Berg-2020-Science}
\bibinfo{author}{Podgorski, J.}, \bibinfo{author}{Berg, M.},
  \bibinfo{year}{2020}.
\newblock \bibinfo{title}{Global threat of arsenic in groundwater}.
\newblock \bibinfo{journal}{Science} \bibinfo{volume}{368},
  \bibinfo{pages}{845--850}.
\newblock \DOIprefix\doi{10.1126/science.aba1510}.
\bibitem[{Reboredo et~al.(2016)Reboredo, Rivera-Castro and
  Ugolini}]{Reboredo-RiveraCastro-Ugolini-2016-JBankFinanc}
\bibinfo{author}{Reboredo, J.C.}, \bibinfo{author}{Rivera-Castro, M.A.},
  \bibinfo{author}{Ugolini, A.}, \bibinfo{year}{2016}.
\newblock \bibinfo{title}{Downside and upside risk spillovers between exchange
  rates and stock prices}.
\newblock \bibinfo{journal}{J. Bank Financ.} \bibinfo{volume}{62},
  \bibinfo{pages}{76--96}.
\newblock \DOIprefix\doi{10.1016/j.jbankfin.2015.10.011}.
\bibitem[{Tang and Xiong(2012)}]{Tang-Xiong-2012-FinancAnalJ}
\bibinfo{author}{Tang, K.}, \bibinfo{author}{Xiong, W.}, \bibinfo{year}{2012}.
\newblock \bibinfo{title}{Index investment and the financialization of
  commodities}.
\newblock \bibinfo{journal}{Financ. Anal. J.} \bibinfo{volume}{68},
  \bibinfo{pages}{54--74}.
\newblock \DOIprefix\doi{10.2469/faj.v68.n6.5}.
\bibitem[{Tian et~al.(2022)Tian, Alshater and
  Yoon}]{Tian-Alshater-Yoon-2022-EnergyEcon}
\bibinfo{author}{Tian, M.}, \bibinfo{author}{Alshater, M.M.},
  \bibinfo{author}{Yoon, S.M.}, \bibinfo{year}{2022}.
\newblock \bibinfo{title}{Dynamic risk spillovers from oil to stock markets:
  {F}resh evidence from {GARCH} copula quantile regression-based {C}o{V}a{R}
  model}.
\newblock \bibinfo{journal}{Energy Econ.} \bibinfo{volume}{115},
  \bibinfo{pages}{106341}.
\newblock \DOIprefix\doi{10.1016/j.eneco.2022.106341}.
\bibitem[{Tiwari et~al.(2022)Tiwari, Abakah, Adewuyi and
  Lee}]{Tiwari-Abakah-Adewuyi-Lee-2022-EnergyEcon}
\bibinfo{author}{Tiwari, A.K.}, \bibinfo{author}{Abakah, E.J.A.},
  \bibinfo{author}{Adewuyi, A.O.}, \bibinfo{author}{Lee, C.C.},
  \bibinfo{year}{2022}.
\newblock \bibinfo{title}{Quantile risk spillovers between energy and
  agricultural commodity markets: {E}vidence from pre and during {COVID}-19
  outbreak}.
\newblock \bibinfo{journal}{Energy Econ.} \bibinfo{volume}{113},
  \bibinfo{pages}{106235}.
\newblock \DOIprefix\doi{10.1016/j.eneco.2022.106235}.
\bibitem[{Wei et~al.(2023)Wei, Gephart, Iizumi, Ramankutty and
  Davis}]{Wei-Gephart-Iizumi-Ramankutty-Davis-2023-NatSustain}
\bibinfo{author}{Wei, D.}, \bibinfo{author}{Gephart, J.A.},
  \bibinfo{author}{Iizumi, T.}, \bibinfo{author}{Ramankutty, N.},
  \bibinfo{author}{Davis, K.F.}, \bibinfo{year}{2023}.
\newblock \bibinfo{title}{Key role of planted and harvested area fluctuations
  in {US} crop production shocks}.
\newblock \bibinfo{journal}{Nat. Sustain.} \bibinfo{volume}{6},
  \bibinfo{pages}{1177--1185}.
\newblock \DOIprefix\doi{10.1038/s41893-023-01152-2}.
\bibitem[{Wright(2011)}]{Wright-2011-ApplEconPerspectPolicy}
\bibinfo{author}{Wright, B.D.}, \bibinfo{year}{2011}.
\newblock \bibinfo{title}{The economics of grain price volatility}.
\newblock \bibinfo{journal}{Appl. Econ. Perspect. Policy} \bibinfo{volume}{33},
  \bibinfo{pages}{32--58}.
\newblock \DOIprefix\doi{10.1093/aepp/ppq033}.
\bibitem[{Xiao et~al.(2020)Xiao, Yu, Fang and
  Ding}]{Xiao-Yu-Fang-Ding-2020-JFuturesMark}
\bibinfo{author}{Xiao, B.}, \bibinfo{author}{Yu, H.}, \bibinfo{author}{Fang,
  L.}, \bibinfo{author}{Ding, S.}, \bibinfo{year}{2020}.
\newblock \bibinfo{title}{Estimating the connectedness of commodity futures
  using a network approach}.
\newblock \bibinfo{journal}{J. Futures Mark.} \bibinfo{volume}{40},
  \bibinfo{pages}{598--616}.
\newblock \DOIprefix\doi{10.1002/fut.22086}.
\bibitem[{Zhang et~al.(2023)Zhang, Hong and Ding}]{Zhang-Hong-Ding-2023-Energy}
\bibinfo{author}{Zhang, H.}, \bibinfo{author}{Hong, H.}, \bibinfo{author}{Ding,
  S.}, \bibinfo{year}{2023}.
\newblock \bibinfo{title}{The role of climate policy uncertainty on the
  long-term correlation between crude oil and clean energy}.
\newblock \bibinfo{journal}{Energy} \bibinfo{volume}{284},
  \bibinfo{pages}{128529}.
\newblock \DOIprefix\doi{10.1016/j.energy.2023.128529}.
\bibitem[{Zhou et~al.(2024)Zhou, Dai, Duong and
  Dai}]{Zhou-Dai-Duong-Dai-2024-JEconBehavOrgan}
\bibinfo{author}{Zhou, W.X.}, \bibinfo{author}{Dai, Y.S.},
  \bibinfo{author}{Duong, K.T.}, \bibinfo{author}{Dai, P.F.},
  \bibinfo{year}{2024}.
\newblock \bibinfo{title}{The impact of the {R}ussia-{U}kraine conflict on the
  extreme risk spillovers between agricultural futures and spots}.
\newblock \bibinfo{journal}{J. Econ. Behav. Organ.} \bibinfo{volume}{217},
  \bibinfo{pages}{91--111}.
\newblock \DOIprefix\doi{10.1016/j.jebo.2023.11.004}.
\bibitem[{Zhu et~al.(2024)Zhu, Dai and Zhou}]{Zhu-Dai-Zhou-2024-JFuturesMark}
\bibinfo{author}{Zhu, H.Y.}, \bibinfo{author}{Dai, P.F.},
  \bibinfo{author}{Zhou, W.X.}, \bibinfo{year}{2024}.
\newblock \bibinfo{title}{Uncovering the {S}ino-{US} dynamic risk spillovers
  effects: Evidence from agricultural futures markets}.
\newblock \bibinfo{journal}{J. Futures Mark.} \bibinfo{volume}{44},
  \bibinfo{pages}{1888--1910}.
\newblock \DOIprefix\doi{10.1002/fut.22551}.
\bibitem[{Zmami and Ben-Salha(2023)}]{Zmami-BenSalha-2023-AgricEcon}
\bibinfo{author}{Zmami, M.}, \bibinfo{author}{Ben-Salha, O.},
  \bibinfo{year}{2023}.
\newblock \bibinfo{title}{What factors contribute to the volatility of food
  prices? {N}ew global evidence}.
\newblock \bibinfo{journal}{Agric. Econ. - Czech} \bibinfo{volume}{69},
  \bibinfo{pages}{171--184}.
\newblock \DOIprefix\doi{10.17221/99/2023-AGRICECON}.

\end{thebibliography}


\appendix
\section{Decomposition results of return series}
\label{S:Appendix}

\setcounter{figure}{0}

\begin{figure}[!ht]
  \centering
  \includegraphics[width=0.985\linewidth]{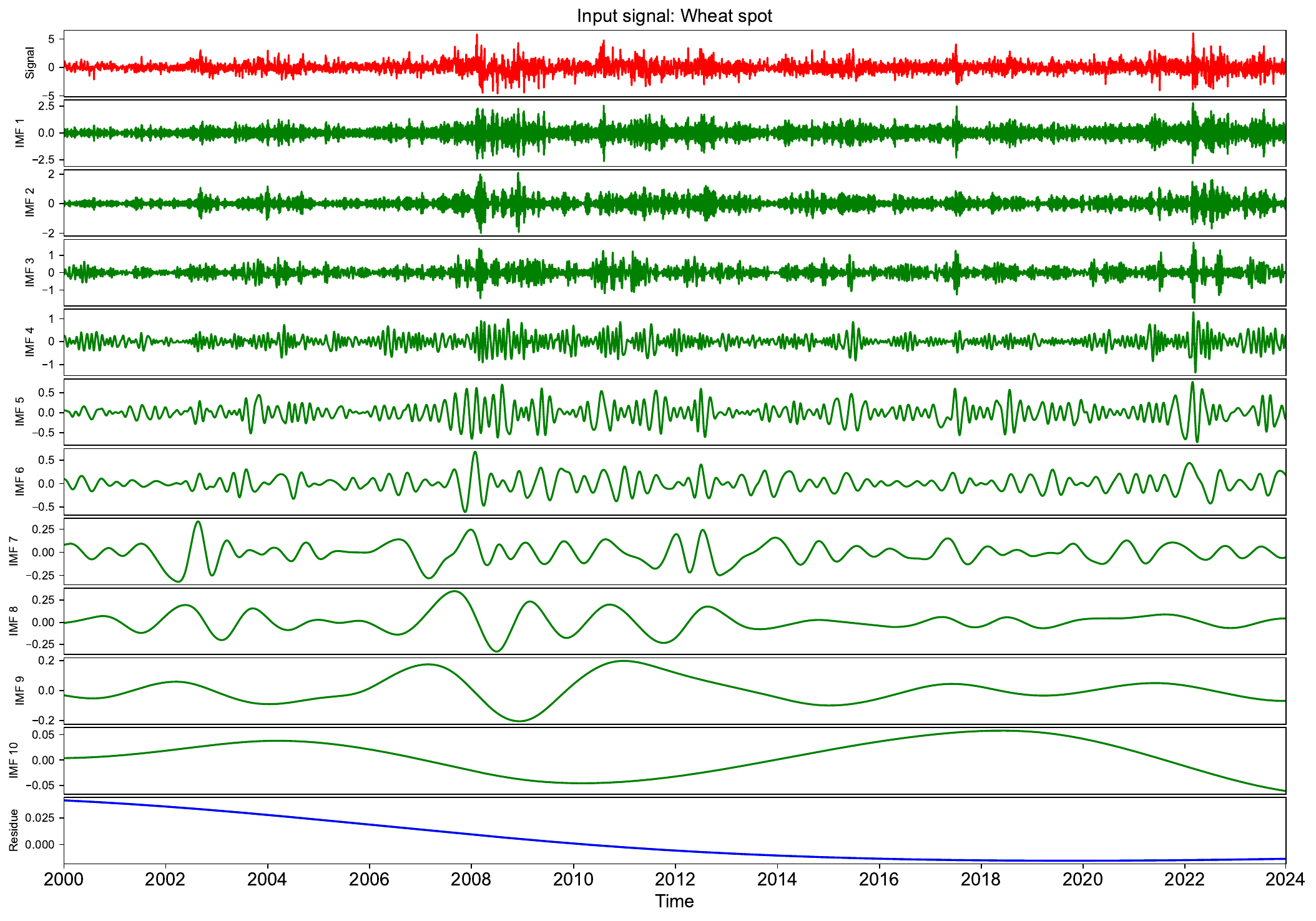}
  \caption{Decomposition results of wheat spot returns.}
\label{Fig:AgroReturn_decomposition_WS}
\end{figure}

\begin{figure}[!ht]
  \centering
  \includegraphics[width=0.985\linewidth]{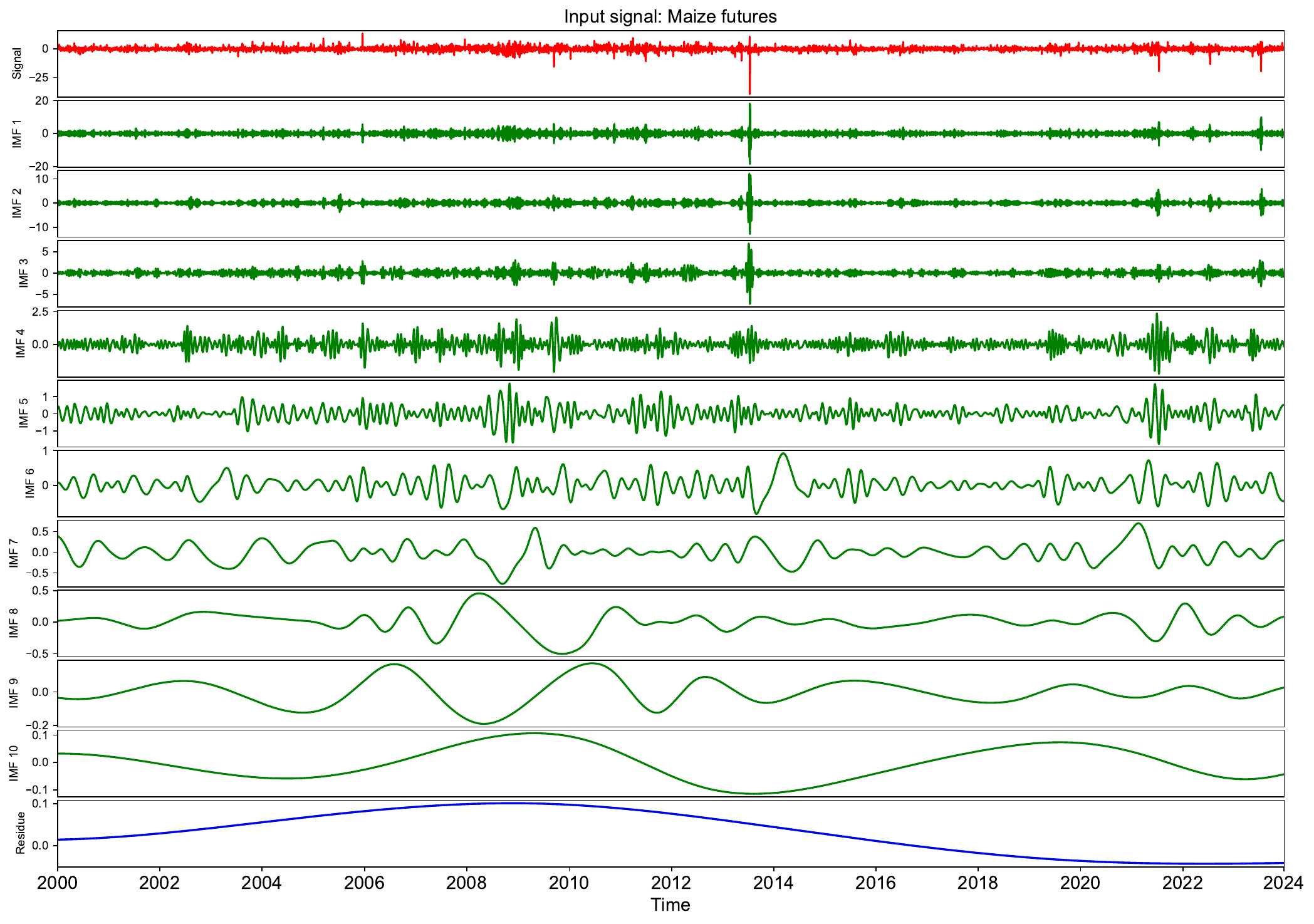}
  \caption{Decomposition results of maize futures returns.}
\label{Fig:AgroReturn_decomposition_MF}
\end{figure}

\begin{figure}[!ht]
  \centering
  \includegraphics[width=0.985\linewidth]{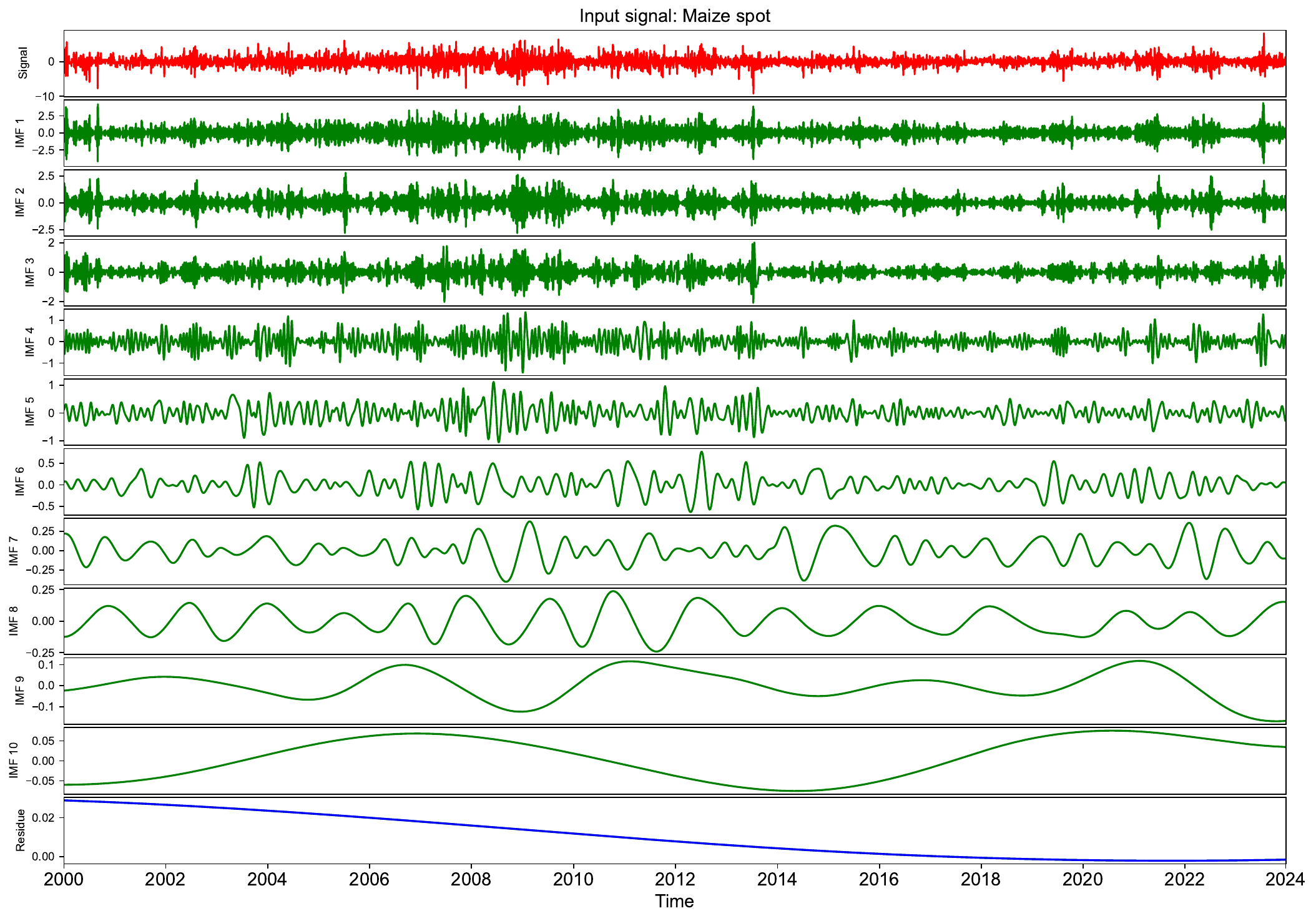}
  \caption{Decomposition results of maize spot returns.}
\label{Fig:AgroReturn_decomposition_MS}
\end{figure}

\begin{figure}[!ht]
  \centering
  \includegraphics[width=0.985\linewidth]{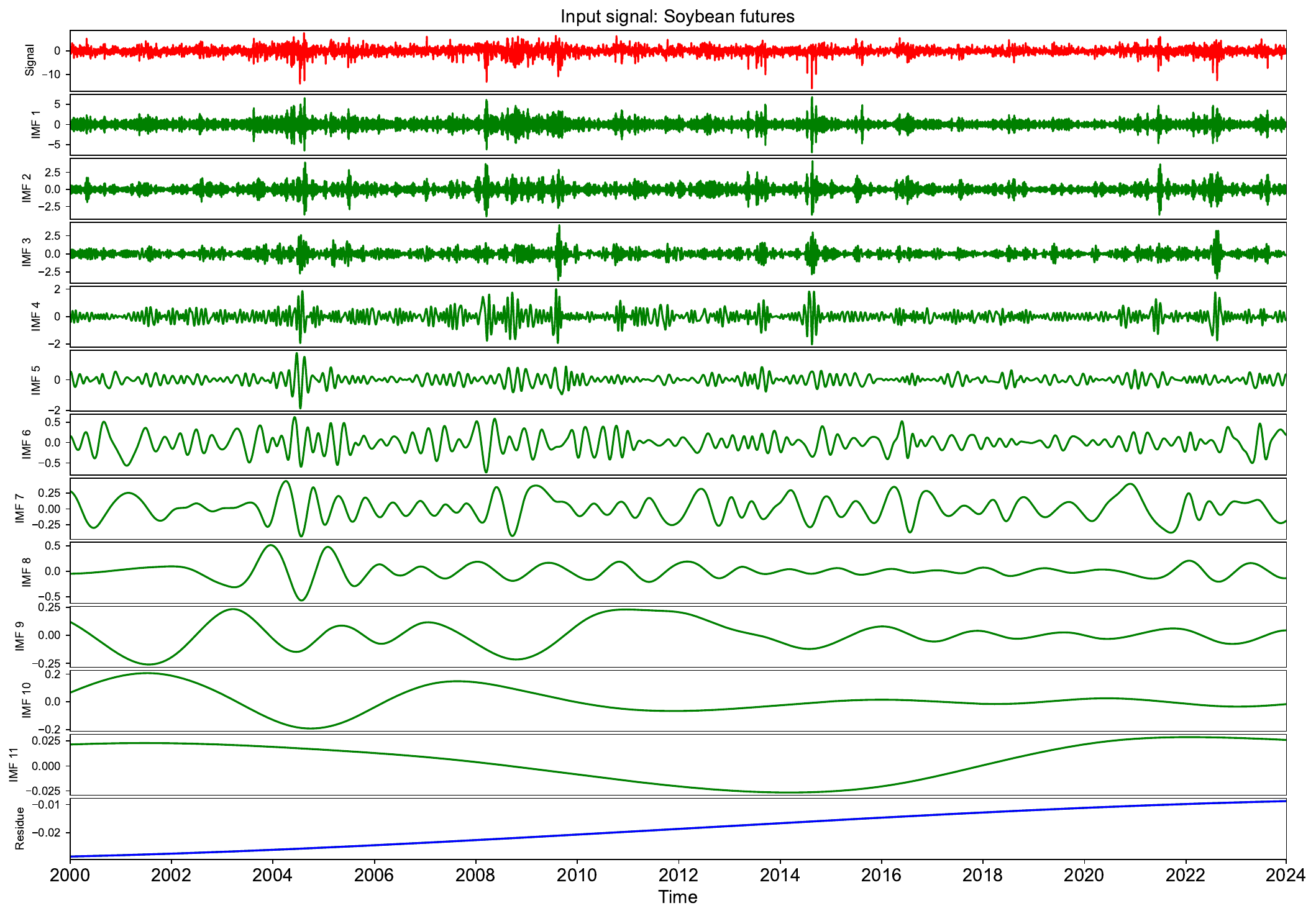}
  \caption{Decomposition results of soybean futures returns.}
\label{Fig:AgroReturn_decomposition_SF}
\end{figure}

\begin{figure}[!ht]
  \centering
  \includegraphics[width=0.985\linewidth]{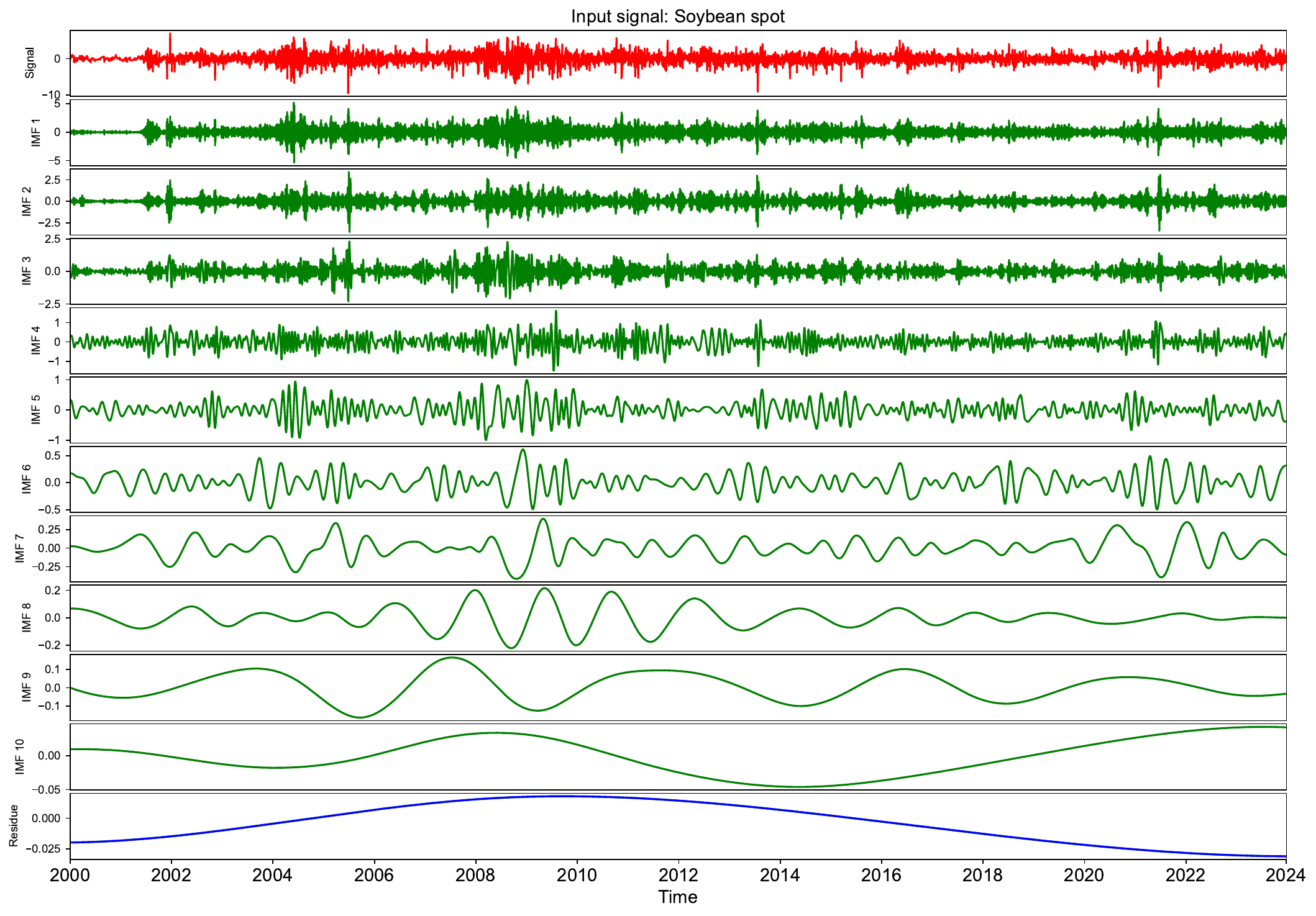}
  \caption{Decomposition results of soybean spot returns.}
\label{Fig:AgroReturn_decomposition_SS}
\end{figure}

\begin{figure}[!ht]
  \centering
  \includegraphics[width=0.985\linewidth]{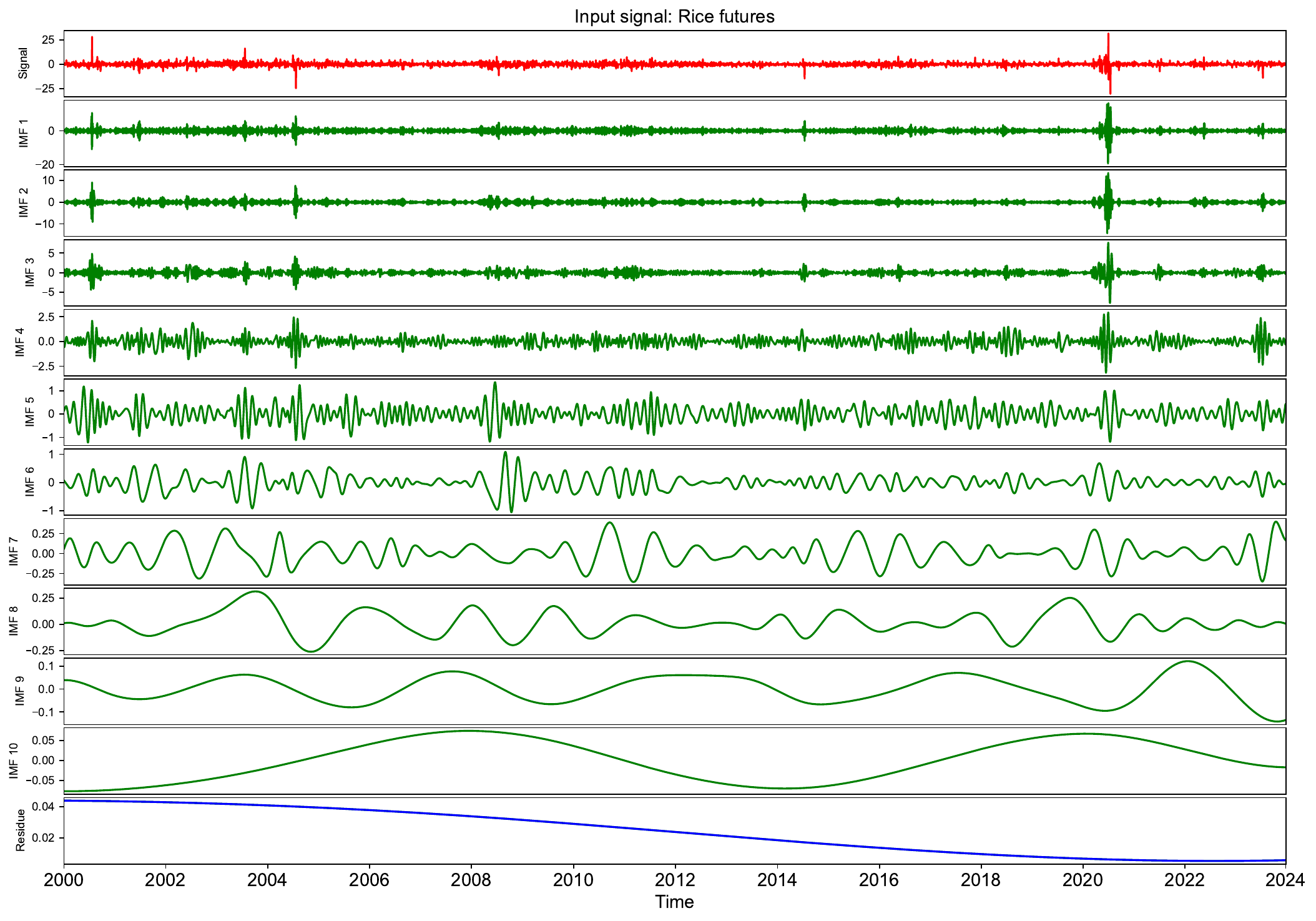}
  \caption{Decomposition results of rice futures returns.}
\label{Fig:AgroReturn_decomposition_RF}
\end{figure}

\begin{figure}[!ht]
  \centering
  \includegraphics[width=0.985\linewidth]{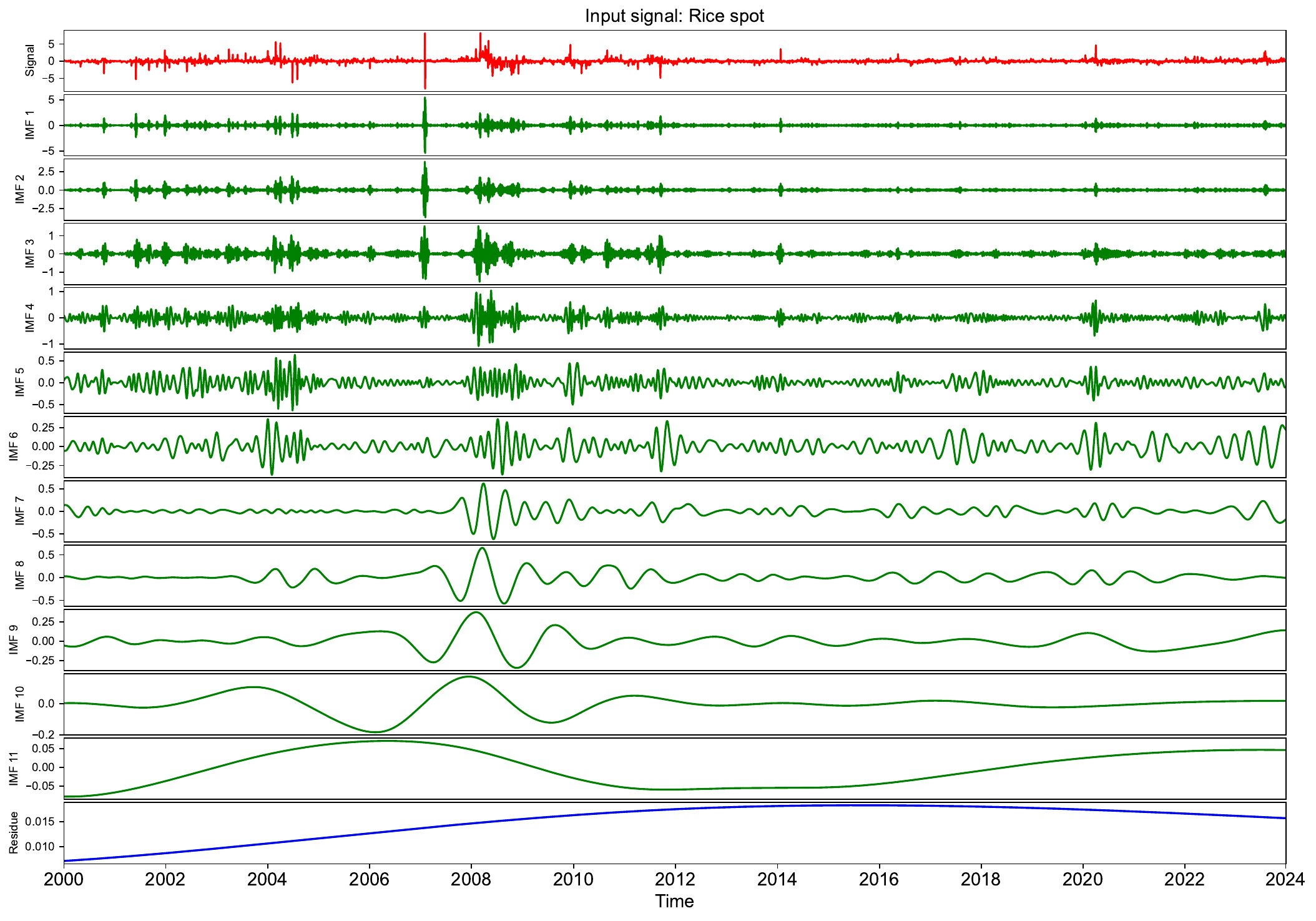}
  \caption{Decomposition results of rice spot returns.}
\label{Fig:AgroReturn_decomposition_RS}
\end{figure}

\end{document}